\documentclass[12pt]{article}
\usepackage{color}
\usepackage{latexsym}
\usepackage{epsfig,amssymb,euscript, mathrsfs}
\usepackage{amsmath}
\definecolor{MyDarkBlue}{rgb}{0.15,0.15,0.45}
\usepackage[linktocpage=true]{hyperref}
\hypersetup{
colorlinks=true,
citecolor=MyDarkBlue,
linkcolor=MyDarkBlue,
urlcolor=MyDarkBlue,
pdfauthor={Luis F Alday, Martin Fluder, Caroline Matte-Gregory, Paul Richmond, James Sparks},
pdftitle={Gravity Dual of Supersymmetric Gauge Theories on a Squashed 5-sphere},
pdfsubject={hep-th}
}
\newsavebox{\ns}
\newsavebox{\dbrane}
\newsavebox{\dbshort}

\def\be{\begin{equation}}
\def\ee{\end{equation}}
\def\bea{\begin{eqnarray}}
\def\eea{\end{eqnarray}}

\newcommand{\nn}{\nonumber}

\newcommand\R{\mathbb{R}}

\newcommand\C{\mathbb{C}}

\newcommand\diff{\mathrm{d}}

\newcommand{\dd}{\mathrm{d}}

\newcommand{\ii}{\mathrm{i}}

\newcommand{\ex}{\mathrm{e}}
\newcommand{\vol}{\mathrm{vol}}

\newcommand{\Real}{\mathrm{Re}\, }

\newlength{\sswidth}
\newcommand{\sla}[1]{
   \settowidth{\sswidth}{$#1$}
   \mbox{$\not{\hspace*{-0.3\sswidth}#1}$}}

\newcommand{\Free}{\mathcal{F}}
\newcommand{\elly}{\mathscr{L}}
\newcommand{\NN}{\mathcal{N}}
\newcommand{\MM}{\mathcal{M}}

\numberwithin{equation}{section}       

\begin{document}

\bibliographystyle{utphys}

\begin{titlepage}

\begin{center}

\today

\vskip 2.3 cm 

\vskip 5mm

{\Large \bf Supersymmetric gauge theories on squashed five-spheres and their gravity duals}

\vskip 15mm

{Luis F. Alday, Martin Fluder, Carolina Matte Gregory,}\\
{ Paul Richmond and James Sparks}

\vspace{1cm}
\centerline{{\it Mathematical Institute, University of Oxford,}}
\centerline{{\it Andrew Wiles Building, Radcliffe Observatory Quarter,}}
\centerline{{\it Woodstock Road, Oxford, OX2 6GG, UK}}

\end{center}

\vskip 2 cm

\begin{abstract}
\noindent  We construct the gravity duals of large $N$ supersymmetric gauge theories defined 
on squashed five-spheres with $SU(3)\times U(1)$ symmetry. These five-sphere backgrounds 
are continuously connected to the round sphere, and we find a one-parameter family of 3/4 BPS deformations 
and a two-parameter family of (generically) 1/4 BPS deformations. The gravity duals are 
constructed in Euclidean Romans $F(4)$ gauged supergravity in six dimensions, and uplift to massive type IIA supergravity. 
We holographically renormalize the Romans theory, and use our general result to compute the renormalized on-shell actions 
for the solutions. The results agree perfectly with the large $N$ limit of the dual gauge theory partition function, which we 
compute using large $N$ matrix model techniques. In addition we compute BPS Wilson loops in these backgrounds, 
both in supergravity and in the large $N$ matrix model, again finding precise agreement. Finally, we 
conjecture a general formula for the partition function on 
any five-sphere background, which for fixed gauge theory depends only on a certain supersymmetric Killing vector.
\end{abstract}

\end{titlepage}

\pagestyle{plain}
\setcounter{page}{1}
\newcounter{bean}
\baselineskip18pt
\tableofcontents

\allowdisplaybreaks


\section{Introduction}\label{sec:intro}

Over the last few years there has been increasing interest in defining and studying 
supersymmetric gauge theories on curved backgrounds. Such constructions lead
to interesting classes of observables that can be computed exactly, which may in turn 
be used to test and explore conjectured dualities. In this paper we focus on
the case of five-dimensional gauge theories. These have been defined 
on round spheres \cite{Kallen:2012cs, Hosomichi:2012ek, Kallen:2012va, Kim:2012ava, Jafferis:2012iv}, 
as well as on certain continuous deformations thereof \cite{Imamura:2012xg, Imamura:2012bm}, referred to as squashed five-spheres.  The main observable  
that can be computed exactly in these theories is the partition function $Z$, which depends 
non-trivially on the background geometry. A particular class of  five-dimensional  superconformal gauge theories, with gauge group $USp(2N)$ and arising from a $D4-D8$-system, is expected to have a large $N$ description in terms of massive type IIA supergravity \cite{Ferrara:1998gv, Brandhuber:1999np, Bergman:2012kr}.  In \cite{Jafferis:2012iv} the large $N$ limit of the partition function of these theories on the {\it round} sphere was computed and successfully compared to the entanglement entropy of the dual warped AdS$_6 \times S^4$ supergravity solution. 

In this  paper we shall present the first construction of gravity 
duals to gauge theories on non-conformally flat backgrounds (specifically, 
certain families of squashed five-spheres). As we shall explain, 
we may effectively work in six-dimensional Romans $F(4)$ supergravity 
\cite{Romans:1985tw}, which is a consistent truncation of massive IIA supergravity 
on $S^4$ \cite{Cvetic:1999un}. In particular the computation of \cite{Jafferis:2012iv} effectively 
determines the six-dimensional Newton constant. 
 Having constructed supergravity 
solutions that have squashed five-sphere conformal boundaries, 
we compute the holographic free energy $\Free=-\log Z$ by 
holographically renormalizing the on-shell Euclidean action. 
More specifically, we construct families of solutions with different numbers of preserved supercharges. Two of these families are shown to be dual to the 1/4 BPS and 3/4 BPS gauge theories defined in \cite{Imamura:2012bm}.  The perturbative partition function for these theories has been computed in \cite{Imamura:2012xg} and we explicitly show that the large $N$ limit of these partition functions is in precise agreement with the holographic free energies of  our supergravity solutions. We also present more general solutions (and in particular a 1/2 BPS solution) which have  not previously been considered from the gauge theory side. 

From the Killing spinors of a supersymmetric supergravity solution one can always construct a certain Killing vector $K$. For all solutions found in this paper the free energy is only sensitive to this Killing vector $\Free=\Free(K)$, and not to other parameters of the solution. It is natural to conjecture that this is also the case for more general solutions, extending what happens in four dimensions \cite{Farquet:2014kma}. 
In addition we compute the expectation values of BPS Wilson loops in these backgrounds, 
both in supergravity and in the large $N$ matrix model, finding precise agreement. Again the expectation value depends only on the Killing vector $K$.

The rest of this paper is organized as follows. In section \ref{sec:gaugetheory} we discuss supersymmetric gauge theories defined on squashed five-spheres, their exact partition function and the large $N$ limit. 
In section \ref{SecRomansSUGRA} we change focus and describe the Romans $F(4)$ supergravity theory we will work with. 
Then in section \ref{SUGRA} we present our supergravity solutions dual to the squashed five-sphere backgrounds. 
In section \ref{SecFreeEnergy} we apply holographic renormalization to the Romans $F(4)$ supergravity theory and use this to compute the holographic free energy of our solutions. 
In section \ref{SecSUSYatBoundary} we examine the supersymmetry  conditions which arise at the conformal boundary for the Romans supergravity theory.
Another exact observable that can be computed both in supersymmetric gauge theories and in supergravity are Wilson loops, which 
are the subject of 
 section \ref{Wilson}. 
Finally, we end in section \ref{Fernandosection} with some discussion and possible future problems to explore. We also include appendices \ref{AppSUSYConditions}, \ref{AppSolutions} and \ref{asymptotes}, which expand upon some of the elements in the main body of the paper.


\section{Supersymmetric gauge theories on squashed five-spheres}\label{sec:gaugetheory}

We begin in section \ref{sec:squashed} by describing the squashed five-sphere backgrounds of interest \cite{Imamura:2012xg}. 
One can define a supersymmetric gauge theory with general matter content on such a background, 
and in \cite{Imamura:2012bm} the perturbative partition function was computed via a twisted reduction 
of the supersymmetric index in six dimensions\footnote{See also \cite{Lockhart:2012vp}.}, that we summarize in section  \ref{sec:twisted}.
 A particular class of five-dimensional gauge theories, with gauge group 
$USp(2N)$ and arising from a $D4-D8$ system in massive type IIA string theory, 
is expected to have a large $N$ limit with a gravity dual. In section \ref{sec:largeN} 
we compute the large $N$ limit of the partition function for these theories using 
matrix model techniques.

\subsection{$SU(3)\times U(1)$ squashed five-sphere}\label{sec:squashed}

The squashed $S^5$ backgrounds of interest are homogeneous spaces with 
 symmetry $SU(3)\times U(1)$. 
In particular this is the isometry group of the metric
\bea\label{S5}
\diff s^2_5 &=& \frac{1}{s^2}(\diff\tau+C)^2+ 
\diff\sigma^2 + \frac{1}{4}\sin^2\sigma(\diff\theta^2+\sin^2\theta\dd\varphi^2)\nn\\
&&+\frac{1}{4}\cos^2\sigma\sin^2\sigma (\dd\psi+\cos\theta\dd\varphi)^2~,
\eea
where we have defined the (local) one-form
\bea\label{C}
C &=& - \frac{1}{2}\sin^2\sigma(\dd\psi+\cos\theta\dd\varphi)~.
\eea
We refer to the parameter $s$ as a squashing parameter, and note that $s=1$ is the round sphere. 
The coordinates in (\ref{S5}) realize the five-sphere as the total space of the Hopf circle bundle 
over $\mathbb{CP}^2$, where $\tau$ is a $2\pi$-period coordinate along the circle fibre.  
The coordinates $\sigma, \psi, \theta, \varphi$ are 
then coordinates on the base $\mathbb{CP}^2$, with $\psi$ having period $4\pi$, 
$\varphi$ having period $2\pi$,  while $\sigma\in [0,\tfrac{\pi}{2}]$, $\theta\in[0,\pi]$. 
The local one-form $C$ in (\ref{C}) satisfies
\bea\label{dC}
\diff C &\equiv & 2\omega~ \ = \ -\sin\sigma\cos\sigma\diff\sigma\wedge (\dd\psi+\cos\theta\dd\varphi)+\frac{1}{2}\sin^2\sigma\sin\theta\dd\theta\wedge\dd\varphi,
\eea
where $\omega$ is the K\"ahler two-form on $\mathbb{CP}^2$.

In order to preserve supersymmetry one must also turn on other backgrounds fields. 
In particular in \cite{Imamura:2012xg} it was shown that one can define 
general supersymmetric gauge theories on the above squashed five-sphere, provided 
one turns on a background $SU(2)_R$ gauge field
\bea\label{AR}
\mathcal{A} &=& \frac{(1+Q\sqrt{1-s^2})\sqrt{1-s^2}}{s^2}(\diff \tau +C)~,
\eea
where we have embedded $U(1)_R\subset SU(2)_R$. More precisely, 
writing the $SU(2)_R\sim SO(3)_R$ gauge field as a triplet of one-forms $\mathcal{A}^i$, 
$i=1,2,3$, we have $\mathcal{A}^1=\mathcal{A}^2=0$, while $\mathcal{A}^3=\mathcal{A}$ is given by 
(\ref{AR}).
For supersymmetric backgrounds the parameter $Q$ takes the values $Q=1$ and $Q=-3$, which lead
to 3/4 BPS and 1/4 BPS solutions, respectively. Notice that the gauge field (\ref{AR}) is also invariant 
under $SU(3)\times U(1)$, and
is real when $|s|<1$ but complex for $|s|>1$. 

A supersymmetric background of course admits an appropriate Killing spinor,
which then enters the supersymmetry transformations of a supersymmetric 
gauge theory defined on the background.
Recall that a Killing spinor $\chi$ on the round $S^5$ with $s=1$, 
solving $\nabla_m\chi = -\tfrac{\ii}{2}\gamma_m\chi$ 
where $\gamma_m$ generate the Clifford algebra 
Cliff$(5,0)$ in an orthonormal frame, transforms in the 
$\mathbf{4}$ of the $SU(4)\sim SO(6)$ isometry. 
The squashing breaks this symmetry to $SU(3)\times U(1)$, 
and for $Q=1$ the resulting Killing spinor transforms 
as $\mathbf{3_{+1}}$, while for $Q=-3$ the resulting 
Killing spinor instead transforms as $\mathbf{1_{-3}}$. 
Similarly, solutions to $\nabla_m\chi = \tfrac{\ii}{2}\gamma_m\chi$ transform in the 
$\mathbf{\bar{4}}$ of $SU(4)$, which is broken to  $\mathbf{\bar{3}_{-1}}$ and $\mathbf{1_{+3}}$
in the two cases, respectively. 

The corresponding  Killing spinor equation for the squashed $S^5$ was obtained in \cite{Imamura:2012xg} 
via a twisted reduction (described in the next subsection) of a standard Killing spinor 
equation in six dimensions. In order to write this down, we first introduce an orthonormal frame
for the metric (\ref{S5})
\bea\label{S5frame}
e^1_{(5)} &=& \frac{1}{s}(\diff\tau+C)~, \qquad e^2_{(5)} \ = \ \diff\sigma~, \qquad e^3_{(5)} \ = \ \frac{1}{2}\sin\sigma\cos\sigma \tau_3~,\nn\\
e^4_{(5)} & = & \frac{1}{2}\sin\sigma\tau_2~, \qquad e^5_{(5)} \ = \ \frac{1}{2}\sin\sigma\tau_1~,
\eea
where $\tau_i$, $i=1,2,3$, are left-invariant one-forms on $SU(2)$. These are parametrized in terms of the Euler angles as
\bea\label{leftinvariant}
\tau_1+\ii\tau_2 &=& \ex^{-\ii\psi}(\diff\theta+\ii\sin\theta\diff\varphi)~, \qquad \tau_3 \ = \ \diff\psi+\cos\theta\diff\varphi~.
\eea
The Killing spinor equation then reads
\bea\label{5dKSE}
\nabla_m\chi_I + \tfrac{\ii}{2}\mathcal{A}^i_m(\sigma^i)_I^{\ J}\chi_J & =&  -\frac{\ii\left(1+Q\sqrt{1-s^2}\right)}{2s}(\sigma^3)_I^{\ J}\gamma_m\chi_J \nn\\&&+\frac{\sqrt{1-s^2}}{4s}\left(3\gamma_m\sla\omega - \sla\omega\gamma_m\right)\chi_I~,
\eea
which is supplemented by the following algebraic equation
\bea\label{5ddilatino}
Q\sqrt{1-s^2}\chi_I &=& -\sqrt{1-s^2}\gamma_1\chi_I-\ii \sqrt{1-s^2}(\sigma^3)_I^{\ J}\sla{\omega}\chi_J~.
\eea
Here $\chi_I$, $I=1,2$, form a doublet under the $SU(2)_R$ symmetry, $\gamma_m$ generate the Clifford algebra 
Cliff$(5,0)$ in the orthonormal frame (\ref{S5frame}), and $(\sigma^i)_I^{\ J}$ denote the Pauli matrices. 
Recall also that $\omega$ denotes the K\"ahler form on $\mathbb{CP}^2$, given by (\ref{dC}), 
and if $\alpha$ is a $p$-form we denote $\sla{\alpha}\equiv \frac{1}{p!}\alpha_{m_1\cdots m_p}\gamma^{m_1\cdots m_p}$.

Of course in the case at hand we have that the $SU(2)_R$ gauge field $\mathcal{A}^i$ is 
only turned on in the $i=3$ direction, with $\mathcal{A}^3=\mathcal{A}$ given by (\ref{AR}), and we may also write 
(\ref{5dKSE}) and (\ref{5ddilatino}) as
\begin{align}\label{5dKSEpm}
\nabla_m\chi_\pm \pm \tfrac{\ii}{2}\mathcal{A}_m\chi_\pm \ =& \ \mp \frac{\ii\left(1+Q\sqrt{1-s^2}\right)}{2s}\gamma_m\chi_\pm +\frac{\sqrt{1-s^2}}{4s}\left(3\gamma_m\sla\omega - \sla\omega\gamma_m\right)\chi_\pm~,\\\label{5ddilatinopm}
Q\sqrt{1-s^2}\chi_\pm \ =& \  -\sqrt{1-s^2}\gamma_1\chi_\pm\mp\ii \sqrt{1-s^2}\sla{\omega}\chi_\pm~,
\end{align}
where $\chi_+=\chi_1$, $\chi_-=\chi_2$. Provided the background fields are 
real, meaning in particular that the metric and $\mathcal{A}$ are real and 
$|s|<1$, then notice that the equations for $\chi_-$ are simply the charge 
conjugates of the $\chi_+$ equations, where we define the charge conjugate as
\bea
\chi^c & \equiv & \mathcal{C}_5\chi^*~,
\eea
and the charge conjugation matrix $\mathcal{C}_5$ satisfies $\mathcal{C}_5^{-1}\gamma_m\mathcal{C}_5=\gamma_m^*$. In particular it is then 
consistent to impose the symplectic Majorana condition $\chi_-=\chi_+^c$, or equivalently $\varepsilon_{I}^{\ J}\chi_J = \mathcal{C}_5\chi_I^*$, as we shall see below. 

Notice that in setting $s=1$ to obtain the round sphere one has that (\ref{5ddilatino}) is trivially satisfied, while 
the Killing spinor equation (\ref{5dKSE}) implies that $\chi_1$ and $\chi_2$ transform in the 
$\mathbf{4}$ and $\mathbf{\bar{4}}$ of the enhanced $SU(4)\sim SO(6)$ symmetry, respectively. 
In order to present the general solution to (\ref{5dKSE}), (\ref{5ddilatino}) (which is not written in \cite{Imamura:2012xg}), 
we first introduce the following basis of Cliff$(5,0)$
\bea
\gamma_1 \ = \  \left(\begin{array}{cc}1_2 & 0\\  0 & -1_2 \end{array}\right)~,\quad \ \ \, \gamma_2 &=& \left(\begin{array}{cc}0 & 1_2\\ 1_2& 0\end{array}\right)~,\quad \gamma_3 \ = \ \left(\begin{array}{cc}0 & \ii \sigma^3 \\ -\ii \sigma^3 & 0\end{array}\right)~,\nn \\
 \quad \gamma_4 \ = \ \left(\begin{array}{cc} 0 & \ii \sigma^2 \\ -\ii \sigma^2 & 0\end{array}\right)~, \quad \gamma_5 &=& \left(\begin{array}{cc}0 & \ii \sigma^1 \\ -\ii \sigma^1 & 0\end{array}\right)~,
\eea
where as above $\sigma^i$, $i=1,2,3$ denote the Pauli matrices, and $1_2$ is the $2\times 2$ identity matrix. A choice of the charge conjugation matrix in this basis is 
\bea
\mathcal{C}_5 &=&  \left(\begin{array}{cc}-\ii \sigma_2 & 0\\  0 & -\ii \sigma_2 \end{array}\right)~.
\eea
Then for the 1/4 BPS background we find the general solution to (\ref{5dKSE}), (\ref{5ddilatino})  (or equivalently  (\ref{5dKSEpm}), (\ref{5ddilatinopm})) is given by
\bea
\label{5dKS1/4}
\chi_+ &=& c_+\ex^{-\tfrac{3\ii\tau}{2}} \left(\begin{array}{c}0 \\ 1 \\ 0 \\ 0\end{array}\right)~,\qquad \chi_-  \ = \ c_- \ex^{\tfrac{3\ii\tau}{2}}\left(\begin{array}{c}-1 \\ 0 \\ 0 \\ 0\end{array}\right)~,
\eea
where $c_\pm$ are integration constants. In particular then notice that the symplectic 
Majorana condition $\chi_-=\chi_+^{c}$ simply imposes $c_-=c_+^*$.

For the 3/4 BPS background the solution is a little more complicated. 
One finds
\bea
\label{5dchip}
\chi_+ & =&  a^{(1)}_+ \, \ex^{\ii \frac{\tau}{2}} \left( \begin{array}{c} 
\cos \sigma + \ii \lambda_+ (s) \ex^{\ii \frac{\psi}{2}} S_{+}^{(1)} \sin \sigma\\ 0 \\ \ii \lambda_- (s) \sin \sigma - \ex^{\ii \frac{\psi}{2}} S_{+}^{(1)} \cos \sigma \\ -\ii \ex^{-\ii \frac{\psi}{2}} S_{+}^{(2)} \end{array}\right)~,
\eea
where
\bea
 S_{\pm}^{(1)} \ = \ S^{(1)}_{\pm}(\theta, \varphi) &= &a_\pm^{(3)} \ex^{\pm \ii \frac{\varphi}{2}}\cos \frac{\theta}{2} - a_\pm^{(2)} \ex^{\mp \ii \frac{\varphi}{2}} \sin \frac{\theta}{2}~, \nn\\
S_{\pm}^{(2)}  \ = \ S^{(2)}_{\pm}(\theta, \varphi) &= & a^{(2)}_{\pm} \ex^{\mp \ii \frac{\varphi}{2}} \cos\frac{\theta}{2}+a_{\pm}^{(3)} \ex^{\pm \ii \frac{\varphi}{2}} \sin \frac{\theta}{2}~,
\eea
and where we have introduced $ \lambda_\pm (s) \equiv  (\pm 1 + \sqrt{1-s^{2}} )/s$. As expected, the solution depends on three integration constants $a^{(1)}_+,a^{(2)}_+,a^{(3)}_+$. Similarly, one finds
\bea
\label{5dchim}
 \chi_- & =&  a^{(1)}_- \, \ex^{- \ii \frac{\tau}{2}} \left( \begin{array}{c} 0 \\ \cos \sigma - \ii \lambda_+ (s) \ex^{-\ii \frac{\psi}{2}} S_{-}^{(1)} \sin \sigma \\  - \ii \ex^{\ii \frac{\psi}{2}} S_{-}^{(2)} \\ - \ii \lambda_- (s) \sin \sigma - \ex^{- \ii \frac{\psi}{2}} S_{-}^{(1)} \cos \sigma\end{array}\right)~,
\eea
where $a^{(i)}_-$ are integration constants. One can once again impose the symplectic Majorana condition, which leads to the relation  $(a_-^{(i)} )^{*}=a_+^{(i)}$ for $i=1,2,3$. 

\subsection{Twisted reduction and the partition function}\label{sec:twisted}

The backgrounds above may be obtained via 
a twisted reduction of $ \R\times S^5$, starting from the 
\emph{round} metric on $S^5$. This is important, 
as the perturbative partition function on the squashed five-spheres was computed in 
\cite{Imamura:2012bm} indirectly, by taking a limit 
of the supersymmetric index of a corresponding six-dimensional 
theory on $ \R\times  S^5$. 

We thus begin with the product metric on $\R$ times the round $S^5$ 
\bea\label{S5R}
\diff s^2_{ \R\times S^5} &=& \diff t^2+\sum_{i=1}^3|\diff w_i|^2~,
\eea
where the complex coordinates $w_i$ on $\C^3\cong \R^6$, $i=1,2,3$, satisfy the constraint
$\sum_{i=1}^3|w_i|^2=1$. We then compactify this space by identifying
\bea\label{twist}
(t,w_i)\sim (t+\beta,\ex^{\ii \mu_i\beta}w_i)~,
\eea
where $\beta>0$ and the $\mu_i$ are also sometimes referred to as squashing parameters. 
Notice that (\ref{twist}) is an isometry for $\mu_i\in\R$. We may then change coordinates
\bea
\rho_i\ex^{\ii\varphi_i} &\equiv & \ex^{-\ii \mu_i t}w_i~, 
\eea
where $\rho_i\geq 0$ and the $\varphi_i$ have period $2\pi$. In terms of these new coordinates 
the identification (\ref{twist}) reads $(t,\rho_i,\varphi_i)\sim(t+\beta,\rho_i,\varphi_i)$. We then dimensionally 
reduce along the $t$-direction to obtain the five-dimensional metric
\bea\label{squashedS5general}
\diff s^2_5&=& \sum_{i=1}^3(\diff\rho_i^2+\rho_i^2\diff\varphi_i^2)-\frac{1}{1+\sum_{i=1}^3\mu_i^2\rho_i^2}\left(\sum_{i=1}^3\mu_i\rho_i^2\diff\varphi_i\right)^2~.
\eea
Notice that, via the constraint $\sum_{i=1}^3\rho_i^2=1$, the first term in (\ref{squashedS5general}) 
is the round metric on $S^5$.

One then makes contact with the previous section by choosing
\bea\label{mus}
-\mu_1 &=& \mu_2 \ = \ \mu_3 \ = \ \ii\sqrt{1-s^2}~,\qquad\ \ \, \quad \mbox{3/4 BPS}~,\nn\\
\mu_1 &=& \mu_2 \ = \ \mu_3 \ = \ -\ii\sqrt{1-s^2}~,\qquad \quad \mbox{1/4 BPS}~.
\eea
Notice these are real only if $| s | \geq 1$. The metric (\ref{squashedS5general}) then agrees with the 
metric (\ref{S5}) on making the standard polar coordinate identifications
\bea\label{rhos}
\rho_1 &=& \cos\sigma~, \qquad \rho_2 \ = \ \sin\sigma\cos\frac{\theta}{2}~, \qquad  \rho_3 \ = \ \sin\sigma\sin\frac{\theta}{2}~,
\eea
together with
\bea\label{varphis}
\varphi_1 &=& -\tau~, \quad \varphi_2 \ = \ \tau - \frac{1}{2}(\psi+\varphi)~, \quad \varphi_3 \ = \ \tau - \frac{1}{2}(\psi-\varphi)~, \quad \mbox{3/4 BPS}~,\nn\\
\varphi_1 &=& \tau~, \quad\ \ \, \varphi_2 \ = \ \tau - \frac{1}{2}(\psi+\varphi)~, \quad \varphi_3 \ = \ \tau - \frac{1}{2}(\psi-\varphi)~, \quad \mbox{1/4 BPS}~.
\eea
The Killing spinor equation (\ref{5dKSE}) and algebraic equation (\ref{5ddilatino}) were then obtained 
in \cite{Imamura:2012xg} by dimensionally reducing a standard Killing spinor equation on 
the $\R\times S^5$ background (\ref{S5R}). 

In practice the perturbative contribution to the squashed $S^5$ partition function, with more general squashed metric 
(\ref{squashedS5general}), was computed in \cite{Imamura:2012bm} by dimensionally reducing 
the superconformal index of a corresponding six-dimensional theory on the $\R\times S^5$ background (\ref{S5R}) with 
twisted identification (\ref{twist}), and then taking the limit $\beta\rightarrow 0$, so that the radius of the 
circle we reduced on to obtain (\ref{squashedS5general}) is sent to zero. For a gauge theory with gauge group $G$, 
prepotential $\mathscr{F}$, which is a cubic polynomial in the scalar $\sigma$ in the vector multiplet, 
and matter in the real representation $\mathbf{R}\oplus\mathbf{\bar{R}}$ of $G$, 
the result is
\bea
{Z}_{\text{pert}} \, = \,  C(\mathbf{b})\prod_{a =1}^{\mathrm{rank} \ G}  \int_{-\infty}^{\infty} \mathrm{d} \sigma_a \, \ex^{ - \frac{(2 \pi)^{3}}{b_1 b_2 b_3} \mathscr{F}(\sigma)}\frac{ \prod_\alpha S_3 \left( - \ii \alpha (\sigma) \mid \mathbf{b} \right)}{\prod_\rho S_3 \left( - \ii \rho (\sigma) + \tfrac{1}{2}(b_1+b_2+b_3)
\mid \mathbf{b}  \right)} \, . \label{Z}
\eea
Here we have introduced
\bea\label{bdef}
\mathbf{b} &=& (b_1,b_2,b_3)~, \qquad \mbox{where} \qquad b_i \ = \ 1 + \ii \mu_i~,
\eea
and the prefactor $C(\mathbf{b})$ in (\ref{Z}) depends only on $(b_1,b_2,b_3)$, and 
in particular will not contribute to the large $N$ limit of interest in the next section.\footnote{The precise
formula for $C(\mathbf{b})$ may be found in \cite{Imamura:2012bm}.}  The perturbative partition function 
thus localizes onto field configurations in which the only non-zero field is a constant mode for the scalar $\sigma$ in the vector multiplet, 
and this is then integrated over in (\ref{Z}). As usual in such expressions the product over $\alpha$ in the numerator 
is over roots of $G$, while the product over $\rho$ in the denominator is over weights 
in a weight space decomposition of $\mathbf{R}$. Finally,
$S_3 \left( z \mid \mathbf{b} \right)$ is the triple sine function, which is a special case of the multiple sine functions defined by
\begin{align}
S_\NN \left( z \mid \mathbf{b} \right) \ \equiv \ & \, \Gamma_{\NN} (z \mid \mathbf{b})^{-1} \ \Gamma_{\NN} (b_{\text{tot}} - z \mid \mathbf{b})^{(-1)^{\NN}}\\
=\ & \prod_{n_1, \ldots, n_\NN=0}^{\infty} \left[ \sum_{i=1}^{\NN} n_i b_i +z \right] \prod_{n_1, \ldots, n_\NN =1}^{\infty} \left[\sum_{i=1}^{\NN} n_ib_i - z\right]^{(-1)^{\NN-1}} \, ,
\end{align}
where we have written $\mathbf{b}=(b_1,\ldots, b_\NN)$ and defined 
$b_{\text{tot}}= \sum_{i=1}^{\NN} b_i$. The function $ \Gamma_{\NN} (z \mid \mathbf{b})$ is the so-called Barnes' multiple gamma function
\bea
 \Gamma_{\NN} (z \mid \mathbf{b})& \equiv & \prod_{n_1, \ldots, n_\NN=0}^{\infty} \left[ \sum_{i=1}^{\NN} n_i b_i +z \right]^{-1} ~.
\eea

We conclude this section by noting from (\ref{mus}) and (\ref{bdef}) that for the $SU(3)\times U(1)$ squashed five-spheres
in section \ref{sec:squashed}
\bea\label{bsquashed}
b_1 &=& 1+ \sqrt{1-s^2}~, \qquad b_2 \ = \ b_3 \ = \ 1-\sqrt{1-s^2}~, \qquad \mbox{3/4 BPS}~,\nn\\
b_1 &=& b_2 \ = \ b_3 \ = \ 1 + \sqrt{1-s^2}~, \qquad \qquad \qquad   \qquad  \qquad  \mbox{1/4 BPS}~.
\eea
In particular it is straightforward to see \cite{Imamura:2012bm} that in the 1/4 BPS case the perturbative partition function (\ref{Z}) is 
independent of the squashing parameter $s$. 

It is interesting to note that (\ref{twist}) is an isometry of the original six-dimensional $\R\times S^5$ background
only for real $\mu_i$, which via (\ref{mus}) one sees corresponds to $|s|\geq 1$. 
On the other hand from (\ref{bsquashed}) we see that the parameters $b_i$ 
are real (and then positive) only if $|s|\leq 1$. The dual six-dimensional supergravity backgrounds 
we shall construct in section \ref{SUGRA} will correspondingly  be real for $|s|\leq 1$.

\subsection{The large $N$ limit}\label{sec:largeN}

The result for the perturbative partition function (\ref{Z}) 
in the previous section is valid for a general supersymmetric gauge theory in five dimensions, 
but we  now focus on a 
particular class of theories with gauge group 
$G=USp(2N)$, that arises from a system of $N$ D4-branes and some number of D8-branes and orientifold 
planes in massive type IIA string theory. 
These theories are expected to have a large $N$ limit that has a dual description in massive type IIA supergravity
\cite{Ferrara:1998gv, Brandhuber:1999np, Bergman:2012kr}. Indeed, 
in \cite{Jafferis:2012iv} the large $N$
limit of the partition function of these theories on the
\emph{round} five-sphere was computed and successfully compared to
the entanglement entropy of the dual warped AdS$_6\times S^4$
supergravity solution. Here the gauge theories flow to a UV superconformal 
fixed point, and 
in particular the localization computation in the IR supersymmetric Yang-Mills 
theory coupled to matter theory  successfully reproduces
the expected $N^{5/2}$ scaling of the number of degrees of freedom.

In general one certainly expects non-perturbative contributions to the full partition function $Z$, in addition 
 to the perturbative 
result (\ref{Z}). In particular in the localization computation of \cite{Kallen:2012va} on the 
round five-sphere
one finds that the gauge multiplet localizes onto instanton configurations on $\mathbb{CP}^2$. 
There is thus a non-perturbative contribution to $Z$ involving 
a sum over the instanton number. For fixed instanton number $n\neq 0$ and fixed choice of instanton, in addition to the classical 
instanton action there will also be one-loop 
determinant contributions around that instanton, plus an integral over the instanton moduli space with fixed $n$. 
In general this expression will be very difficult to evaluate. However, in \cite{Jafferis:2012iv} it was argued that 
in the large $N$ limit these instanton contributions should be suppressed. We shall also assume this to be the case
on the squashed five-sphere, although clearly this issue deserves further study. In particular, 
for general choice of the vector $\mathbf{b}=(b_1,b_2,b_3)$ we expect 
to find instantons not  on $\mathbb{CP}^2$, but rather instantons 
transverse to the Killing vector $K=\sum_{i=1}^3b_i\partial_{\varphi_i}$, as in  \cite{Qiu:2013pta}. 
 These \emph{contact instantons} 
were discussed in the latter reference in the context of the partition function on Sasaki-Einstein manifolds.
In any case, we leave this issue open for future investigation.

Our task thus reduces to computing the large $N$ limit of the perturbative result (\ref{Z}), for the 
$USp(2N)$ gauge theories of interest. This may be carried out using the matrix model saddle point method
originally introduced in  \cite{Herzog:2010hf}, and subsequently applied to the round ${S}^{5}$ partition function in \cite{Jafferis:2012iv}. As in the latter reference, we  also set the Chern-Simons level 
for the theory $k=0$ (thus setting the cubic terms in the prepotential $\mathscr{F}(\sigma)$ to zero). 
The quadratic and linear terms of $\mathscr{F}(\sigma)$ will only contribute to subleading order in the large $N$ limit. This is because the leading contribution to the free energy arises from the scaling $\sigma=\mathcal{O}(N^{1/2})$. Such a behaviour for $\sigma$ leads to an $\mathcal{O}(N^{2})$ contribution for the classical parts in the perturbative partition function \eqref{Z}. Thus in the limit of large $N$ we only have to analyse the behaviour of the two one-loop determinants from the vector and matter multiplets. In particular, for a given theory we will have to find the expansion of the logarithm of the triple sine function entering (\ref{Z}).

The $USp(2N)$ gauge theories have $N_f$ matter fields in the fundamental and a single hypermultiplet in the antisymmetric representation of the gauge group.
 Let us denote an element in the Cartan subalgebra for $USp(2N)$ as $\left\{\lambda_1, \ldots, \lambda_N \right\}$, 
 so that $\sigma=\mathrm{diag}(\lambda_1,\ldots,\lambda_N,-\lambda_1,\ldots,-\lambda_N)$. The Weyl group 
 acts as $\lambda_i\rightarrow -\lambda_i$ for each $i$, and also permutes the $\lambda_i$. 
   If the normalized weights of the fundamental representation are given by $\pm e_i$, where $\left\{e_1, \ldots, e_N \right\}$ is a basis of $\mathbb{R}^{N}$, then the antisymmetric representation has 
weights $\left\{e_i \pm e_j\right\}_{i \neq j}$ and the adjoint representation has weights $\left\{e_i \pm e_j\right\}_{i \neq j}\cup \left\{\pm 2 e_i\right\}_{i=1}^{N}$. Therefore we can write the free energy for this theory as
\begin{align}
 \Free(\lambda_i) \, = \, & \sum_{\substack{i,j=1 \\ i \neq j}}^{N} G_{V} (\lambda_i + \lambda_j \mid \mathbf{b})+G_{V} (\lambda_i - \lambda_j \mid  \mathbf{b})+G_{H} (\lambda_i + \lambda_j \mid  \mathbf{b})+G_{H} (\lambda_i - \lambda_j \mid  \mathbf{b})  \nn \\
 \, + & \sum_{i=1}^{N} G_{V} \left( 2 \lambda_i\mid \mathbf{b} \right) + G_{V} \left(- 2 \lambda_i\mid  \mathbf{b} \right) + N_f \left[ G_{H}\left(\lambda_i\mid \mathbf{b} \right) + G_{H}\left( - \lambda_i\mid \mathbf{b} \right)\right] \, , \label{freeenergy}
\end{align}
where $G_V$ and $G_H$ are the logarithms of the triple sine functions in the numerator and denominator of \eqref{Z} for the vector and the hypermultiplets, respectively. We are interested in their asymptotics for large $\lambda_i$ only, because we assume that the eigenvalues scale with $N^{\alpha}$ for some $\alpha >0$. These asymptotics are explicitly computed in appendix \ref{asymptotes}, and here we simply quote the results:
\begin{align}
 G_{V} (x \mid  \mathbf{b})+G_{V} (-x \mid \mathbf{b}) \ = \ & - \log  S_3 \left( - \ii x \mid  \mathbf{b} \right)-\log  S_3 \left( \ii x \mid  \mathbf{b} \right)  \nn \\ 
 \sim \ & \, \frac{\pi}{3 \, b_1b_2b_3} |x|^{3} - \frac{\pi \left(b_{\text{tot}}^{2} +b_1 b_2+b_1 b_3+b_2b_3 \right)}{6 \,b_1b_2b_3} |x| \, , \label{expansionvector}
\end{align}
where we have expanded in the limit $|x| \rightarrow \infty$. Here we have assumed that $b_i>0$ for each $i=1,2,3$, 
as this is the case of interest -- see equation (\ref{bsquashed}) and the discussion after it.
Similarly, for the free energy contribution of the hypermultiplet we obtain 
\begin{equation}
G_{H} (x \mid  \mathbf{b}) \ = \ \log S_3 \left(\tfrac{1}{2}b_{\text{tot}} - \ii x 
\mid  \mathbf{b} \right) \  \sim \  \, - \frac{\pi}{6 \, b_1b_2b_3} |x|^{3} - \frac{\pi \left( b^{2}_1+b^{2}_2 +b^{2}_3 \right)}{24 \, b_1b_2b_3} |x|~, \label{expansionhyper}
\end{equation}
in the asymptotic limit $|x| \rightarrow \infty$.

Using the Weyl symmetry of $USp(2N)$ we may take 
 $\lambda_i \geq 0$, and we shall furthermore
assume that these eigenvalues scale as $\lambda_i = N^{\alpha} x_i$ to leading order in the large $N$ limit, with $\alpha >0$. We next introduce the density 
\bea
\rho (x) &=&  \frac{1}{N} \sum^{N}_{i=1} \delta \left( x-x_i \right) ~,
\eea
which becomes an $\mathcal{L}^{1}$ function with 
\be
\int \rho(x) \mathrm{d} x \ = \ 1~, \label{rhoint1}
\ee
once we take $N \rightarrow \infty$. In that limit, the discrete sums in \eqref{freeenergy} become Riemann integrals 
\bea
 \frac{1}{N} \sum_{i=1}^{N} \  \longrightarrow  \ \int_{0}^{x_{\star}} \rho(x) \mathrm{d} x ~.
 \eea
Hence taking the large $N$ limit of \eqref{freeenergy}, we obtain to leading order
\bea
 \Free &\approx & \ N^{2} \int_{0}^{x_\star} \rho(x) \int_{0}^{x_\star} \rho(y) \Big[ G_{V} (\lambda(x) \pm \lambda(y) \mid \mathbf{b})+G_{H} (\lambda(x) \pm \lambda(y) \mid  \mathbf{b}) \Big] \mathrm{d}y \, \mathrm{d}x  \nn \\
 &&+ N \int_{0}^{x_\star} \rho (x) \Big[ G_{V} (\pm 2 \lambda(x) \mid  \mathbf{b}) +N_f \, G_{H} (\pm \lambda(x) \mid \mathbf{b})   \Big]\mathrm{d}x \, . \label{fapproxint}
\eea
By assumption we have $\lambda (x)= N^{\alpha} x$ to leading order in the continuum limit, and hence we may use the above expansions for the vector and hypermultiplet contributions \eqref{expansionvector}, \eqref{expansionhyper} respectively. Then the leading order term in the first line of \eqref{fapproxint} scales as $N^{2 + \alpha}$, because the cubic terms in the asymptotic expansion of $G_H$ and $G_V$ cancel. The leading order term of the second line in \eqref{fapproxint} however does not cancel, and is given by $N^{1+3 \alpha}$. In order to obtain a non-trivial saddle point, both terms must contribute and we deduce that $\alpha = 1/2$. Putting everything together we obtain
\bea
\Free & = &  \ - N^{5/2} \int_{0}^{x_\star} \rho(x) \int_{0}^{x_\star} \rho(y)  \Bigg[ \frac{\pi b_{\text{tot}}^{2}}{8  b_1 b_2 b_3} \left( |x+y|+|x-y| \right) \nn\\
&& - \frac{(8-N_f) \pi}{3 \ b_1b_2b_3} |x|^{3}\Bigg] \mathrm{d}y \ \mathrm{d}x+ \mathcal{O}\left( N^{3/2} \right) \, .
\eea

It thus remains to solve a  simple variational problem for $\rho (x)$ extremizing the free energy. We add a Lagrange multiplier term to impose the constraint \eqref{rhoint1}, namely $\mu \left( \int_0^{x_\star} \rho(x) \mathrm{d}x-1 \right)$, 
and then solve $\frac{\partial F}{\partial \rho} = 0$ for $\rho (x)$. Doing so we find (with $N_f<8$)
\bea\label{rho}
 \rho (x) & =&  \frac{4 (8-N_f)}{b_{\text{tot}}^{2}} |x| \, ,
\eea
inside the interval $[0,x_\star]$, with $\rho$ identically zero outside this interval, and where
extremizing $\Free$ over the end-point $x_\star$ gives
\bea\label{xstar}
x_\star^{2} &= &\frac{b_{\text{tot}}^{2}}{2 (8 - N_f)} \, .
\eea
We may then evaluate the free energy by substituting these saddle point configurations back into \eqref{fapproxint} to obtain
\bea
 \Free & = & - \frac{\sqrt{2} \pi b_{\text{tot}}^{3}}{15 \sqrt{8-N_f} \ b_1b_2b_3} N^{5/2} + \mathcal{O}\left( N^{3/2} \right) ~,
\eea
which may be rewritten as (where recall we have assumed that $b_i>0$ for each $i=1,2,3$)
\bea \Free & = & \  \frac{ (b_1+b_2+b_3)^3}{27 b_1 b_2 b_3} \Free_{{S}^{5}_{\mathrm{round}}} \, , \label{freeenergyfinal}
\eea
where $\Free_{{S}^{5}_{\mathrm{round}} }$ is the large $N$ limit of the free energy on the round five-sphere computed in reference \cite{Jafferis:2012iv}
\bea
 \Free_{{S}^{5}_{\mathrm{round}} } & = &-\frac{9 \sqrt{2} \pi N^{5/2}}{5 \, \sqrt{8-N_f}} + \mathcal{O}\left( N^{3/2} \right) \, .
\eea
We note that the above result has a very similar structure to that obtained in three dimensions \cite{Martelli:2012sz}. Also notice that we get the same result, \eqref{freeenergyfinal}, for the orbifold theories discussed in \cite{Jafferis:2012iv, Bergman:2012kr}.

We conclude this section by noting that for the $SU(3)\times U(1)$ squashed five-spheres, with 
the vector $\mathbf{b}=(b_1,b_2,b_3)$ given by (\ref{bsquashed}), we obtain
the large $N$ free energies
\bea\label{Fernando}
\Free &=& \begin{cases}\ \displaystyle \frac{1}{27s^2}\frac{(3-\sqrt{1-s^2})^3}{1-\sqrt{1-s^2}}\, \Free_{{S}^{5}_{\mathrm{round}} }~, & \qquad \quad \mbox{3/4 BPS \, ,}
 \\ \  \Free_{{S}^{5}_{\mathrm{round}} }~, & \qquad \quad \mbox{1/4 BPS \, .} \end{cases}~
\eea


\section{Romans $F(4)$ supergravity}\label{SecRomansSUGRA}

When the $USp(2N)$ superconformal theories discussed in section \ref{sec:gaugetheory} are put on the 
round ${S}^{5}$, they are conjectured to be dual in the large $N$ limit to the AdS$_6\times {S}^4$ 
solution of massive type IIA supergravity \cite{Ferrara:1998gv, Brandhuber:1999np, Bergman:2012kr}. 
In order to find gravity duals to the same superconformal theories put on different background 
five-manifolds, it is then natural to work in the six-dimenional Romans $F(4)$ supergravity theory 
 \cite{Romans:1985tw}. The key here is that, as shown in  \cite{Cvetic:1999un}, the Romans theory is a consistent 
 truncation of massive type IIA supergravity on ${S}^4$. In the next subsection we 
 shall review this uplift to ten dimensions, and then present the Romans theory in Euclidean 
 signature in section \ref{sec:euclidean}.  

\subsection{Uplift to massive type IIA}\label{sec:uplift}

The Romans theory \cite{Romans:1985tw} is a six-dimensional gauged supergravity that admits an AdS$_6$ vacuum. 
The bosonic fields  consist of the metric, a dilaton $\phi$, a two-form 
potential $B$, a one-form potential $A$, together with an $SU(2)\sim SO(3)$ gauge field $A^i$, $i=1,2,3$. 
It is convenient to introduce the scalar field $X\equiv \exp(-\phi/2\sqrt{2})$, and  
we define the field strengths as $H=\diff B$, $F = \dd A + \frac{2}{3}g B$, $F^i = \diff A^i-\frac{1}{2}g\varepsilon_{ijk}A^j\wedge A^k$. 
Here $g$ denotes the gauge coupling constant. Notice that $B$ appears in the field strength for $A$. 

As shown in \cite{Cvetic:1999un}, this Romans theory is a consistent 
 truncation of massive type IIA supergravity on ${S}^4$. This means that 
 any solution to the Romans theory automatically uplifts, via the non-linear Kaluza-Klein 
 ansatz of \cite{Cvetic:1999un} presented in (\ref{uplift}) below, to a solution of massive type IIA. Moreover, 
 the AdS$_6\times {S}^4$ solution of the latter is the uplift of the AdS$_6$ vacuum 
of the Romans theory.

We shall later need some details of how the six-dimensional solutions uplift to ten dimensions. 
The gauge coupling constant $g$ is related to the ten-dimensional 
mass parameter by $m_{\mathrm{IIA}}=\frac{\sqrt{2}}{3}g$, while the 
remaining fields uplift via
\bea
\diff s^2_{10}&=& (\sin\xi)^{\frac{1}{12}}X^{\frac{1}{8}}\left[\Delta^{\frac{3}{8}}\diff s^2_6 + 
2g^{-2}\Delta^{\frac{3}{8}}X^2\diff\xi^2+\tfrac{1}{2}g^{-2}\Delta^{-\frac{5}{8}}X^{-1}\cos^2\xi\sum_{i=1}^3
(\hat\tau^i-g A^i)^2\right]~,\nn\\
F_{(4)} &=&   -\tfrac{\sqrt2}{6} g^{-3} s^{1/3} c^3 \Delta^{-2}
U \, \dd\xi\wedge\vol_3 -\sqrt2 g^{-3} s^{4/3} c^4 \Delta^{-2}
X^{-3} \, \dd X\wedge \vol_3 \nn\\
&&+\sqrt2 g^{-1} 
s^{1/3} c X^4\, {*H}\wedge \dd\xi
-\tfrac{1}{\sqrt2} s^{4/3} X^{-2} {*F}+\tfrac{1}{\sqrt2} g^{-2}
s^{1/3} c\, F^i  h^i\wedge \dd\xi  \nn\\
&&  -\tfrac{1}{4\sqrt2} g^{-2}
s^{4/3} c^2 \Delta^{-1} X^{-3}  F^i \wedge
h^j\wedge h^k\, \varepsilon_{ijk}~,\nn\\
F_{(3)} &=& s^{2/3} H + g^{-1} s^{-1/3} c \, F\wedge \dd\xi~,\nn\\
F_{(2)} &=& \tfrac{1}{\sqrt2} s^{2/3} F~, \qquad
\ex^{\Phi} \ = \  s^{-5/6} \Delta^{1/4} X^{-5/4}~,
\label{uplift}
\eea
where
\bea
\Delta &\equiv & X\cos^2\xi +X^{-3} \sin^2 \xi~,\nn\\
U &\equiv & X^{-6} s^2 - 3 X^2 c^2 + 4 X^{-2} c^2 - 6 X^{-2}~.
\eea
Here $\diff s^2_{10}$ is the ten-dimensional metric in Einstein frame, $\Phi$ is the ten-dimensional dilaton, $F_{(3)}$ is the NS-NS three-form 
field strength, while $F_{(4)}$ and $F_{(2)}$ are the RR four-form and two-form field
strengths, respectively. The $\hat\tau^i$, $i=1,2,3$, are left-invariant one-forms 
on a copy of $SU(2)\cong S^3$. These are defined precisely as in (\ref{leftinvariant}), 
except here this $S^3$ is in the internal space (hence the hats).
We have also defined $h^i\equiv \hat\tau^i-gA^i$, 
$\vol_3\equiv h^1\wedge h^2\wedge h^3$, and $s=\sin\xi$ and $c=\cos\xi$.
The Hodge duals in (\ref{uplift}) are computed with respect to the six-dimensional 
metric $\diff s^2_6$. This is defined on some six-manifold $M_6$, 
and the ten-dimensional metric in (\ref{uplift}) then describes a 
warped product $M_6\times S^4$. More precisely, the solution 
only describes ``half'' of a four-sphere, where 
 the coordinate $\xi\in (0,\tfrac{\pi}{2}]$ is a polar coordinate for which constant 
 $\xi\in (0,\tfrac{\pi}{2})$ slices are three-spheres, parametrized by Euler angles on $S^3$ as in 
 (\ref{leftinvariant}). 
 The solution is smooth at the north pole $\xi=\tfrac{\pi}{2}$, where 
the $S^3$ slices of $S^4$ collapse to zero size, but singular 
on the equator $\xi=0$. Nevertheless, it is argued in \cite{Brandhuber:1999np,Bergman:2012kr} that the supergravity solution (\ref{uplift}) can be trusted away from this singularity.

\subsection{Euclidean theory}\label{sec:euclidean}

The equations of motion and action for the Romans theory in Lorentz signature appear in 
 \cite{Romans:1985tw, Cvetic:1999un}. However, the gravity duals to the large $N$ field theories 
on the squashed five-sphere of section \ref{sec:gaugetheory} will be constructed in Euclidean 
signature. The corresponding 
 Wick rotation is not entirely straightforward because the Romans theory contains Chern-Simons-type 
 couplings, that become purely imaginary in Euclidean signature in order that the theory is gauge invariant. 
 The associated factors of $\ii$ are also crucial for supersymmetry in Euclidean signature.
The Euclidean equations of motion for the Romans supergravity fields are
\bea\label{FullEOM}
\diff\left(X^4 * H\right) &=& \tfrac{\ii}{2}F\wedge F + \tfrac{\ii}{2}F^i\wedge F^i +\tfrac{2}{3} g X^{-2}*F~,\nn\\
\diff(X^{-2}*F) &=& - \ii F\wedge H~, \nn \\
D(X^{-2}*F^i) & = & - \ii F^i\wedge H~,\nn\\
\diff\left(X^{-1}*\dd X\right) &=& - g^2 \left(\tfrac{1}{6}X^{-6}-\tfrac{2}{3}X^{-2}+\tfrac{1}{2}X^2\right)*1 \nn \\
&&-\tfrac{1}{8}X^{-2}\left(F\wedge *F+F^i\wedge *F^i\right) + \tfrac{1}{4}X^4H\wedge *H ~.
\eea
Here $D\omega^i=\dd\omega^i - g\varepsilon_{ijk}A^j\wedge \omega^k$ is the 
$SO(3)$ covariant derivative, and our convention for the Hodge duality operator is fixed
via
\bea
\alpha \wedge * \beta &=&  \frac{1}{p!} \alpha_{\mu_1 \cdots \mu_p} \beta^{\mu_1 \cdots \mu_p} * 1~,
\eea
where $\alpha$ and $\beta$ are $p$-forms.\footnote{In particular this convention 
differs from that in \cite{Cvetic:1999un}.} 
The Einstein equation is
\bea\label{Einstein}
R_{\mu\nu} &=& 4X^{-2}\partial_\mu X\partial_\nu X + g^2\left(\tfrac{1}{18}X^{-6}-\tfrac{2}{3}X^{-2}-\tfrac{1}{2}X^2\right) g_{\mu\nu} + \tfrac{1}{4}X^4\left(H^2_{\mu\nu}-\tfrac{1}{6}H^2g_{\mu\nu}\right) \nn\\
&& +  \tfrac{1}{2}X^{-2}\left(F^2_{\mu\nu}-\tfrac{1}{8}F^2g_{\mu\nu}\right) +  \tfrac{1}{2}X^{-2}\left((F^i)^2_{\mu\nu}-\tfrac{1}{8}(F^i)^2g_{\mu\nu}\right)~,
\eea
where $F^2_{\mu\nu} = F_{\mu\rho} F_\nu{}^\rho$, $H^2_{\mu\nu}=H_{\mu\rho\sigma}H_{\nu}^{\ \rho\sigma}$.

The Euclidean action which gives rise to these field equations is
\begin{align}\label{Fullaction}
I_E \ =  \ -\frac{1}{16\pi G_N}\int \Big[ &R*1-4X^{-2}\diff X\wedge *\diff X -g^2\left(\tfrac{2}{9}X^{-6}-\tfrac{8}{3}X^{-2}-2X^2\right)*1\nn\\
&-\tfrac{1}{2}X^{-2} \left(F\wedge * F+ F^i \wedge * F^i \right)-\tfrac{1}{2}X^4H\wedge *H \\ 
&- \ii B \wedge \big( \tfrac{1}{2} \dd A \wedge \dd A + \tfrac{1}{3} B \wedge \dd A +\tfrac{2}{27} g^2 B\wedge B + \tfrac{1}{2} F^i \wedge F^i \big)\Big]~.\nn
\end{align}
In particular notice that the final term is a Chern-Simons-type coupling, and is 
accompanied by a factor of $\ii$. This is required for gauge-invariance in
the path integral with Euclidean measure $\exp(-I_E)$. It is also implied by 
supersymmetry. Indeed, a solution to the above equations of motion is 
supersymmetric provided the following Killing spinor equation and dilatino 
equation hold:
\bea
D_\mu \epsilon_I & =&  \frac{\ii}{4\sqrt{2}} g ( X + \tfrac{1}{3} X^{-3} ) \Gamma_\mu \Gamma_7 \epsilon_I - \frac{\ii}{16\sqrt{2}} X^{-1} F_{\nu\rho} ( \Gamma_\mu{}^{\nu\rho} - 6 \delta_\mu{}^\nu \Gamma^\rho ) \epsilon_I \label{KSE}  \\
&& - \frac{1}{48} X^2 H_{\nu\rho\sigma} \Gamma^{\nu\rho\sigma} \Gamma_\mu \Gamma_7 \epsilon_I
+ \frac{1}{16\sqrt{2}}X^{-1} F_{\nu\rho}^i ( \Gamma_\mu{}^{\nu\rho} - 6 \delta_\mu{}^\nu \Gamma^\rho ) \Gamma_7 ( \sigma^i )_I{}^J \epsilon_J ~,\nn\\
0 & = & - \ii X^{-1} \partial_\mu X \Gamma^\mu \epsilon_ I + \frac{1}{2\sqrt{2}} g \left( X - X^{-3} \right) \Gamma_7 \epsilon_I + \frac{\ii}{24} X^2 H_{\mu\nu\rho} \Gamma^{\mu\nu\rho} \Gamma_7 \epsilon_I \nonumber \\
&&- \frac{1}{8\sqrt{2}} X^{-1} F_{\mu\nu} \Gamma^{\mu\nu} \epsilon_I - \frac{\ii}{8\sqrt{2}} X^{-1} F^i_{\mu\nu} \Gamma^{\mu\nu} \Gamma_7 ( \sigma^i )_I{}^J \epsilon_J~.\label{dilatino}
\eea
Here $\epsilon_I$, $I=1,2$, are two Dirac spinors, $\Gamma_\mu$ generate the Clifford 
algebra $\mathrm{Cliff}(6,0)$ in an orthonormal frame, and we have defined the chirality operator
 $\Gamma_7 = \ii \Gamma_{012345}$, which satisfies $\Gamma_7^2=1$. 
The $SO(3)\sim SU(2)$ gauge field $A^i$ is an R-symmetry gauge field, 
 with the spinor $\epsilon_I$ transforming in the two-dimensional representation via
  the Pauli matrices $(\sigma^i)_I{}^J$. Thus the covariant derivative acting on the spinor is $D_\mu\epsilon_I=\nabla_\mu\epsilon_I+\frac{\ii}{2} g A_\mu^i(\sigma^i)_I{}^J\epsilon_J$.

Returning to the equations of motion (\ref{FullEOM}), 
notice that the exterior derivative of the first 
equation (the equation of motion for $B$) implies the second 
equation on using the Bianchi identities for $F$ and $F^i$, where note that $\diff F=\frac{2}{3}gH$. 
This is related to the fact that the theory possesses a gauge invariance 
$A\rightarrow A+\frac{2}{3}g\lambda$, $B\rightarrow B-\diff\lambda$, where 
$\lambda$ is an arbitrary one-form. Using this freedom one can then 
gauge away $A=0$, leaving $F=\frac{2}{3}gB$. The kinetic term for $F$ 
in the action (\ref{Fullaction}) then becomes a mass term for the $B$-field; that is, the $B$-field ``eats'' the $U(1)$ gauge field $A$ in a Higgs-like mechanism.
Notice that there is also a cubic Chern-Simons coupling for $B$ in (\ref{Fullaction}), making 
it a somewhat exotic field. 
We may also make a simple rescaling of the fields via $g_{\mu\nu}\rightarrow \frac{1}{g^2}g_{\mu\nu}$, $B\rightarrow \frac{1}{g^2}B$, $A\rightarrow \frac{1}{g}A$, $A^i\rightarrow \frac{1}{g}A^i$, 
after which one sees that the coupling constant $g$ only appears in the action as an overall constant $1/g^4$ 
factor. Thus we may without loss of generality set $g=1$, which we henceforth will do. 

In appendix \ref{AppSUSYConditions} we compute
the integrability conditions for the Killing spinor equation (\ref{KSE})
and dilatino equation (\ref{dilatino}), and show that these 
are compatible with the equations of motion (\ref{FullEOM}), (\ref{Einstein}).

\subsection{Killing vector bilinear}\label{Fernandoothersection}

Given a supersymmetric solution to the Euclidean Romans theory, one can 
verify that the bilinear
\bea\label{FernandoK}
K_\mu & \equiv & \varepsilon^{IJ}\epsilon_I^T \mathcal{C}\Gamma_\mu \epsilon_J~,
\eea
is a Killing one-form. Here $\mathcal{C}$ is the charge conjugation matrix, 
satisfying $\Gamma_\mu^T=\mathcal{C}^{-1}\Gamma_\mu \mathcal{C}$ 
and in our conventions is  antisymmetric satisfying $\mathcal{C}^2=-1$. 
If we also impose a symplectic Majorana condition 
\bea\label{FernandoSM}
\mathcal{C}\epsilon_I^* &=& \varepsilon_I^{\ J}\epsilon_J~,
\eea
then this Killing one-form may  be rewritten  as
\bea
K_\mu &=& \epsilon_I^\dagger \Gamma_\mu \epsilon_I~,
\eea
which is then manifestly real. In particular we will be able to 
impose this symplectic Majorana condition for the solutions 
we construct in section \ref{SUGRA}. In this ``real'' case 
the Killing spinors $\epsilon_I$ define an $SU(2)$ structure 
on $M_6$.
One could similarly analyse the differential conditions 
on the corresponding $SU(2)$ structure  bilinears, but we shall leave this for the future.


\section{Supergravity solutions}\label{SUGRA}

In this section we present supergravity duals to the $SU(3)\times U(1)$ 
squashed five-sphere backgrounds of section \ref{sec:gaugetheory}. 
Via the consistent truncation to the Romans theory in the previous section, 
this effectively becomes a filling problem in six-dimensional 
gauged supergravity: one seeks a smooth, asymptotically locally 
Euclidean AdS$_6$ supersymmetric supergravity solution, with conformal 
boundary data given by the squashed five-sphere background in 
section \ref{sec:gaugetheory}. In particular this means the 
bulk supergravity solution is equipped with an $SU(2)_R$ doublet 
of Killing spinors $\epsilon_I$, $I=1,2$, solving (\ref{KSE}) and (\ref{dilatino}), which should suitably approach the 
boundary Killing spinors in section \ref{sec:squashed}. 
We shall indeed find such fillings for both the 3/4 BPS and 1/4 BPS 
solutions. In the process shall extend the 1/4 BPS 
solution to a two-parameter family of solutions, containing a 
one-parameter 1/2 BPS subfamily of new solutions.

\subsection{$SU(3)\times U(1)$ invariant ansatz}\label{ansatz}

The squashed five-sphere backgrounds of section \ref{sec:squashed} 
have $SU(3)\times U(1)$ symmetry, and one expects this 
symmetry to be preserved by the bulk supergravity filling. Indeed, 
for asymptotically locally Euclidean AdS solutions of the \emph{vacuum} Einstein equations 
this is a theorem \cite{2007arXiv0704.3373A}. This leads to the following 
ansatz for the Romans supergravity fields
\begin{eqnarray}\label{ansatz}
\dd s^2_6 &=& \alpha^2(r)\dd r^2+\gamma^2(r)(\dd\tau+C)^2+\beta^2(r)\Big[\dd \sigma^2 + \frac{1}{4}\sin^2\sigma(\dd \theta^2+\sin^2\theta \dd \varphi^2)\nn\\
&&+\frac{1}{4}\cos^2\sigma\sin^2\sigma (\dd \psi+\cos\theta \dd\varphi)^2\Big]~,\nn\\
B&=& p(r)\dd r\wedge (\dd\tau+C) + \frac{1}{2}q(r)\dd C~,\nn\\
A^i &=& f^i(r)(\dd\tau +C)~, \quad i\ = \ 1,2,3~,
\end{eqnarray}
together with $X=X(r)$. Recall here that we have used the gauge freedom to 
set the $U(1)$ gauge field (which is really a Stueckelberg field) to $A=0$. 
The additional coordinate $r$ is a radial coordinate, and we shall choose 
a parametrization in which the conformal boundary is at $r=\infty$. 
For fixed $r$, provided $\gamma(r)$ and $\beta(r)$ are non-zero 
the constant $r$ surfaces in (\ref{ansatz}) are squashed five-spheres. 
We shall seek solutions with the topology of a ball, so that 
$r\in [r_0,\infty)$ with $r=r_0$ being the origin. At this point the squashed 
five-spheres must become \emph{round} in order that the metric 
extends smoothly to the origin of the ball. 
Similarly, in order for the gauge fields $B$, $A^i$ 
in (\ref{ansatz}) to be non-singular at the origin they must tend to zero sufficiently quickly
at $r=r_0$. In writing the ansatz (\ref{ansatz}) we have used the fact that the 
only $SU(3)\times U(1)$ invariant one-form on the squashed five-sphere is 
the global angular form $\diff \tau + C$ for the Hopf fibration $S^1\hookrightarrow S^5\rightarrow \mathbb{CP}^2$, while the only 
invariant two-form is the pull-back $\frac{1}{2}\diff C = \omega$ of the K\"ahler form 
on $\mathbb{CP}^2$.

Substituting the cohomogeneity one ansatz (\ref{ansatz}) into the equations 
of motion (\ref{FullEOM}) and Einstein equation (\ref{Einstein}) leads to 
a rather complicated coupled system of ODEs. 
The equations of motion for the background $SU(2)_R$ gauge field imply
$f^i(r)  =  \kappa_i f(r)$, $i=1,2,3$. The equations for the other fields then depend only on the $SU(2)\sim SO(3)$ invariant $\kappa_1^2+\kappa_2^2+\kappa_3^2$, which we can set to one by rescaling $f(r)$. The equations of motion then result in the coupled ODEs 
 for the functions $\alpha(r)$, $\beta(r)$, 
$\gamma(r)$, $p(r)$, $q(r)$, $f(r)$, $X(r)$, which can be found in appendix \ref{Appeqns}. 

Since the solutions we find are continuously connected to Euclidean  AdS$_6$, 
we first present the latter in these coordinates:
\begin{eqnarray}\label{AdS6}
\alpha(r) & =&  \frac{3\sqrt{3}}{\sqrt{6r^2-1}}~, \qquad 
\beta(r) \ = \ \gamma(r) \ = \  \frac{3 \sqrt{6r^2-1}}{\sqrt{2}}~,\nn\\
p(r) & =& q(r)  \ = \ f(r) \ =\ 0~, \qquad X(r) \ =\ 1~.
\end{eqnarray}
Here only the metric is non-trivial, and (\ref{AdS6}) realizes Euclidean 
AdS$_6$ as a hyperbolic ball with radial coordinate $r\in [\frac{1}{\sqrt{6}},\infty)$, 
with the conformal boundary at infinity $r=\infty$. Thus the origin is 
at $r_0=\frac{1}{\sqrt{6}}$. 
Notice in particular that the conformal boundary at $r=\infty$ is equipped 
with a \emph{round} metric on $S^5$, which is conformally flat.
 We would like to find families of solutions that generalize (\ref{AdS6}) by allowing 
for a squashed five-sphere boundary, keeping the metric 
asymptotically locally Euclidean AdS near $r=\infty$. That is, 
near $r=\infty$ the metric should approach
\bea
\dd s^2_6 & \simeq & \frac{9\diff r^2}{2r^2}+ 27r^2\diff s^2_5~,
\eea
where $\diff s^2_5$ is the squashed five-sphere (\ref{S5}).
 For such solutions we may thus define the squashing parameter by
\begin{equation}
\lim_{r \rightarrow \infty} \frac{\gamma(r)}{r} \ =\  3 \sqrt{3} ~\frac{1}{s}~,
\end{equation}
so that $s=1$ for the round sphere. Even though we did not manage to find supersymmetric solutions in closed form, the solutions can nevertheless be given as expansions around different limits. In general notice that we can use reparametrization invariance to set
\begin{equation}\label{betafixedapp}
\beta(r) \ = \ \frac{3 \sqrt{6r^2-1}}{\sqrt{2}}~,
\end{equation}
which we assume henceforth. In particular this fixes the origin of the ball
to be at $r_0=\frac{1}{\sqrt{6}}$.

In the following we summarize the various families of supersymmetric solutions 
we have constructed with the ansatz (\ref{ansatz}). Details 
of the computations may be found in appendix \ref{AppSolutions}.

\subsection{3/4 BPS solutions}\label{34solution}

There is  a one-parameter family of 3/4 BPS solutions parametrized by the squashing parameter $s$. The solution expanded around the conformal boundary is given by
\begin{eqnarray}\label{3rinfinity}
\alpha(r)&=&\frac{3}{\sqrt{2}} \frac{1}{r}+\frac{8+s^2}{36 \sqrt{2} s^2}\frac{1}{r^3}+\ldots~,~~~\\
\gamma(r)&=&\frac{3 \sqrt{3}}{s} r+\frac{-16+7 s^2}{12 \sqrt{3} s^3 }\frac{1}{r}-\frac{-1280+1120 s^2+241 s^4}{2592 \sqrt{3} s^5}\frac{1}{r^3}+\ldots~,\nn\\
X(r)&=&1+\frac{1-s^2-3  \sqrt{1-s^2}}{54 s^2}\frac{1}{r^2}+\frac{s^2 \sqrt{1-s^2} \kappa}{12 \left(1- s^2+\sqrt{1-s^2}\right)}\frac{1}{r^3}+\ldots~,\nn\\
 p(r) &=& -\frac{ \ii \sqrt{\frac{2}{3}} \left(s^2+3\sqrt{1-s^2}-1\right)}{s^3}\frac{1}{r^2}+\ldots~,\nn\\
 q(r) &=& -\frac{3 \ii \left(\sqrt{6} \sqrt{1-s^2}\right)}{s} r 
 + \frac{\sqrt{\frac{2}{3}} \ii \sqrt{1-s^2} \left(5 s^2+9  \sqrt{1-s^2}-5\right)}{3 s^3}\frac{1}{r}+\ldots~,\nn\\
f(r) &=& \frac{1-s^2+\sqrt{1-s^2}}{s^2} +\frac{2 \left(-2+2 s^2- (2+s^2) \sqrt{1-s^2}\right)}{9 s^4}\frac{1}{r^2}+\frac{\kappa}{r^3}+\ldots~,\nn
\end{eqnarray}
where we have computed this expansion up to $\mathcal{O}(1/r^9)$.
The extra parameter $\kappa$ is fixed by requiring regularity at the origin 
$r=\frac{1}{\sqrt{6}}$ (see (\ref{kappaexp}) below). Notice that the $SU(2)_R$ gauge field at the conformal boundary 
agrees with the gauge field (\ref{AR}) with $Q=1$. 
We may also expand the solution around Euclidean AdS$_6$, which has $s=1$:
\begin{eqnarray}
\alpha(r)&=&\frac{3 \sqrt{3}}{\sqrt{6 r^2-1}}\nn\\
&&+\tfrac{\left(-5 \sqrt{6}+330 \sqrt{6} r^2-3744 r^3+1620 \sqrt{6} r^4+8640 r^5-7560 \sqrt{6} r^6+5184 \sqrt{6} r^8\right)}{9 \sqrt{2} r^2 \left(6 r^2-1\right)^{9/2}} (1-s)+\ldots~,\nn\\
\gamma(r)&=&\frac{3 \sqrt{6 r^2-1}}{\sqrt{2}}\nn\\
&&-\tfrac{\left(55 \sqrt{2}-384 \sqrt{3} r+1080 \sqrt{2} r^2+768 \sqrt{3} r^3-5400 \sqrt{2} r^4+11232 \sqrt{2} r^6-11664 \sqrt{2} r^8\right) }{6 \left(6 r^2-1\right)^{7/2}} (1-s)+\ldots~,\nn\\
X(r) &=& 1-\frac{\left(\sqrt{2} \left(1-2 \sqrt{6} r+6 r^2\right)\right)}{3 \left(6 r^2-1\right)^2}\sqrt{1-s} +\ldots~,\nn\\
p(r) &=& \frac{18 \ii \sqrt{2} \left(\sqrt{6}-16 r+12 \sqrt{6} r^2-12 \sqrt{6} r^4\right) }{\left(6 r^2-1\right)^3}\sqrt{1-s}+\ldots~,\nn\\
q(r) &=& -\frac{3 \ii \sqrt{2} \left(-4+9 \sqrt{6} r-24 r^2-12 \sqrt{6} r^3+36 \sqrt{6} r^5\right) }{\left(6 r^2-1\right)^2}\sqrt{1-s}+\ldots~,\nn\\
f(r) &=& \frac{\sqrt{2} \left(-3+8 \sqrt{6} r-36 r^2+36 r^4\right)}{\left(6 r^2-1\right)^2} \sqrt{1-s} +\ldots~.
\end{eqnarray}
In particular one can check that these functions lead to a regular solution 
at the origin $r=\frac{1}{\sqrt{6}}$, although this is not manifest in the formulas presented 
above. Indeed, we have computed this expansion up to sixth order, and 
by comparing the two expansions we find that regularity 
at the origin fixes the parameter $\kappa$ in (\ref{3rinfinity}) 
via
\bea\label{kappaexp}
\frac{3\sqrt{3}}{4} \kappa & =& \delta +\frac{\sqrt{2}}{3}\delta ^2+\frac{113 }{36}\delta ^3+\frac{25}{9 \sqrt{2}} \delta ^4+\frac{1127}{288}\delta ^5+\frac{35 }{9 \sqrt{2}}\delta ^6+\ldots~,
\eea
where we have introduced 
\bea
\delta^2 & \equiv &  \frac{1}{s}-1~.\eea

The explicit solution $\epsilon_I$ to the Killing spinor (\ref{KSE}) and dilatino equation (\ref{dilatino})
for this solution may be found in appendix \ref{AppSolutions}. 
In particular there are three independent constants of integration 
after imposing the symplectic Majorana condition (\ref{FernandoSM}).
Using this solution one can compute the Killing vector bilinear (\ref{FernandoK}). 
Requiring that this Killing vector lies in the Lie algebra of the maximal torus $U(1)^3\subset SU(3)\times U(1)$ 
fixes the constants of integration, up to an overall irrelevant scaling. 
In this case we obtain
\bea
K &=& b_1\partial_{\varphi_1}+b_2\partial_{\varphi_2}+ b_3\partial_{\varphi_3}~,
\eea
where $b_1=1+\sqrt{1-s^2}$, $b_2=b_3=1-\sqrt{1-s^2}$ and 
the coordinates $\varphi_i$ are related to $\tau$, $\psi$ and $\varphi$
via (\ref{varphis}). 

\subsection{1/4 BPS solutions}\label{14solution}

We also find a two-parameter family of 1/4 BPS solutions, parametrized by the squashing parameter $s$ and the background $SU(2)_R$ field at the conformal boundary, which is 
parametrized by $f_0$. The solution 
expanded around the conformal boundary is given by
\begin{eqnarray}
\alpha(r) &=& \frac{3}{\sqrt{2}}\frac{1}{r}-\frac{f_0^2 s^2+9 \left(-2+s^2\right)-6 f_0 \left(-1+s^2\right)}{36 \sqrt{2} }\frac{1}{r^3}+\ldots~,\nn\\
\gamma(r) &=& \frac{3 \sqrt{3}}{s} r+\frac{2 f_0^2 s^2-12 f_0 \left(-1+s^2\right)+9 \left(-3+2 s^2\right)}{12 \sqrt{3} s}\frac{1}{r}+\ldots~,\nn\\
X(r) &=& 1+\frac{18-3 f_0-18 s^2+12 f_0 s^2-2 f_0^2 s^2}{54} \frac{1}{r^2}+\ldots~,\nn\\
p(r) &=& \frac{\ii \sqrt{\frac{2}{3}} (-3+f_0) \left(3+(-3+f_0) s^2\right)}{s}\frac{1}{r^2}+\ldots~,\nn\\
q(r) &=& -\frac{3 \ii \sqrt{6} \left(3+(-3+f_0) s^2\right)}{s} r\nn\\
&&+\frac{\ii \left(3+(-3+f_0) s^2\right) \left(f_0^2 s^2+9 \left(-1+s^2\right)-6 f_0 \left(1+s^2\right)\right)}{6 \sqrt{6} s } \frac{1}{r}+\frac{\xi_1}{r^2}+\ldots~,\nn\\
f(r) &=& f_0+\frac{2 (-3+f_0) f_0}{9}\frac{1}{r^2}+\frac{\xi_2}{r^3}+\ldots~.
\end{eqnarray}
Again, we have found this solution up to $\mathcal{O}(1/r^9)$.
The constants $\xi_1$ and $\xi_2$ are again fixed by requiring regularity at the origin.

There are a number of interesting special cases. First, we obtain the one-parameter family of 1/4 BPS 
squashed five-spheres of section \ref{sec:squashed} by choosing the constant $f_0$ 
so as to reproduce (\ref{AR}) with $Q=-3$.  That is, $f_0=(1-3\sqrt{1-s^2})\sqrt{1-s^2}/s^2$. 
We show  explicitly in appendix \ref{AppSolutions} that the 
supergravity Killing spinor matches onto the five-dimensional spinors in section 
\ref{sec:squashed}. Another interesting case is $f_0=0$.
In this case the $SU(2)_R$ background gauge field is completely switched off, but the solution is still supersymmetric with a squashed five-sphere at the
conformal boundary. This solution has enhanced supersymmetry -- 
as we show in appendix \ref{AppSolutions}  it is 1/2 BPS.
On the other hand we may also set $s=1$, so that the conformal boundary 
is the \emph{round} five-sphere, but keep the parameter $f_0$. 
This shows that one can define non-trivial Killing spinors on the \emph{round}
$S^5$ by turning on other fields. 

We may also expand the solution around Euclidean AdS$_6$ with $s=1$:
\begin{eqnarray} 
\alpha(r)&=& \frac{3\sqrt{3}}{\sqrt{6r^2-1}} +\frac{\sqrt{3} \left(1-54 r^2+96 \sqrt{6} r^3-324 r^4+216 r^6\right)}{2 r^2 \left(6 r^2-1\right)^{7/2}}(1-s)+\ldots~,\nn\\
\gamma(r)&=& \frac{3 \sqrt{6r^2-1}}{\sqrt{2}} +\frac{\left(15-48 \sqrt{6} r+270 r^2-540 r^4+648 r^6\right)}{\sqrt{2} \left(6 r^2-1\right)^{5/2}}(1-s)+\ldots~,\nn\\
X(r)&=&1+\frac{\left(1-2 \sqrt{6} r+6 r^2\right) (4+\omega )}{\left(6 r^2-1\right)^2} (1-s) +\ldots~,\nn\\
p(r)&=&-\frac{18 \ii \sqrt{2} \left(-\sqrt{3}+8 \sqrt{2} r-12 \sqrt{3} r^2+12 \sqrt{3} r^4\right) (6+\omega )}{\left(6 r^2-1\right)^3} (1-s) +\ldots~,\nn\\
q(r)&=& -\frac{3 \ii \left(-4+9 \sqrt{6} r-24 r^2-12 \sqrt{6} r^3+36 \sqrt{6} r^5\right) (6+\omega )}{\left(6 r^2-1\right)^2} (1-s)+\ldots~,\nn\\
f(r)&=& \frac{\left(-3+8 \sqrt{6} r-36 r^2+36 r^4\right) \omega}{\left(6 r^2-1\right)^2} (1-s)+\ldots~,
\end{eqnarray}
where we have introduced the parameter $\omega$ via  $(1-s)\omega= f_0$.
As before it can be  checked explicitly that the solution is regular at $r=\frac{1}{\sqrt{6}}$, 
and we have checked this up to fourth order in the expansion variable 
\bea
\delta &\equiv & \frac{1}{s}-1~.
\eea 
Comparing this expansion with the expansion around the conformal boundary we deduce
\begin{eqnarray}\label{Fernando14}
\xi_1&=&2 \ii (6+\omega ) \delta -\frac{\ii \left(144+98 \omega +13 \omega ^2\right)}{5}  \delta^2 \nn \\ 
&+&\frac{\ii \left(307719+209547 \omega +41094 \omega ^2+1282 \omega ^3\right)}{9450} \delta ^3\nn \\
&-&\frac{\ii \left(26693550+21683700 \omega +6126111 \omega ^2+771474 \omega ^3+51568 \omega ^4\right)}{623700}  \delta ^4+\ldots~,\nn\\
\xi_2&=&\frac{2}{3} \sqrt{\frac{2}{3}} \omega  \delta -\frac{2\left(-\sqrt{6} \omega +2 \sqrt{6} \omega ^2\right)}{45}  \delta ^2+\frac{\left(-999 \sqrt{6} \omega -594 \sqrt{6} \omega ^2+244 \sqrt{6} \omega ^3\right)}{42525}\delta ^3 \nn\\
&+&\frac{\left(32724 \sqrt{6} \omega +26082 \sqrt{6} \omega ^2+6105 \sqrt{6} \omega ^3+935 \sqrt{6} \omega ^4\right) }{1403325}\delta ^4+\ldots~.
\end{eqnarray}

The explicit solution $\epsilon_I$ to the dilatino and Killing spinor equation (\ref{dilatino}), (\ref{KSE}) 
for this solution may also be found in appendix \ref{AppSolutions}. 
In this case there is a single integration constant (for generic $f_0$, or equivalently $\omega$). 
The Killing vector automatically lies in the Lie algebra of the torus $U(1)^3\subset SU(3)\times U(1)$, and 
with an appropriate scaling we obtain
\bea
K &=& \partial_{\tau} \ =\ b_1\partial_{\varphi_1}+b_2\partial_{\varphi_2}+ b_3\partial_{\varphi_3}~,
\eea
where $b_1=b_2=b_3=1$ and 
the coordinates $\varphi_i$ are related to $\tau$, $\psi$ and $\varphi$
via (\ref{varphis}). 


\section{Holographic free energy}\label{SecFreeEnergy}

In this section we describe how the on-shell action for the Euclidean Romans theory detailed in section \ref{SecRomansSUGRA} can be computed, 
and for asymptotically locally Euclidean AdS solutions 
holographically renormalized by adding boundary counterterms \cite{Emparan:1999pm,deHaro2000xn,Taylor:2000xw}. For the supersymmetric solutions presented in section \ref{SUGRA} we evaluate the renormalized on-shell action and determine the holographic free energies.

\subsection{On-shell action}\label{onshell}

We will work in the gauge $A=0$. Starting from the Euclidean action \eqref{Fullaction} and using the equations of motion \eqref{FullEOM} together with the Einstein equation \eqref{Einstein} and its trace, we find the following for the on-shell action defined on a manifold $M_6$ with boundary $\partial M_6$ 
\bea
I_{\mathrm{on-shell}} &=& I_{\mathrm{bulk}} + I_{\mathrm{boundary}}  \, ,
\eea
where
\begin{align}
I_{\mathrm{bulk}} \ =& \ \frac{1}{16\pi G_N}\int_{M_6} \tfrac{4}{9} X^{-2} \left( 2+3X^4\right) *1 + \tfrac{1}{3} X^{-2} F^i \wedge * F^i + \tfrac{\ii}{3} B \wedge F^i \wedge F^i \, , \label{OSbulk} \\
I_{\mathrm{boundary}} \ =& \ \frac{1}{16\pi G_N}\int_{\partial M_6} \tfrac{2}{3} \left(X^{-1}*\dd X\right) + \tfrac{1}{3} ( B\wedge X^4*H ) \, . \label{OSboundary}
\end{align}
Here we have used Stokes' theorem to write a total derivative as a boundary integral. In particular this assumes that the 
potentials $B$ and $A^i$ are globally defined, which is the case for our supergravity solutions.
The Hodge duals in (\ref{OSboundary}) are defined on $M_6$, and then restricted to the boundary. The on-shell action is divergent due to the infinite volume of $M_6$ and $\partial M_6$, and from divergences in the supergravity fields as the conformal boundary $r \rightarrow \infty$ is approached. Consequently, $I_{\mathrm{bulk}}$ should be understood as integrated up to a finite cut-off which is then sent to infinity only after adding counterterms which regularize the divergences. In addition, because of the presence of boundary terms in the on-shell action, one should add a Gibbons-Hawking term \cite{Gibbons:1976ue} 
\bea
I_{\mathrm{GH}} &=& -\frac{1}{8\pi G_N}\int_{\partial M_6}\mathcal{K}\sqrt{\det h}\,\diff^5 x~.
\eea
This involves the trace $\mathcal{K}$  of the extrinsic curvature of the boundary, and where $h_{mn}$ is the induced boundary metric, and also leads to divergences. Hence the finite on-shell action is
\bea
I_{\mathrm{renormalized}} &=& I_{\mathrm{on-shell}} + I_{\mathrm{GH}} + I_{\mathrm{counterterms}} \, .
\eea
In the next subsection we determine the precise form of the counterterms.

\subsection{Boundary counterterms}\label{counterterms}

The counterterms needed to regularize the action of the Euclidean Romans $F(4)$ theory were stated without derivation in \cite{Alday:2014rxa}. Here we provide a full account of their construction. We assume a general expansion of the fields for an asymptotically locally Euclidean AdS$_6$ solution. In particular, we take the metric to be given in Fefferman-Graham form \cite{Fefferman,2007arXiv0710.0919F}
\bea\label{metric}
\diff s^2_6 &=& \frac{\ell^2}{z^2} \diff z^2+\frac{1}{z^2} \gamma_{mn} ( z,x ) \dd x^m \dd x^n~,
\eea
where $\ell=3/\sqrt{2}$ is the AdS$_6$ radius, and in turn 
\bea
\gamma_{mn} ( z,x )&=& \gamma_{mn}^{0}+z^2\gamma_{mn}^{2}+z^4\gamma_{mn}^{4}+\mathcal{O}(z^5)~.
\eea
Here $\gamma_{mn}^{0} (x)$ is the  metric induced on the conformal boundary which, due to the radial coordinate transformation $r \rightarrow \frac{1}{z}$, is now at $z=0$. The Gibbons-Hawking term is then
\bea
I_{\mathrm{GH}} &=& \frac{1}{8\pi G_N} \int_{\partial M_6} \frac{z}{\ell} \partial_z  \sqrt{ \det h} \ \dd^5 x \, ,
\eea
and $h_{mn} = \frac{1}{z^2} \gamma_{mn}$ is the induced metric on the boundary.

The Ricci tensor of the six-dimensional metric \eqref{metric} is
\bea
R_{zz} &=&- \frac{5}{z^2}-\frac{1}{2}\left[\mathrm{Tr}\left(\gamma^{-1}\partial^2_z\gamma\right)-\frac{1}{z}\mathrm{Tr}\left(\gamma^{-1}\partial_z\gamma\right)-\frac{1}{2}\mathrm{Tr}\left(\gamma^{-1}\partial_z\gamma\right)^2\right]~,\nn\\
R_{mn} &=& -\frac{5}{\ell^2z^2}\gamma_{mn}-\frac{1}{\ell^2}\Big[\frac{1}{2}\partial^2_z\gamma-\frac{2}{z}\partial_z\gamma-\frac{1}{2}(\partial_z\gamma)\gamma^{-1}(\partial_z\gamma)+\frac{1}{4}(\partial_z\gamma)\mathrm{Tr}\left(\gamma^{-1}\partial_z\gamma\right)\nn\\
&&-\ell^2R(\gamma)-\frac{1}{2z}\gamma\mathrm{Tr}\left(\gamma^{-1}\partial_z\gamma\right)\Big]_{mn}~,\nn\\
R_{zm} &=& \tfrac{1}{2}(\gamma^{-1})^{np}\left[\nabla_m\gamma_{np,z}-\nabla_p\gamma_{mn,z}\right]~,
\eea
with $\nabla$ being the covariant derivative for $\gamma(z,x)$. We also assume an asymptotic expansion for bulk scalar and gauge fields, namely
\bea
X &=& 1 + z X_1+ z^2 X_2 + \cdots \, , \nn \\
B &=& \frac{1}{z}b+\dd z\wedge A_0 + B_0+z\dd z\wedge A_1 + zB_{1}+\cdots \, , \nn \\
H = \dd B &=& - \frac{1}{z^2} \dd z \wedge b + \frac{1}{z} \dd b - \dd z \wedge \dd A_0 + \dd B_0 + \dd z \wedge B_1 - z \dd z \wedge \dd A_1 \cdots \, , \nn \\
F^i &=& f^i + \dd z \wedge A^i_0 + z \dd z \wedge A^i_1 + z F^i_1 + \cdots \, .
\eea
The $1/z$ term appearing in the $B$-field expansion is non-standard but is justified by being compatible with the equations of motion as we will see below.

It is useful to establish some formulas. We write (in general)
\bea
\alpha\wedge *\alpha &=& \|\alpha\|^2\vol~,
\eea
to define the norm $\|\cdot\|$ of a $p$-form. The inner product of two $p$-forms $\alpha,\beta$ is denoted $\langle \alpha , \beta \rangle$.
First we compute
\bea
*\alpha_p &=& -\ell z^{2p-6} \left(*_\gamma \alpha_p \right)\wedge \dd z~,\nn\\
*(\dd z\wedge \alpha_{p-1}) &=& \frac{1}{\ell}z^{2p-6}*_\gamma \alpha_{p-1}~,
\eea
where $\alpha_p$ represents a general $p$-form that is orthogonal to $\partial_z$. Here the volume 
forms are related as
\bea
\vol_6 &=& \frac{\ell}{z^6}\dd z\wedge \vol_\gamma \ = \ \frac{\ell}{z^6}\dd z\wedge \sqrt{\det \gamma} \ \dd x^1\wedge \cdots \wedge \dd x^5~.
\eea
We will need the expansion of the determinant and Hodge dual for $\gamma_{mn}$. The former is
\begin{align}
\sqrt{\det \gamma} = \sqrt{\det \gamma^0}\Big[&1 + \tfrac{z^2}{2} \mathrm{Tr} \left[ \gamma^2 (\gamma^0)^{-1} \right] + \tfrac{z^4}{2} \mathrm{Tr} \left[ \gamma^4 (\gamma^0)^{-1} \right] \nn \\
&- \tfrac{z^4}{4} \mathrm{Tr} \left[ \gamma^2 (\gamma^0)^{-1} \right]^2 + \tfrac{z^4}{8} \left( \mathrm{Tr} \left[ \gamma^2 (\gamma^0)^{-1} \right] \right)^2 +\mathcal{O}(z^5) \Big]~,
\end{align}
whilst the latter may be computed similarly as
\bea
*_\gamma\alpha_p &=& *_{\gamma^0}\alpha_p + z^2\left[\tfrac{1}{2}\mathrm{Tr} \left[\gamma^2(\gamma^0)^{-1}\right]*_{\gamma^0}\alpha_p- p *_{\gamma^0}(\gamma^2\circ \alpha_p)\right] + \mathcal{O}(z^4)~.
\eea
 Here we have defined the $p$-form
\bea
(\gamma^2\circ\alpha_p)_{m_1\cdots m_p} &\equiv & (\gamma^2)_{[m_1}{}^{n}(\alpha_p)_{|n| m_2\cdots m_p]}~,
\eea
and indices are always raised with $\gamma^0$, so $(\gamma^2)_{m}{}^{n}\equiv (\gamma^2)_{mp}(\gamma^0)^{pn}$.

The idea now is to substitute these expansions into the Romans field equations  and then on-shell action. We first look at the lowest order term in $z$ in each of the $X$, $B$ and Einstein equations. The leading order term in the $X$ equation of motion dictates
\bea
X_1 &=& 0~.
\eea
Specifically, the term $\frac{1}{z^5}\diff z\wedge \vol_{\gamma^0}$ has a coefficient proportional to $X_1$ times a non-zero number, thus forcing $X_1=0$. 
Next one finds that the leading order term in the $B$ equation of motion, which is proportional to $\frac{1}{z^3}\diff z\wedge *_{\gamma^0}b$, has a coefficient that is zero if and only if $\ell^2=9/2$. Similarly, the leading order term in the $mn$ component of the Einstein equation, which is $\mathcal{O}(1/z^2)$, is satisfied if and only if $\ell^2=9/2$. We will substitute $\ell = 3/\sqrt{2}$ from now on. 

The first divergence we encounter, which is at order $\mathcal{O}(1/\epsilon^5)$ where $z=\epsilon$ is the finite cut-off, comes from expanding the $\tfrac{4}{9} X^{-2} \left( 2+3X^4\right) *1$ integrand in $I_{\mathrm{bulk}}$ and the Gibbons-Hawking term. It is
\bea
I^{\mathrm{div}}_{\mathcal{O}(1/\epsilon^5)} &=& \frac{1}{8\pi G_N} \frac{1}{\epsilon^5} \int_{\partial M_6} -\frac{4\sqrt{2}}{3} \sqrt{\det \gamma^0} \ \dd^5 x \, ,
\eea
and is simply cancelled by adding the counterterm
\bea
I^{\mathrm{counterterm}}_5 &=& \frac{1}{8\pi G_N}\cdot \frac{4\sqrt{2}}{3}\int_{\partial M_6}\sqrt{\det h} \ \dd^5x~.
\eea
We write the counterterm action in terms of the induced boundary metric $h_{mn}$ as the divergences most naturally appear in this form \cite{Balasubramanian:1999re}.
There is no divergence at $\mathcal{O}(1/\epsilon^4)$ as a consequence of $X_1=0$. The divergence at $\mathcal{O}(1/\epsilon^3)$ has contributions from each of $I_{\mathrm{bulk}}$, $I_{\mathrm{boundary}}$, $I_{\mathrm{GH}}$ and the expansion of $I^{\mathrm{counterterm}}_5$, and is
\begin{align}
I^{\mathrm{div}}_{\mathcal{O}(1/\epsilon^3)} \ =& \ \ \frac{1}{8\pi G_N} \frac{1}{\epsilon^3} \int_{\partial M_6} \left[ \frac{4\sqrt{2}}{9} \mathrm{Tr} \left[\gamma^2(\gamma^0)^{-1}\right] + \frac{1}{9\sqrt{2}} \| b \|^2_{\gamma^0} \right] \sqrt{ \det \gamma^0 } \ \dd^5 x \, . 
\end{align}
Clearly we will need some control on $\gamma^2$, and this comes from the $\mathcal{O}(1)$ term in the $mn$ direction of the Einstein equation. Carefully expanding we find this fixes
\bea
\gamma^2_{mn} &=& -\frac{3}{2}\left[R(\gamma^0)_{mn}-\frac{1}{8}R(\gamma^0)\gamma^0_{mn}\right]+\frac{1}{2}b^2_{mn}-\frac{3}{16}\|b\|^2_{\gamma^0}\gamma^0_{mn}~.\label{gammatwo}
\eea
Here $R(g)_{mn}=\mathrm{Ric}(g)_{mn}$ denotes the Ricci tensor of a metric $g_{mn}$, with $R(g)$ the Ricci scalar. The curvature terms in $\gamma^2_{mn}$ are standard \cite{Emparan:1999pm}, while the terms involving $b$ are specific to the Romans theory and boundary conditions we are considering. Taking the trace of \eqref{gammatwo}, or alternatively examining the $zz$ component of the Einstein equation at order $\mathcal{O}(1)$, gives
\bea
\mathrm{Tr}\left[\gamma^2(\gamma^0)^{-1}\right] &=& -\frac{9}{16}R(\gamma^0)+\frac{1}{16}\|b\|^2_{\gamma^0}~. \label{Trgammatwo}
\eea
This expression will need to be used extensively due to its appearance in the Hodge dual and metric determinant. Substituting $\mathrm{Tr}\left[\gamma^2(\gamma^0)^{-1}\right]$ into the right hand side of $I^{\mathrm{div}}_{\mathcal{O}(1/\epsilon^3)}$ leads to
\bea
I^{\mathrm{div}}_{\mathcal{O}(1/\epsilon^3)} &=& \frac{1}{8\pi G_N} \frac{1}{\epsilon^3} \int_{\partial M_6} \left[ - \frac{1}{2\sqrt{2}} R(\gamma^0) + \frac{1}{6\sqrt{2}} \| b \|^2_{\gamma^0} \right] \sqrt{ \det \gamma^0 } \ \dd^5 x \, , 
\eea
and the appropriate counterterm is therefore
\bea
I^{\mathrm{counterterm}}_3 &=& \frac{1}{8\pi G_N}\int_{\partial M_6} \left[\frac{1}{2\sqrt{2}}R(h)-\frac{1}{6\sqrt{2}}\|B\|^2_h\right]\sqrt{\det h} \ \dd^5x~.
\eea

{\it A priori} there is also an $\mathcal{O}(1/\epsilon^2)$ divergence, but one easily sees from the various expansions that only the scalar field contributes to it. This term (temporarily reinstating the AdS length scale) is 
\bea
I^{\mathrm{div}}_{\mathcal{O}(1/\epsilon^2)} &=& \frac{1}{8\pi G_N}\left(\frac{4\ell}{9}\cdot \frac{1}{2}-\frac{1}{\ell}\right)\frac{1}{\epsilon^2}\int_{\partial M_6}X_3\sqrt{\det \gamma^0} \ \dd^5 x \ = \ 0~,
\eea
where the first term comes from expanding the bulk integral \eqref{OSbulk}, while the second (which cancels it) comes from the boundary $X^{-1} * \diff X$ term in \eqref{OSboundary}. Thus this potential divergence is zero, without needing a counterterm or indeed even needing to use any of the equations of motion.

Continuing we find there are many terms that contribute at $\mathcal{O}(1/\epsilon)$ including $A_1$ and $B_0$ from the asymptotic expansion of the $B$-field. It is prudent to look at higher orders of $z$ in the equations of motion for simplifications along the lines of $X_1=0$. Indeed by looking at the $z^{-2}\dd z\wedge \alpha_3$ coefficient of the $B$-field equation of motion we find
\bea
B_0 &=& 0 \, .
\eea
The $z^{-1}\alpha_4$ coefficient similarly implies
\bea
A_1 &=& 0 \, .
\eea
With these simplifications the $\mathcal{O}(1/\epsilon)$ divergence becomes
\bea
I^{\mathrm{div}}_{\mathcal{O}(1/\epsilon)} &=& \frac{1}{8\pi G_N} \frac{1}{\epsilon} \int_{\partial M_6} \bigg[ \frac{29\sqrt{2}}{9} (X_2)^2 + \frac{2\sqrt{2}}{9} X_4 + \frac{2\sqrt{2}}{9} X_2 \mathrm{Tr}\left[ \gamma^2 (\gamma^0)^{-1} \right] + \frac{\sqrt{2}}{4} \| f^i \|^2_{\gamma^0} \nn \\
&&\qquad - \frac{\sqrt{2}}{72} \mathrm{Tr}\left[\gamma^2(\gamma^0)^{-1}\right] \| b \|^2_{\gamma^0} + \frac{\sqrt{2}}{18} \langle b , \gamma^2\circ b \rangle + \frac{2\sqrt{2}}{9} X_2 \, \| b \|^2_{\gamma^0} + \frac{\sqrt{2}}{18} \langle b , \dd A_0 \rangle \nn \\
&&\qquad + \frac{4\sqrt{2}}{3}\mathrm{Tr} \left[\gamma^4(\gamma^0)^{-1}\right] - \frac{2\sqrt{2}}{3} \mathrm{Tr}\left[\gamma^2(\gamma^0)^{-1}\right]^2 + \frac{\sqrt{2}}{3} \left( \mathrm{Tr}\left[\gamma^2(\gamma^0)^{-1}\right] \right)^2 \nn \\
&&\qquad- \frac{\sqrt{2}}{4}R(\gamma^0)_{ij} (\gamma^2)^{ij} + \frac{\sqrt{2}}{8}R(\gamma^0) \mathrm{Tr}\left[\gamma^2(\gamma^0)^{-1}\right] \bigg] \sqrt{\det \gamma^0} \ \dd^5x \, . \label{div5intermediate}
\eea
We now seek to determine $A_0$, $X_4$ and $\gamma^4$ in terms of lower order boundary quantities such as $b$. Examination of the $z^{-2}\alpha_4$ coefficient of the $B$-field equation of motion gives
\bea
\diff *_{\gamma^0}b &=& -\frac{\ii\sqrt{2}}{3}b\wedge b -\frac{4}{9}*_{\gamma^0}A_0~,
\eea
which we should regard as fixing $A_0$ in terms of the boundary field $b$. Specifically, since $*_5^2=1$ on any form, we solve this as
\bea
A_0 &=& -\frac{9}{4}*_{\gamma^0}\left(\diff *_{\gamma^0}b + \frac{\ii\sqrt{2}}{3}b\wedge b\right)~.
\eea
Note we may also write $*_{\gamma^0}\diff *_{\gamma^0}b = \delta_{\gamma^0}b$ 
in terms of the adjoint $\delta_{\gamma^0}$ of $\diff$ with respect to $\gamma^0$. The $z^{-1}\dd z\wedge \alpha_3$ coefficient determines $B_1$ to be
\begin{align}
B_1 \ =& \ \ *_{\gamma^0}\left(\frac{9}{4}\diff *_{\gamma^0}\diff b-\frac{\ii\sqrt{2}}{3}b\wedge A_0\right)
+2 b X_2 -\frac{1}{2}\mathrm{Tr}\left[\gamma^2(\gamma^0)^{-1}\right]b+2\gamma^2\circ b~,%
\end{align}
which may be rewritten as
\bea
B_1 &= & \frac{9}{4}*_{\gamma^0}\left[\diff *_{\gamma^0}\diff b + \frac{\ii\sqrt{2}}{3}b\wedge\delta_{\gamma^0} b - \frac{2}{9}b\wedge *_{\gamma^0}(b\wedge b)\right]  +2 b X_2 \nn\\&&-\frac{1}{2}\mathrm{Tr}\left[\gamma^2(\gamma^0)^{-1}\right]b+2\gamma^2\circ b~.
\eea

The next coefficient we need is $X_4$, the coefficient of $z^4$ in the expansion of $X(z,x^m)$ and is found from the $z^{-2}\dd z\wedge \vol_{\gamma^0}$ terms in the X field equation
\bea
X_4 &=&- \frac{9}{4} \Delta_{\gamma^0} X_2 - X_2\mathrm{Tr} \left[ \gamma^2 (\gamma^0)^{-1} \right] - \frac{11}{2} (X_2)^2 + \frac{3}{4} X_2 \| b \|^2_{\gamma^0} \nn \\
&&+ \frac{9}{16} \| \dd b \|^2_{\gamma_0} - \frac{1}{36} \| A_0 \|^2_{\gamma^0} - \frac{1}{2} \langle B_1 , b \rangle + \frac{1}{4} \langle b , \dd A_0 \rangle - \frac{9}{32} \| f^i \|^2_{\gamma^0} \, .
\eea
Here $\Delta_{\gamma^0}=\delta_{\gamma^0}\dd$ acting on functions but will not contribute for a compact boundary (after integrating by parts).

We also need $\gamma^4_{mn}$, which comes from expanding the $zz$ component of the Einstein equation at $\mathcal{O}(z^2)$:
\bea
\mathrm{Tr} \left[ \gamma^4 (\gamma^0)^{-1} \right] &=&+ \frac{1}{4} \mathrm{Tr}\left[ \gamma^2 (\gamma^0)^{-1} \right]^2 - \frac{5}{2} (X_2)^2 - \frac{1}{24} \| A_0 \|^2_{\gamma^0} + \frac{9}{32} \| \dd b \|^2_{\gamma^0} - \frac{3}{8} X_2 \| b \|^2_{\gamma^0} \nn \\
&&+ \frac{1}{4} \langle b , B_1 \rangle - \frac{1}{8} \langle b , \dd A_0 \rangle + \frac{9}{64} \| f^i \|^2_{\gamma^0} \, .
\eea

Next we record some intermediate formulae which follow from the expression for $\gamma^2_{mn}$ in \eqref{gammatwo}:
\bea
\mathrm{Tr}\left[\gamma^2(\gamma^0)^{-1}\right]^2 &=& \frac{9}{4} \left[
R(\gamma^0)_{mn}R(\gamma^0)^{mn}-\frac{11}{64}R(\gamma^0)^2\right]+\frac{1}{4}\mathrm{Tr}_{\gamma^0}b^4\\
&&-3\langle \mathrm{Ric}(\gamma^0)\circ b,b\rangle_{\gamma^0}+\frac{75}{128}
R(\gamma^0)\|b\|^2_{\gamma^0}-\frac{51}{256}\|b\|^4_{\gamma^0}~,\nn\\
R(\gamma^0)_{mn}(\gamma^2)^{mn} &=& -\frac{3}{2}R(\gamma^0)_{mn}R(\gamma^0)^{mn} +\frac{3}{16}R(\gamma^0)^2+\langle \mathrm{Ric}(\gamma^0)\circ b,b\rangle_{\gamma^0}\nn\\&&-\frac{3}{16}R(\gamma^0)\|b\|^2_{\gamma^0}~,\nn\\
\langle \gamma^2\circ b,b\rangle &=& - \frac{3}{2}\langle \mathrm{Ric}(\gamma^0)\circ b,b\rangle_{\gamma^0}+\frac{1}{4}\mathrm{Tr}_{\gamma^0}b^4+\frac{3}{16}R(\gamma^0)\|b\|^2_{\gamma^0}-\frac{3}{16}\|b\|^4_{\gamma^0}~.\nn
\eea
Here we have defined $\mathrm{Tr}_{\gamma^0}b^4 \equiv b_m{}^{n} b_n{}^{p}b_p{}^{q}b_q{}^{m}$.
Notice that $\mathrm{Tr}_{\gamma^0}b^2 = -2\|b\|^2_{\gamma^0}$, with this notation.

We now have all that we need to compute the $\mathcal{O}(1/\epsilon)$ counterterm. Inserting all our intermediate results along with the newfound expressions for $X_4$ {\it etc}\ into $I^{\mathrm{div}}_{\mathcal{O}(1/\epsilon)}$ in \eqref{div5intermediate} leads to
\bea
I^{\mathrm{div}}_{\mathcal{O}(1/\epsilon)} &=& \frac{1}{8\pi G_N} \frac{1}{\epsilon} \int_{\partial M_6} \Bigg\{\bigg[ - \frac{3}{4\sqrt{2}}R(\gamma^0)_{mn}R(\gamma^0)^{mn} + \frac{15}{64\sqrt{2}}R(\gamma^0)^2 \nn \\
&&+\frac{3}{4\sqrt{2}} \| f^i \|^2_{\gamma^0}- \frac{1}{12\sqrt{2}}\mathrm{Tr}_{\gamma^0}b^4 + \frac{13}{192\sqrt{2}} \|b\|^4_{\gamma^0} + \frac{1}{\sqrt{2}} \| \dd b \|^2_{\gamma^0} \nn \\
&&- \frac{5}{8\sqrt{2}} \| \diff *_{\gamma^0}b + \tfrac{\ii\sqrt{2}}{3}b\wedge b \|^2_{\gamma^0} + \frac{1}{4\sqrt{2}} \langle b , \dd \delta_{\gamma^0}b + \tfrac{\ii\sqrt{2}}{3} \dd [ *_{\gamma^0}b\wedge b ] \rangle \nn \\
&&- \frac{4\sqrt{2}}{3} ( X_2 )^2 + \frac{1}{\sqrt{2}}\langle \mathrm{Ric}(\gamma^0)\circ b,b\rangle_{\gamma^0} - \frac{9}{32\sqrt{2}} R(\gamma^0) \| b \|^2_{\gamma^0} \bigg] \sqrt{\det \gamma^0} \ \dd^5x \nn \\
&&+ \frac{1}{4\sqrt{2}} \langle b , *_{\gamma^0}\big[\diff *_{\gamma^0}\diff b + \tfrac{\ii\sqrt{2}}{3}b\wedge\delta b - \tfrac{2}{9}b\wedge *_{\gamma^0}(b\wedge b)\big]\rangle \Bigg\} \, .
\eea
The corresponding counterterm is hence
\bea
I^{\mathrm{counterterm}}_1 &=& \frac{1}{8\pi G_N}\int_{\partial M_6}\Bigg\{\bigg[ \frac{3}{4\sqrt{2}}R(h)_{mn}R(h)^{mn}-\frac{15}{64\sqrt{2}}R(h)^2 \nn \\
&&-\frac{3}{4\sqrt{2}}\|F^i\|^2_h+\frac{1}{12\sqrt{2}}\mathrm{Tr}_{h}B^4 -\frac{13}{192\sqrt{2}}\|B\|^4_{h}-\frac{1}{\sqrt{2}}\|\dd B\|^2_{h}\nn \\
&&+\frac{5}{8\sqrt{2}}\|\diff *_{h}B+\tfrac{\ii\sqrt{2}}{3}B\wedge B\|^2_{h}-\frac{1}{4\sqrt{2}}\langle B,\diff\delta_{h}B+\tfrac{\ii \sqrt{2}}{3}\dd*_{h}B\wedge B\rangle_{h}\nn\\
&&+\frac{4\sqrt{2}}{3}(1-X)^2-\frac{1}{\sqrt{2}}\langle \mathrm{Ric}(h)\circ B,B\rangle_{h}+\frac{9}{32\sqrt{2}}R(h)\|B\|^2_{h} \bigg]\sqrt{\det h}\,\dd^5x\nn\\
&&-\frac{1}{4\sqrt{2}}B\wedge \bigg[\diff *_h\diff B + \frac{\sqrt{2}\ii}{3}B\wedge\delta_h B  - \frac{2}{9}B\wedge *_{h}(B\wedge B)\bigg]\Bigg\}~.
\eea
Once again the pure gravity terms found in the first line agree with the literature \cite{Emparan:1999pm}.

{\it A priori} the bulk integral in \eqref{OSbulk} is logarithmically divergent. Of course a log divergence should not appear, as the boundary is odd-dimensional and on general grounds one does not expect local anomalies. In keeping with this argument the equations of motion at even higher order in $z$ constrain the fields such that the potential log divergence cancels without the need for a counterterm.

Collating all the expressions for the counterterms we finally arrive at \cite{Alday:2014rxa}
\bea
I_{\mathrm{counterterms}} &=& \frac{1}{8\pi G_N}\int_{\partial M_6}\Bigg\{\bigg[ \frac{4\sqrt{2}}{3}+\frac{1}{2\sqrt{2}}R(h)-\frac{1}{6\sqrt{2}}\|B\|^2_h \nn \\
&&+\frac{3}{4\sqrt{2}}R(h)_{mn}R(h)^{mn}-\frac{15}{64\sqrt{2}}R(h)^2 \nn \\
&&-\frac{3}{4\sqrt{2}}\|F^i\|^2_h+\frac{1}{12\sqrt{2}}\mathrm{Tr}_{h}B^4 -\frac{13}{192\sqrt{2}}\|B\|^4_{h}-\frac{1}{\sqrt{2}}\|\dd B\|^2_{h}\nn \\
&&+\frac{5}{8\sqrt{2}}\|\diff *_{h}B+\tfrac{\ii\sqrt{2}}{3}B\wedge B\|^2_{h}-\frac{1}{4\sqrt{2}}\langle B,\diff\delta_{h}B+\tfrac{\ii \sqrt{2}}{3}\dd*_{h}B\wedge B\rangle_{h}\nn\\
&&+\frac{4\sqrt{2}}{3}(1-X)^2-\frac{1}{\sqrt{2}}\langle \mathrm{Ric}(h)\circ B,B\rangle_{h}+\frac{9}{32\sqrt{2}}R(h)\|B\|^2_{h} \bigg]\sqrt{\det h}\,\dd^5x\nn\\
&&-\frac{1}{4\sqrt{2}}B\wedge \bigg[\diff *_h\diff B + \frac{\sqrt{2}\ii}{3}B\wedge\delta_h B  - \frac{2}{9}B\wedge *_{h}(B\wedge B)\bigg]\Bigg\}~.\label{Icounterterms}
\eea

\subsection{Free energy of the solutions}

The renormalized on-shell action determined in the previous subsection holds for  all
Romans supergravity solutions which are asymptotically locally AdS.  
In particular we may use these results to compute the holographic free energy 
for the supersymmetric solutions of section \ref{SUGRA}. 
In order to present the results, we first split the renormalized action as
\bea
I_{\mathrm{renormalized}} &=& I_{\mathrm{bulk}} + I_{\mathrm{non-bulk}}~,
\eea
where $I_{\mathrm{bulk}}$ is the bulk integral given by (\ref{OSbulk}), while 
\bea
I_{\mathrm{non-bulk}} &=& I_{\mathrm{boundary}} + I_{\mathrm{GH}} + I_{\mathrm{counterterms}}~,
\eea
where $I_{\mathrm{boundary}}$ is the boundary contribution to the 
on-shell action (\ref{OSboundary}), $I_{\mathrm{GH}}$ is the Gibbons-Hawking term, 
while $I_{\mathrm{counterterms}}$ is the full counterterm (\ref{Icounterterms}). 
For our $SU(3)\times U(1)$ ansatz (\ref{ansatz}), with $f^1(r)\equiv f^2(r)\equiv 0$ and $f^3(r)=f(r)$, we have in particular
\bea
I_{\mathrm{bulk}} &=& \frac{\pi^2}{36 G_N}\int_{r=\frac{1}{\sqrt{6}}}^{\Lambda}\Bigg[3X^2(r)\alpha(r)\beta^4(r)\gamma(r)+ 6 \ii 
f(r)\left[f(r)p(r)+q(r)f'(r)\right]\nn\\
&&+ \frac{24f^2(r)\alpha^2(r)\gamma^2(r)+8\alpha^2(r)\beta^4(r)\gamma^2(r)+3\beta^4(r)(f'(r))^2}{4X^2(r)\alpha(r)\gamma(r)}\Bigg]\diff r~,
\eea
where $\Lambda$ is the cut-off for the $r$ coordinate. 

\subsubsection*{$3/4$ BPS solution} 

For the one-parameter family of  3/4 BPS solutions in section \ref{34solution} we obtain
\bea
I_{\mathrm{bulk}} &=& \frac{\pi^2}{36G_N}\Bigg[\frac{6561 \sqrt{\frac{3}{2}} }{s}\Lambda ^5-\frac{243 \sqrt{\frac{3}{2}} \left(3+12 s^2+\sqrt{1-s^2}\right) }{s^3}\Lambda ^3\nn \\
&&-\frac{2187\sqrt{6} \kappa \left(-1 + \sqrt{1-s^2}  \right) }{8 s}\Lambda ^2\\
&&+\frac{27 \left[\sqrt{\frac{3}{2}} \left(74+66 s^4-14 \sqrt{1-s^2}-s^2 \left(5+4 \sqrt{1-s^2}\right)\right)\right] }{4 s^5} \Lambda\nn\\
&&-243+\frac{81 \delta }{2 \sqrt{2}}-1377 \delta ^2-\frac{1467 \delta ^3}{8 \sqrt{2}}-\frac{6693 \delta ^4}{2}-\frac{44073 \delta ^5}{64 \sqrt{2}}-4482 \delta ^6+\mathcal{O}(\delta^7)\Bigg]\nn\\
&&+\mathcal{O}\left(\frac{1}{\Lambda}\right)~,\nn\eea
together with
\bea
I_{\mathrm{non-bulk}} &=& \frac{\pi^2}{36G_N}\Bigg[ -\frac{6561 \sqrt{\frac{3}{2}} }{s}\Lambda ^5+\frac{243 \sqrt{\frac{3}{2}} \left(3+12 s^2+\sqrt{1-s^2}\right) }{s^3}\Lambda ^3\nn \\
&&+\frac{2187\sqrt{6} \kappa \left(-1 + \sqrt{1-s^2}  \right) }{8 s}\Lambda ^2\\
&&-\frac{27 \left[\sqrt{\frac{3}{2}} \left(74+66 s^4-14 \sqrt{1-s^2}-s^2 \left(5+4 \sqrt{1-s^2}\right)\right)\right] }{4 s^5} \Lambda\nn\\
&&+\frac{81 \sqrt{\frac{3}{2}} \left(-16+16 \sqrt{1-s^2}+13 s^2 \left(1+3 \sqrt{1-s^2}\right)\right) \kappa }{8 s^3}\Bigg]+\mathcal{O}\left(\frac{1}{\Lambda}\right)~,\nn\eea
where recall that $\kappa$ is given as a series in $\delta$ in (\ref{kappaexp}). 
Adding the two contributions and taking the cut-off $\Lambda\rightarrow\infty$, the 
divergences cancel and we are left with the following finite result
\bea
\label{Iren}
I_{\mathrm{renormalized}} \ =\  -\frac{27 \pi^2}{4 G_N} \left( 1+\frac{8}{3}  \delta^2 +\frac{16 \sqrt{2}}{27} \delta^3+\frac{68 }{27}\delta^4 +\frac{28 \sqrt{2}}{27} \delta ^5+\frac{32 }{27}\delta ^6 +\ldots\right)~,
\eea
where the six-dimensional Newton constant is given by\footnote{This was effectively calculated in \cite{Jafferis:2012iv} by identifying the holographic free energy of Euclidean AdS$_6$ with an entanglement entropy. The $N^{5/2}$ scaling of the free energy had previously been 
predicted in \cite{Brandhuber:1999np}.}
\begin{equation}
G_N~=~\frac{15  \pi \sqrt{8-N_f}}{4 \sqrt{2} N^{5/2}}. 
\end{equation}
The holographic free energy is identified with $I_{\mathrm{renormalized}}$ 
and agrees precisely with the series expansion of the large $N$ field theory 
result (\ref{Fernando})
\bea
\cal{F} &=& \frac{1}{27s^2}\frac{(3-\sqrt{1-s^2})^3}{1-\sqrt{1-s^2}} \mathcal{F}_{\mathrm{round}\, S^5}~,
\eea
where recall that $s=1/(1+\delta^2)$.

\subsubsection*{$1/4$ BPS Solution} 

We may similarly compute the holographic free energy of the two-parameter family of 1/4 BPS 
solutions in section \ref{14solution}. Again we obtain two divergent contributions whose 
divergences cancel. The finite piece may be computed as an expansion in 
$\delta=\frac{1}{s}-1$ using  
 the series expansions 
of the parameters $\xi_1$, $\xi_2$ in (\ref{Fernando14}). Putting everything together 
we obtain
\bea
I_{\mathrm{renormalized}} \ =\  -\frac{27 \pi^2}{4 G_N} \left( 1+\mathcal{O}(\delta^5)\right)~.
\eea
This again agrees with large $N$ field theory result (\ref{Fernando}). Of course 
the latter field theory result was computed for a one-parameter subfamily of boundary 
conditions in section \ref{sec:gaugetheory}, while here we have a more general two-parameter family. We shall elaborate on this in section \ref{Fernandosection}.


\section{Boundary supersymmetry conditions}\label{SecSUSYatBoundary}

In this section we determine the form of the Euclidean Romans supersymmetry conditions, given in section \ref{SecRomansSUGRA}, 
near the five-dimensional conformal boundary. Closely related work has appeared in \cite{Nishimura:2000wj}. Our conventions are the following: we use $x^\mu=(r,x^m)$ to denote six-dimensional coordinates, so that the indices $\mu, \nu , \ldots \in \{0,1,2,3,4,5\}$. Six-dimensional frame indices are indexed by $A,B, \ldots \in \{0,1,2,3,4,5\}$ and five-dimensional frame indices by early Roman letters $a,b$ {\it etc}. 

We continue to use the Fefferman-Graham coordinates outlined in subsection \ref{counterterms}, although compared to that section we change coordinates $z \rightarrow 1/r$ so that the conformal boundary is now at $r = \infty$. We can then scale the $r$ coordinate $r \rightarrow \lambda r$ without changing the position of the conformal boundary or modifying the five-dimensional boundary metric $\gamma^0$. After this scaling the asymptotic six-dimensional metric is now
\bea
\diff s^2_6 &=& \frac{\ell^2}{r^2} \diff r^2+ \lambda^2 r^2 \gamma_{mn}\dd x^m \dd x^n~, 
\eea
where
\bea
\gamma_{mn} &=& \gamma_{mn}^{0}+ \frac{1}{\lambda^2 r^2} \gamma_{mn}^{2}+ \frac{1}{\lambda^4 r^4} \gamma_{mn}^{4}+ \frac{1}{\lambda^5 r^5} \gamma_{mn}^{5}+\mathcal{O}\left( \frac{1}{r^6} \right)~.
\eea
We introduce a six-dimensional vielbein $e^A$ such that
\bea
\diff s^2_6 &=& e^A e^A \ = \ e^0 e^0 + e^a e^a \, .
\eea
If we denote by $e^a_{(5)}$ the vielbein for $\gamma^0$, then the six-dimensional frame components may be written as 
\bea
e^0_r \ = \ \frac{\ell}{r} \, , \qquad e^0_m  \ =\  0 \, , \qquad e^a_r \ = \ 0 \, , \qquad e^a_m (r,x) \ = \ \lambda r e^a_{(5) m} (x) + \cdots \, , \label{VBexp}
\eea
where the ellipsis denotes subleading powers of $r$ which will not play a part in what follows. The inverse frame is
\bea
e_0^r &=& \frac{r}{\ell} \, , \qquad e_0^m \ = \ 0 \, , \qquad e_a^r \ = \ 0 \, , \qquad ( e^a_m )^{-1} \ = \ e_a^m \ = \ \frac{1}{\lambda r} e_{(5)a}^m + \ldots \, . \label{VBexpinv}
\eea 
The six-dimensional spin connection is given by $\omega_\mu{}^{AB} = e^{\nu [A} \partial_{\mu} e_{\nu}{}^{B]} - e^{\nu [A} \partial_{\nu} e_{\mu}{}^{B]} - e_\nu^{[A} e^{B]}_\sigma e_\mu^C \partial^\nu e^\sigma_C$ and from this expression it is easy to show that
\begin{align}
\omega_r{}^{bc} \ =&\  0 \ =\  \omega_r{}^{0b}\, , \qquad \omega_a{}^{0c}\  =\  - \frac{1}{\ell} \delta_a^c + \ldots \, , \qquad \omega_a{}^{bc}\  =\  \frac{1}{\lambda r} \omega^{(5d)}_a{}^{bc} + \ldots \, ,
\end{align}
where $\omega^{(5d)}_a{}^{bc}$ is the spin connection associated with the 5d boundary metric $\gamma^0$.

Incorporating some of the results from the holographic renormalization in subsection \ref{counterterms}, the asymptotic bulk field expansions in the local six-dimensional coordinates are\footnote{In this section we use a calligraphic font $\mathcal{A}^i$ to denote the $SU(2)$ gauge field so that there is no confusion with other notation.}
\bea
X &=& 1+ \frac{1}{r^2} X_2 +\cdots~, \nn \\
F = \tfrac{2}{3} B &=& \frac{2r}{3} b - \frac{2}{3r^2} \dd r \wedge A_0 +\cdots~, \nn \\
H = \dd B &=& \dd r \wedge b + r \dd b +\cdots~, \nn \\
F^i &=& f^i + \cdots~, \nn \\
\mathcal{A}^i &=& a^i + \cdots~.
\eea
Note that not all the fields appearing on the right hand side are independent. For example $f^i = \dd a^i - \tfrac{1}{2} \varepsilon^{ijk} a^j \wedge a^k$ and $A_0$ was found in subsection \ref{counterterms} to be given by
\bea
A_0 &=& -\frac{9}{4}*_{\gamma^0}\left(\diff *_{\gamma^0}b + \frac{\ii\sqrt{2}}{3}b\wedge b\right)~.
\eea
However, for simplicity we keep $A_0$ and substitute in terms of $b$ only at the end of our computation. Converting the bulk field expansions first into the six-dimensional frame and then into the 5d frame using \eqref{VBexpinv} we can read off the following  components for the asymptotic fields
\bea
H_{0 ab} &=& \frac{r}{\ell ( \lambda r)^2} b_{ab} + \mathcal{O} \left( \frac{1}{r^3} \right) \, , \qquad H_{abc} \ = \ \frac{r}{(\lambda r)^3} (\dd b)_{abc} + \mathcal{O} \left( \frac{1}{r^4} \right) \, , \nn \\
F_{0 a} &=& - \frac{2}{3\ell \lambda r^2} ( A_0 )_a + \mathcal{O} \left( \frac{1}{r^3} \right) \, , \qquad F_{ab} \ = \ \frac{2r}{3 (\lambda r)^2} b_{ab} + \mathcal{O} \left( \frac{1}{r^3} \right) \, , \nn \\
X + \tfrac{1}{3} X^{-3} &=& \frac{4}{3} + \mathcal{O} \left( \frac{1}{r^4} \right) \, , \qquad X- X^{-3} \ = \ \frac{4}{r^2} X_2 + \mathcal{O} \left( \frac{1}{r^3} \right) \, , \nn \\
X^{-1} \partial_0 X &=& - \frac{2}{\ell r^2} X_2 + \mathcal{O} \left( \frac{1}{r^3} \right) \, , \qquad X^{-1} \partial_a X \ = \ \mathcal{O} \left( \frac{1}{r^3} \right) \, , \nn \\
F^i_{ab} &=& \frac{1}{(\lambda r)^2} f^i_{ab} + \mathcal{O} \left( \frac{1}{r^3} \right) \, , \qquad F^i_{0 a} \ = \ \mathcal{O} \left( \frac{1}{r^3} \right) \, , \nn \\ 
\mathcal{A}^i_a &=& \frac{1}{\lambda r} a^i_a + \mathcal{O} \left( \frac{1}{r^3} \right) \, , \qquad \mathcal{A}^i_0 \ = \ \mathcal{O} \left( \frac{1}{r^3} \right) \, . \label{AsymptoticFields}
\eea

The full six-dimensional Killing spinor equation for the Euclidean Romans theory, where all indices are orthonormal frame indices,  is
\begin{align}
D_A \epsilon_I \  =\ &  \frac{\ii}{4\sqrt{2}} ( X + \tfrac{1}{3} X^{-3} ) \Gamma_A \Gamma_7 \epsilon_I - \frac{1}{48} X^2 H_{BCD} \Gamma^{BCD} \Gamma_A \Gamma_7 \epsilon_I  \nn \\
&- \frac{\ii}{16\sqrt{2}} X^{-1} F_{BC} ( \Gamma_A{}^{BC} - 6 \delta_A{}^B \Gamma^C ) \epsilon_I \nn \\
&+ \frac{1}{16\sqrt{2}}X^{-1} F_{BC}^i ( \Gamma_A{}^{BC} - 6 \delta_A{}^B \Gamma^C ) \Gamma_7 ( \sigma^i )_I{}^J \epsilon_J \, ,
\end{align}
where $D_A \epsilon_I = \partial_A \epsilon_I + \frac{1}{4} \omega_A{}^{BC} \Gamma_{BC} \epsilon_I + \frac{\ii}{2} \mathcal{A}^i_A ( \sigma^i )_I{}^J \epsilon_J$.
Taking the free index to be $A=0$ and substituting the field components \eqref{AsymptoticFields} leads to
\bea\label{badger}
\partial_r \epsilon_I &=&+ \frac{\ii}{2r} \Gamma_0 \Gamma_7 \epsilon_I + \mathcal{O} \left( \frac{1}{r^2} \right) \, .
\eea
Similarly, if we take the free index in the Killing spinor equation to be $A=a$ then we find
\begin{align}
\nabla_a \epsilon_I \ =\ & \frac{\lambda}{3\sqrt{2}} r \Gamma_a ( \ii \Gamma_7 -\Gamma_0 ) \epsilon_I - \frac{\ii}{2} a^i_a ( \sigma^i )_I{}^J \epsilon_J \label{KSEpr}\\
&- \frac{\ii}{24 \lambda \sqrt{2}} b_{bc} \Gamma_a{}^{bc} ( 1 + \ii \Gamma_0 \Gamma_7 ) \epsilon_I + \frac{\ii}{4 \lambda \sqrt{2}} b_{ab} \Gamma^{b} \left( 1 + \tfrac{\ii}{3} \Gamma_0 \Gamma_7 \right) \epsilon_I + \mathcal{O} \left( \frac{1}{r} \right) \, , \nn
\end{align}
with $\nabla_a$ being the covariant derivative with respect to the 5d spin connection.

Now we decompose the six-dimensional gamma matrices and spinors. We take our coordinate independent Cliff$(6,0)$ gamma matrices to be
\begin{align}
\Gamma_0 \ =\ & \left(\begin{array}{cc}0 & 1_4 \\1_4 & 0\end{array}\right) \, , \qquad \Gamma_{a} \ =\  \left(\begin{array}{cc}0 & \ii \gamma_{a} \\ - \ii \gamma_{a} & 0\end{array}\right) \, , \qquad \Gamma_7 \ =\  \left(\begin{array}{cc}-1_4 & 0 \\0 & 1_4\end{array}\right) \, ,
\end{align}
where $\gamma_{a}$ are a Hermitian basis of Cliff$(5,0)$. The six-dimensional spinor $\epsilon_I$ is decomposed as 
\begin{equation}
\epsilon_I \ =  \ \left(\begin{array}{c} \epsilon_I^+ \\ \epsilon_I^- \end{array}\right) \, ,
\end{equation}
where $\epsilon_I^{\pm}$ are 4-component spinors.

With this basis of gamma matrices and splitting of the spinors, the $r$ direction of the  Killing spinor equation (\ref{badger}), to lowest order in $r$, is 
\begin{align}
\left(\begin{array}{c} \partial_r \epsilon_I^+ \\ \partial_r \epsilon_I^- \end{array}\right) \ =\ & \frac{\ii}{2r} \left(\begin{array}{c} \epsilon_I^- \\ - \epsilon_I^+ \end{array}\right) \, .
\end{align}
The general solution determines the asymptotic dependence on $r$:
\begin{align}
\epsilon_I \ = \ \left(\begin{array}{c} \epsilon_I^+ \\ \epsilon_I^- \end{array}\right) \ = \ \sqrt{r} \left(\begin{array}{c} \chi_I \\ - \ii \chi_I \end{array}\right) + \frac{1}{\sqrt{r}} \left(\begin{array}{c} \varphi_I \\ \ii \varphi_I \end{array}\right) + \cdots \label{Spinorrexp} \, ,
\end{align}
where $\chi_I, \varphi_I$ depend only on the boundary coordinates $x^m$.
Having found the asymptotic dependence on $r$ for the spinors $\epsilon_I$ we can then substitute into the remaining components of the Killing spinor equation \eqref{KSEpr}. Taking only the lowest terms in $r$ gives two copies of
\begin{equation}
\nabla_a \chi_I \ = \ - \frac{\lambda \sqrt{2} \ii}{3} \gamma_a \varphi_I - \frac{\ii}{2} a^i_a ( \sigma^i )_I{}^J \chi_J - \frac{\ii}{12 \lambda \sqrt{2}} b_{bc} \gamma_a{}^{bc} \chi_I + \frac{\ii}{3 \lambda \sqrt{2}} b_{ab} \gamma^{b} \chi_I \, . \label{5dKSEgeneral}
\end{equation}
This is the five-dimensional boundary Killing spinor equation.

Now recall that the six-dimensional dilatino condition in the frame reads
\begin{align}
0 \ =\ & - \ii X^{-1} \partial_A X \Gamma^A \epsilon_ I + \frac{1}{2\sqrt{2}} \left( X - X^{-3} \right) \Gamma_7 \epsilon_I + \frac{\ii}{24} X^2 H_{ABC} \Gamma^{ABC} \Gamma_7 \epsilon_I \nn \\
&- \frac{1}{8\sqrt{2}} X^{-1} F_{AB} \Gamma^{AB} \epsilon_I - \frac{\ii}{8\sqrt{2}} X^{-1} F^i_{AB} \Gamma^{AB} \Gamma_7 ( \sigma^i )_I{}^J \epsilon_J \, . 
\end{align}
We may follow precisely the same steps as for the Killing spinor equation to determine the asymptotic form of the dilatino equation. Doing so we find the five-dimensional constraint 
\begin{align}
0 \ =& \ - \frac{1}{6\sqrt{2}} b_{ab} \gamma^{ab} \varphi_I - \frac{\sqrt{2}}{3} \lambda^2 X_2 \chi_I + \frac{\ii}{24\lambda} (\dd b)_{abc} \gamma^{abc} \chi_I + \frac{\lambda \ii}{8} \nabla^b b_{ab}  \gamma^{a} \chi_I \nn \\
&+ \frac{\lambda}{48\sqrt{2}} b_{ab} b_{cd} \gamma^{abcd} \chi_I + \frac{\ii}{8\sqrt{2}} f^i_{ab} \gamma^{ab} ( \sigma^i )_I{}^J \chi_J \, .
\end{align}

We would prefer to have five-dimensional supersymmetry conditions which are homogeneous in the spinor $\chi_I$ instead of the current dependence on both $\chi_I$ and $\varphi_I$. To remove $\varphi_I$ we contract \eqref{5dKSEgeneral} with $\gamma^a$. This gives
\bea
\varphi_I &=& \frac{\ii}{5} \frac{3}{\lambda\sqrt{2}} \bigg[ \gamma^a \bigg( \delta_I^J  \nabla_a + \frac{\ii}{2} a^i_a ( \sigma^i )_I{}^J + \frac{\ii}{12 \lambda \sqrt{2}} b_{bc} ( \gamma_a{}^{bc} - 4 \delta_a^b \gamma^c ) \delta_I^J \bigg) \bigg] \chi_J\nn \\
&\equiv& \frac{\ii}{5} \frac{3}{\lambda\sqrt{2}} D_I{}^J \chi_J \, .
\eea
We may then write the boundary Killing spinor equation in the form
\begin{equation}
\left( \tilde{\nabla}_I{}^J{}_a - \frac{1}{5} \gamma_a D_I{}^J \right) \chi_J \ =\ 0 \, , \label{CKS}
\end{equation}
where $\tilde{\nabla}_I{}^J{}_a = \delta_I^J  \nabla_a + \frac{\ii}{2} a^i_a ( \sigma^i )_I{}^J + \frac{\ii}{12 \lambda \sqrt{2}} b_{bc} ( \gamma_a{}^{bc} - 4 \delta_a^b \gamma^c ) \delta_I^J$. The boundary dilatino constraint reads
\begin{align}
0\ =\ &- \frac{\ii}{20 \lambda} b_{ab} \gamma^{ab}D_I{}^J \chi_J - \frac{\sqrt{2}}{3} \lambda^2 X_2 \chi_I + \frac{\ii}{24\lambda} (\dd b)_{abc} \gamma^{abc} \chi_I + \frac{\lambda \ii}{8} \nabla^b b_{ab}  \gamma^{a} \chi_I \nn \\
&+ \frac{\lambda}{48\sqrt{2}} b_{ab} b_{cd} \gamma^{abcd} \chi_I + \frac{\ii}{8\sqrt{2}} f^i_{ab} \gamma^{ab} ( \sigma^i )_I{}^J \chi_J \, .
\end{align}

For vanishing $b$-field, solutions of \eqref{CKS} are known as charged conformal Killing spinors (CCKS), or twistor spinors. Within the current context of gauge/gravity duality,  CCKS have been classified for 3-manifolds and 4-manifolds in both Euclidean and Lorentzian signature in \cite{Klare:2012gn,Hristov:2013spa,Cassani:2013dba,Klare:2013dka}. More recently, solutions in five dimensions (with arbitrary signature) have been studied in \cite{2014arXiv1403.2311L}. To our knowledge the more general charged conformal Killing spinor equation, where the charge is with respect to both the triplet of one-forms $a^i$ and the two-form $b$, has not been studied in the literature. It would be interesting to understand the relationship between the five-dimensional conditions found here from the Romans supergravity theory and the rigid limit of five-dimensional $\mathcal{N}=1$ Poincar\'e supergravity \cite{Zucker:1999ej,Kugo:2000hn} studied in \cite{Pan:2013uoa,Imamura:2014ima}.

Finally, whilst we do not yet understand the general properties of a solution to \eqref{CKS}, we are able to state the precise relation between the spinors $\varphi_I$ and $\chi_I$ for our supersymmetric solutions (for which $\lambda = 3 \sqrt{3}$). For the 3/4 BPS solution we find
\begin{equation}
 \varphi_I \ = \ (-1)^{I} \frac{3-\sqrt{1-s^{2}}}{6 \sqrt{6} s} \chi_I - (-1)^{I} \frac{4\sqrt{1-s^{2}}}{6 \sqrt{6} s} \gamma_1 \chi_I \, ,
\end{equation}
and for the two-parameter family of 1/4 BPS solutions
\be
 \varphi_I\ =\ \frac{(f_0-3) s}{6 \sqrt{6}} \chi_I \, .
\ee
In appendix \ref{AppSolutions} we give further details of the explicit six-dimensional Killing spinors and their relation to the five-dimensional spinors of section \ref{sec:gaugetheory}.


\section{Wilson loops}\label{Wilson}

In this section we compute the expectation values of certain BPS Wilson loops, both in the large $N$ 
matrix  model of section \ref{sec:largeN} and also in the supergravity 
dual solutions of section \ref{SUGRA}. More precisely it will be important 
to uplift these solutions to massive type IIA supergravity, where the 
Wilson loop in the fundamental representation is dual to a 
fundamental string. Minus the action of this string precisely 
matches the logarithm of the Wilson loop VEV in the large $N$ limit, 
as a function of the parameters of the solutions.

\subsection{Large $N$ field theory}\label{WilsonFT}

An interesting observable to consider is the VEV of the Wilson loop in a 
representation $\mathbf{R}$ of the gauge group $G$:
\bea\label{VEV}
\langle\, W_{\mathbf{R}}\, \rangle  &=& \frac{1}{\dim \mathbf{R}}\left\langle 
\mathrm{Tr}_{\mathbf{R}}\, \mathcal{P}\exp \int \left(\mathscr{A}_m\dot{x}^m+\sigma |\dot{x}|\right)\diff t \right\rangle~.
\eea
Here $\mathscr{A}$ denotes the dynamical gauge field for the gauge group $G$, 
$\sigma$ is the scalar in the corresponding  vector multipet, and the 
worldline is parametrized by $x^m(t)$. It is straightforward to see that (\ref{VEV}) 
is invariant under the supersymmetry transformations for the squashed 
five-sphere (\ref{squashedS5general}) appearing in section 3.3 of \cite{Imamura:2012xg} 
provided the Wilson loop wraps an orbit of the Killing vector bilinear\footnote{Of course 
we have similarly defined a Killing vector $K$ in the six-dimensional bulk as (\ref{FernandoK}). The
latter restricts to (\ref{5dK}) on the conformal boundary, so this is only a slight abuse of notation.}
\bea\label{5dK}
K_m & = & \varepsilon^{IJ}{\chi_I}^T \mathcal{C}_{(5)}\gamma_m\chi_J~.
\eea
That is, we take $x^m(t)$ to be an integral curve of $K$. The supersymmetry 
variations of the two terms in (\ref{VEV}) then cancel each other.

The large $N$ limit of (\ref{VEV}) for the $USp(2N)$ gauge theories 
described in section \ref{sec:largeN}
was computed for the round five-sphere 
in \cite{Assel:2012nf}. It is straightforward to extend this to the more 
general squashed sphere matrix model in section \ref{sec:largeN}. 
The key point is that the insertion of the Wilson loop into the 
path integral does not affect the leading order saddle point configuration
because its logarithm scales as $N^{1/2}$, while the free energy 
instead scales as $N^{5/2}$. The dynamical gauge field $\mathscr{A}$ 
localizes to zero, so only the constant scalar $\sigma$ contributes to the Wilson loop 
(\ref{VEV}) in the localization computation. Thus the VEV (\ref{VEV}), for the fundamental representation 
of $USp(2N)$, is effectively computed 
in the large $N$ matrix model as
\bea
\langle\, W_{\mathrm{fund}}\, \rangle &=& \int_{0}^{x_\star}\ex^{2\pi \elly \lambda(x)}\, \rho(x)\diff x~,
\eea
where $\rho(x)$ is the saddle point eigenvalue density (\ref{rho}), with the 
eigenvalues supported on $[0,x_\star]$ with $x_\star$ given by (\ref{xstar}). 
We have also denoted by $2\pi\elly=\int |\dot{x}|\, \diff t$ the length of the integral curve of $K$ 
that is wrapped by the Wilson loop, and recall that $\lambda(x)=N^{1/2}x$ to 
leading order. Thus we find the large $N$ result
\bea\label{largeNWilson}
\log\, \langle\, W_{\mathrm{fund}}\, \rangle &=& \frac{(b_1+b_2+b_3)\sqrt{2}\pi \elly }{\sqrt{8-N_f}}N^{1/2}  + o(N^{1/2})~.
\eea
Relative to the round sphere result we thus have
\bea
\log\, \langle\, W_{\mathrm{fund}}\, \rangle &=&\frac{(b_1+b_2+b_3)\elly}{3}\log\, \langle\, W_{\mathrm{fund}}\, \rangle_{\mathrm{round}}~.
\eea
Indeed, recalling that 
\bea
K &=& b_1\partial_{\varphi_1}+b_2\partial_{\varphi_2}+b_3\partial_{\varphi_3}~,
\eea
in terms of the standard $U(1)^3$ action on $S^5\subset \R^2\oplus \R^2\oplus \R^2$, 
then the orbits of $K$ are always closed circles 
at the origins of any two copies of $\R^2$. If we call these $U(1)^3$ invariant circles $S^1_i$, $i=1,2,3$, then 
$\elly=1/b_i$ and we may write 
\bea
\log\, \langle\, W_{\mathrm{fund},\, S^1_i}\, \rangle &=&\frac{(b_1+b_2+b_3)}{3b_i}\log\, \langle\, W_{\mathrm{fund}}\, \rangle_{\mathrm{round}}~.
\eea
Notice that this formula is invariant under a constant rescaling 
$K\rightarrow c\cdot K$.
We now explain how to reproduce this large $N$ result from the dual supergravity solutions.

\subsection{Dual fundamental strings}\label{WilsonSUGRA}

The supergravity dual of the Wilson loop $W_{\mathrm{fund}}$ was studied 
in \cite{Assel:2012nf} for the round five-sphere. The supergravity background 
is in this case the massive type IIA uplift AdS$_6\times S^4$ of the AdS$_6$ vacuum 
of the Romans theory of section \ref{SecRomansSUGRA}. The Wilson loop 
maps to a fundamental string sitting at the north pole $\xi=\frac{\pi}{2}$ 
of the internal $S^4$, in the notation of section \ref{sec:uplift}. 
The string then wraps a copy of $\R^2\subset$ AdS$_6$ parametrized by 
the radial direction $r$ in AdS together with the Wilson loop curve 
$S^1\subset S^5$.

We now generalize this to our supergravity backgrounds in section  
\ref{SUGRA}. Here the type IIA background is a warped and fibred product 
$M_6\times S^4$, together with various non-trivial background fluxes. 
However,  $M_6$ still has the topology of a ball, with a natural radial 
direction $r$. Thus the candidate dual of the Wilson loops computed in 
the previous section is a fundamental string sitting at $\xi=\frac{\pi}{2}$ 
in the internal $S^4$ of (\ref{uplift}), together with the Wilson 
loop curve $S^1\subset S^5_{\mathrm{squashed}}$ and the radial 
direction $r$. This is then a copy of $\Sigma_2\cong\R^2\subset M_6$, and 
we would like to compute the regularized action of a fundamental 
string wrapping this submanifold. 

In order to compute the string action we must first convert to the string frame metric 
in (\ref{uplift}), which introduces a factor of $\ex^{ \Phi/2}$, where $\Phi$ is the 
ten-dimensional dilaton. The induced string frame metric on $M_6$ at the north 
pole $\xi=\frac{\pi}{2}$ of $S^4$ is then
\bea
\diff s^2_{M_6}\mid_{\xi = \frac{\pi}{2},\, \mathrm{string}}&=& X^{-2}\diff s^2_6~,
\eea 
where $\diff s^2_6$ is the Romans supergravity metric. The $B$-field then uplifts 
to the type IIA $B$-field with curvature $F_{(3)}=H=\diff B$ via (\ref{uplift}) at the north pole $\xi=\frac{\pi}{2}$. In section \ref{SecRomansSUGRA} we have set most of the physical 
scaling parameters to specific numerical values -- for example the Romans mass is set to $m_{\mathrm{IIA}}=\frac{\sqrt{2}}{3}$, while the correctly normalized value 
for the supergravity dual to the $USp(2N)$ gauge theories is $(8-N_f)/(2\pi \ell_s)$ 
where $\ell_s$ is the string length. In particular restoring the 
AdS radius to its physical value 
\bea
L^4 &=& \frac{8\pi^2 N}{9(8-N_f)}\ell_s^4~,
\eea
(as in \cite{Assel:2012nf}) the string frame action is
\bea\label{string}
S &=& \frac{N^{1/2}\sqrt{2}}{3\sqrt{(8-N_f)}}\int_{\Sigma_2}X^{-2}\sqrt{\det \gamma}\, \diff^2x + \ii B~,
\eea
where $\gamma_{ab}$ is the metric induced on $\Sigma_2$ via its embedding into 
the Romans metric $\diff s^2_6$ on $M_6$, and we have included the usual Wess-Zumino coupling to the ten-dimensional $B$-field. More precisely, (\ref{string}) is divergent, 
and as usual one may regularize it by cutting off the $r$ integral at some 
$r=\Lambda$, and including a boundary counterterm given by the length of the 
boundary $S^1\subset S^5$ at $r=\Lambda$. Thus the regularized action reads
\bea\label{regstring}
S_{\mathrm{string}} \ = \   \frac{N^{1/2}\sqrt{2}}{3\sqrt{(8-N_f)}}\left[\int_{\Sigma_2}\left(X^{-2}\sqrt{\det \gamma}\, \diff^2x + \ii B\right) - \frac{3}{\sqrt{2}}\mathrm{length}(\partial \Sigma_2)\right]~,
\eea
where this is understood to mean the limit as one takes the cut-off $\Lambda\rightarrow \infty$.
We now compute this for our various solutions.

\subsubsection*{1/4 BPS background}

We begin with the 1/4 BPS background, as in this case the supersymmetric Killing
vector bilinear is simply $K=\partial_\tau$ (up to an irrelevant constant rescaling). Via the $SU(3)$ symmetry of the background 
all orbits of $K$ are equivalent, and thus there is effectively only one Wilson loop to compute. 
This wraps the $\tau$ and $r$ directions at, say, $\sigma=0$ (which is a point 
on the base $\mathbb{CP}^2$ of $S^1_{\mathrm{Hopf}}\hookrightarrow S^5\rightarrow \mathbb{CP}^2$, all points being equivalent under $SU(3)$). The regularized string action (\ref{regstring}) is 
\bea
S_{\mathrm{string}} \ = \  \lim_{\Lambda\rightarrow\infty}\frac{N^{1/2}2\sqrt{2}\pi}{3\sqrt{(8-N_f)}}\left[  \int_{r=\frac{1}{\sqrt{6}}}^{\Lambda} \left[X^{-2}(r)\alpha(r)\gamma(r)+\ii\,  p(r)\right]\diff r - \frac{3}{\sqrt{2}} \gamma(\Lambda)\right]~,
\eea
where we have used that $\tau$ has period $2\pi$. Evaluating this 
for the two-parameter family of 1/4 BPS solutions, as a series in 
the parameter $\delta$, we find
\bea
-S_{\mathrm{string}} &=& \frac{3 \sqrt{2} \pi }{\sqrt{8-N_f}}N^{1/2}+\mathcal{O}(\delta^5)~,
\eea
which agrees precisely with the large $N$ field theory result (\ref{largeNWilson}) 
since $K=\partial_\tau = \partial_{\varphi_1}+\partial_{\varphi_2}+\partial_{\varphi_3}$ 
so that $b_1=b_2=b_3=1$.

\subsubsection*{3/4 BPS background}

For the 3/4 BPS solution recall that the supersymmetric Killing vector $K$ has 
$b_1=1+\sqrt{1-s^2}$, $b_2=b_3=1-\sqrt{1-s^2}$. For generic 
values of the squashing parameter $s$ the generic orbit of $K$ will be open. 
However, the orbits always close over the circles $S^1_i$ defined in section 
\ref{WilsonFT}, which have lengths $\elly=2\pi /b_i$. Since $b_2=b_3$ these circles give rise to two distinct
Wilson loop VEVs:
\bea\label{34VEVs}
\frac{\log\, \langle\, W_{\mathrm{fund},\, S^1_i}\, \rangle}{\log\, \langle\, W_{\mathrm{fund}}\, \rangle_{\mathrm{round}}} &=& \begin{cases}\ \displaystyle
\frac{3-\sqrt{1-s^2}}{3(1+\sqrt{1-s^2})}~, & \qquad  i \ = \ 1 ~,\\[15pt] \ \displaystyle \frac{3-\sqrt{1-s^2}}{3(1-\sqrt{1-s^2})}~, & \qquad i \ = \ 2, 3~.\end{cases}
\eea
We may then compare these results to the regularized string 
action (\ref{regstring}), where for $S^1_i$ the fundamental string wraps 
the circle $\varphi_i$ together with the $r$ direction. 
More precisely, $S^1_1$ is located at $\sigma=0$ in the coordinates 
(\ref{S5}), while $S^1_2$ is located at  $\{\sigma=\frac{\pi}{2}$, $\theta=0\}$, 
as one sees from (\ref{rhos}). The result for $S^1_3$ is the same as that 
for $S^1_2$ due to the $SU(2)\subset SU(3)$ symmetry preserved by the bosonic 
solution and supersymmetric Killing vector.
On the other hand, due to the signs in 
(\ref{varphis}) the relevant string actions to compute are then
\bea
\frac{N^{1/2}2\sqrt{2}\pi}{3\sqrt{(8-N_f)}}\left[  \int_{r=\frac{1}{\sqrt{6}}}^{\Lambda} \left[X^{-2}(r)\alpha(r)\gamma(r)\pm \ii\,  p(r)\right]\diff r - \frac{3}{\sqrt{2}} \gamma(\Lambda)\right]~,
\eea
respectively. Evaluating this 
for the one-parameter family of 3/4 BPS solutions, as a series in 
the parameter $\delta$ up to sixth order where $\delta^2 = \frac{1}{s}-1$, we find
\bea
\frac{S_{\mathrm{string}, S^1_1}}{S_{\mathrm{string}}\mid_{\delta=0}} \ = \ 1-\frac{4\sqrt{2}}{3}\delta + \frac{8}{3}\delta^2 - \frac{5\sqrt{2}}{3}\delta^3 + \frac{4}{3}\delta^4 - \frac{7}{12\sqrt{2}}\delta^5+0\cdot \delta^6+\ldots~,
\eea
while
\bea
\frac{S_{\mathrm{string}, S^1_2}}{S_{\mathrm{string}}\mid_{\delta=0}} \ = \ 1+\frac{2\sqrt{2}}{3}\delta + \frac{4}{3}\delta^2 + \frac{5}{3\sqrt{2}}\delta^3 + \frac{2}{3}\delta^4 + \frac{7}{24\sqrt{2}}\delta^5+0\cdot \delta^6+\ldots~.
\eea
These agree precisely with the series expansions of (\ref{34VEVs}) computed in
field theory. 


\section{Discussion and conjectures}\label{Fernandosection}

In this paper we have constructed supergravity duals to the $USp(2N)$ 
superconformal gauge theories on $SU(3)\times U(1)$ squashed five-spheres. 
These constitute a one-parameter family of 3/4 BPS solutions, and a 
two-parameter family of generically 1/4 BPS. The latter include 
new supersymmetric  squashed five-sphere geometries with 
the background $SU(2)_R$ gauge field turned off, and moreover 
these have enhanced 1/2 BPS supersymmetry. By 
holographically renormalizing the Euclidean Romans supergravity theory, 
we have computed the holographic free energy for our solutions.
We then compared this to the large $N$ limit of the partition function 
of the gauge theories, and found perfect agreement. Given a 
supersymmetric supergravity solution one can construct the Killing vector 
$K^\mu=\varepsilon^{IJ}\epsilon_I^T\mathcal{C}\gamma^\mu\epsilon_J$, 
where $\epsilon_I$, $I=1,2$, is the $SU(2)_R$ doublet of Killing spinors. 
For our solutions the free energy takes the form
\bea\label{free}
\cal{F} &=& \frac{(|b_1|+|b_2|+|b_3|)^3}{27|b_1b_2b_3|} \cal{F}_{\mathrm{AdS_6}}~,
\eea
where we write the supersymmetric Killing vector  as 
$K=\sum_{i=1}^3b_i\partial_{\varphi_i}$, and $\partial_{\varphi_i}$ are standard 
generators of $U(1)^3\subset SU(3)\times U(1)$ acting on $S^5\subset \R^2\oplus \R^2\oplus \R^2$.
 Given the corresponding $4d/3d$ results of \cite{Farquet:2014kma, Alday:2013lba}, it is then natural to conjecture that (\ref{free}) 
holds for \emph{any} supersymmetric supergravity solution 
with the topology of a six-ball and for which the supersymmetric 
Killing vector $K$ may be written as  $K=\sum_{i=1}^3b_i\partial_{\varphi_i}$. 
In the present paper we chose orientation conventions so that $b_i>0$ for $i=1,2,3$. 
More generally we expect the orientations of $\partial_{\varphi_i}$ to be 
fixed as in \cite{Farquet:2014kma}, leading to the modulus signs in (\ref{free}). 
We shall comment further on this below.
We also conjecture that any supersymmetric gauge theory, with 
finite $N$, defined 
on the conformal boundary of such a supergravity solution depends only 
on $b_1, b_2, b_3$.

We have also computed certain BPS Wilson loops, both in supergravity and 
in the large $N$ gauge theories, again finding agreement. In this case 
we find that one can write the Wilson loop VEV as
\bea\label{FernandoWilson}
\log \, \langle\, W \, \rangle &=& \frac{|b_1|+|b_2|+|b_3|}{3|b_i|}\log \, \langle\, W \, \rangle_{\mathrm{AdS}_6}~,
\eea
where the Wilson loop wraps the $\varphi_i$ circle. Again, it is natural to conjecture 
that (\ref{FernandoWilson}) holds for general supergravity backgrounds with 
$U(1)^3$ symmetry and the topology of a six-ball. A general proof of 
the analogous formula to (\ref{FernandoWilson}) for the Wilson loop VEV in four  dimensions appears in \cite{Farquet:2014bda}. 

There are many natural directions which one could follow up. 
Firstly, it would be interesting to study supersymmetric gauge theories 
on a general class of supersymmetric background five-manifolds, 
generalizing the work done in lower dimensions in \cite{Klare:2012gn, Alday:2013lba, Closset:2012ru, Closset:2013vra}. One should then be able to prove (or disprove) 
the conjectures made above. In particular it would be interesting to study 
five-manifolds with different topology. Some work in this direction appears in 
\cite{Qiu:2013pta}, where the authors studied the case 
where the boundary is a Sasaki-Einstein manifold. It would also be 
very interesting to study systematically the geometry of 
Euclidean Romans supergravity backgrounds, as alluded to in section \ref{Fernandoothersection}. Here it is natural to expect that general supersymmetric 
solutions on the six-ball have a canonical complex structure, so that $M_6\cong \C^3$. 
If this is the case, then introducing standard complex coordinates $z_i=\rho_i\ex^{\ii\varphi_i}$, $i=1,2,3$, fixes the relative orientations of $\partial_{\varphi_i}$. 
In analysing the asymptotic expansion of the bulk Killing spinor equation, 
we have obtained a boundary charged conformal Killing
spinor equation, where the charge is with respect to both a one-form and 
also a two-form. To our knowledge, this type of equation has not been studied 
in the literature. In particular, it is an open problem to relate this equation 
to a more standard Killing spinor equation, of the type (\ref{5dKSEgeneral}), in general.


\subsection*{Acknowledgments}

\noindent 
The work of L.~F.~A., M.~F. and P.~R. is supported by ERC STG grant 306260. L.~F.~A. is a Wolfson Royal Society Research Merit Award holder.  J.~F.~S. is supported by the Royal Society. C. M. G. is supported by a CNPq scholarship.


\appendix

\section{Integrability conditions}\label{AppSUSYConditions}

Here we compute the integrability conditions for the Killing spinor equation (\ref{KSE}) 
and dilatino equation (\ref{dilatino}) of the Euclidean Romans theory.

 Recall that a supersymmetric solution must satisfy
\bea
D_\mu \epsilon_I & =&  \frac{\ii}{4\sqrt{2}} g ( X + \tfrac{1}{3} X^{-3} ) \Gamma_\mu \Gamma_7 \epsilon_I - \frac{1}{48} X^2 H_{\nu\rho\sigma} \Gamma^{\nu\rho\sigma} \Gamma_\mu \Gamma_7 \epsilon_I \label{KSEinApp}  \\
&&- \frac{\ii}{16\sqrt{2}} X^{-1} F_{\nu\rho} ( \Gamma_\mu{}^{\nu\rho} - 6 \delta_\mu{}^\nu \Gamma^\rho ) \epsilon_I + \frac{1}{16\sqrt{2}}X^{-1} F_{\nu\rho}^i ( \Gamma_\mu{}^{\nu\rho} - 6 \delta_\mu{}^\nu \Gamma^\rho ) \Gamma_7 ( \sigma^i )_I{}^J \epsilon_J ~,\nn
\eea
\bea
\delta \lambda_I \ = \ 0 & = & - \ii X^{-1} \partial_\mu X \Gamma^\mu \epsilon_ I + \frac{1}{2\sqrt{2}} g \left( X - X^{-3} \right) \Gamma_7 \epsilon_I + \frac{\ii}{24} X^2 H_{\mu\nu\rho} \Gamma^{\mu\nu\rho} \Gamma_7 \epsilon_I \nonumber \\
&&- \frac{1}{8\sqrt{2}} X^{-1} F_{\mu\nu} \Gamma^{\mu\nu} \epsilon_I - \frac{\ii}{8\sqrt{2}} X^{-1} F^i_{\mu\nu} \Gamma^{\mu\nu} \Gamma_7 ( \sigma^i )_I{}^J \epsilon_J~,\label{DilatinoinApp}
\eea
where $\lambda_I$ is the dilatino field.
Let us also record the component form of the Romans field equations in \eqref{FullEOM} and \eqref{Einstein}
\bea
 \left( E_g \right)_{\mu \nu} & \equiv &  R_{\mu \nu} - 4 X^{-2} \partial_\mu X \partial_\nu X - g^2 \left( \tfrac{1}{18} X^{-6} -\tfrac{1}{2} X^2 - \tfrac{2}{3} X^{-2} \right) g_{\mu\nu} \nn \\ & &- \tfrac{1}{4} X^4 ( H_\mu{}^{\rho \sigma} H_{\nu \rho \sigma} - \tfrac{1}{6} g_{\mu\nu} H^{\rho \sigma \tau} H_{\rho \sigma \tau} ) - \tfrac{1}{2} X^{-2} ( F_\mu{}^\rho F_{\nu\rho} - \tfrac{1}{8} g_{\mu\nu} F^{\rho\sigma} F_{\rho \sigma} )\nn \\ && - \tfrac{1}{2} X^{-2} ( F^{i\  \rho}_\mu{} F^i_{\nu \rho} - \tfrac{1}{8} g_{\mu\nu} F^{i\rho\sigma } F^i_{\rho\sigma} ) ~,\nn\\
 \left( E_X \right) &\equiv &  \nabla^\mu ( X^{-1} \partial_\mu X ) + g^2 \left( \tfrac{1}{2} X^2 - \tfrac{2}{3} X^{-2} + \tfrac{1}{6} X^{-6} \right) - \tfrac{1}{24} X^4 H^{\mu\nu\rho} H_{\mu\nu\rho} \nn\\ &&+ \tfrac{1}{16} X^{-2} ( F^{\mu\nu} F_{\mu\nu} + F^{i\mu\nu} F^i_{\mu\nu} )~, \nn\\
\left( E_A \right)^{\mu} &\equiv & \nabla_\nu ( X^{-2} F^{\nu\mu} ) - \tfrac{\ii}{12} \varepsilon^{\mu\nu\rho\sigma\tau\kappa} F_{\nu\rho} H_{\sigma\tau\kappa}~,\nn\\
\left( E_{A^i} \right)^{\mu}& \equiv &  D_\nu ( X^{-2} F^{i\nu\mu } ) - \tfrac{\ii}{12}  \varepsilon^{\mu\nu\rho\sigma\tau\kappa} F^i_{\nu\rho}H_{\sigma\tau\kappa}~,\nn\\
 \left( E_{B} \right)^{\mu \nu} & \equiv &  \nabla_\rho ( X^4 H^{\rho\mu\nu} ) - \tfrac{2}{3} g X^{-2} F^{\mu\nu} - \tfrac{\ii}{8} \varepsilon^{\mu\nu\rho\sigma\tau\kappa} ( F_{\rho\sigma} F_{\tau\kappa} + F^i_{\rho\sigma} F^i_{\tau\kappa} )~.
\eea
The equations of motion are then $E_{\mathrm{field}}=0$.
In addition, the gauge fields satisfy Bianchi identities $B_{\mathrm{field}}=0$, where we define
\begin{eqnarray}
 \left( B_{F} \right)_{\mu\nu\rho}  &\equiv & \nabla_{[\mu} F_{\nu\rho]} - \frac{2}{9} g H_{\mu\nu\rho}~,\nn \\
 \left( B_{F^i} \right)_{\mu\nu\rho}& \equiv &  D_{[\mu} F^i_{\nu\rho]}~,\nn \\ 
\left( B_{H} \right)_{\mu\nu\rho\sigma} &\equiv & \nabla_{[ \mu} H_{\nu\rho\sigma ]}~.
\end{eqnarray}

Taking the commutator of the Killing spinor equation \eqref{KSEinApp} we find the integrability condition to be
\bea
\mathcal{I}_{\mu\nu I}{}^J \epsilon_J \ = \ 0 ~,
\eea
where
\bea \label{Integrability}
\mathcal{I}_{\mu\nu I}{}^J \epsilon_J &= & \tfrac{1}{4} R_{\mu\nu\rho\sigma} \Gamma^{\rho\sigma} \epsilon_I + \tfrac{\ii}{2} g F^i_{\mu\nu} ( \sigma^i )_I{}^J \epsilon_J + \Big[- \tfrac{\ii}{4\sqrt{2}}  g ( 1 - X^{-4} ) \partial_\mu X \Gamma_\nu \Gamma_7 \epsilon_I\nn\\
&&+ \tfrac{1}{24} X \partial_\mu X H^{\rho\sigma\tau} \Gamma_{\rho\sigma\tau} \Gamma_\nu \Gamma_7 \epsilon_I + \tfrac{1}{48} X^2 \nabla_\mu H^{\rho\sigma\tau} \Gamma_{\rho\sigma\tau} \Gamma_\nu \Gamma_7 \epsilon_I \nonumber \\
&&- \tfrac{\ii}{16\sqrt{2}} X^{-2} \partial_\mu X F_{\rho\sigma}{J_{\nu}}^{\rho\sigma}  \epsilon_I + \tfrac{\ii}{16\sqrt{2}} X^{-1} \nabla_\mu F_{\rho\sigma}{J_{\nu}}^{\rho\sigma}  \epsilon_I \nn \\
&&+\tfrac{1}{16\sqrt{2}} X^{-2} \partial_\mu X F_{\rho\sigma}^i{J_{\nu}}^{\rho\sigma}  \Gamma_7 ( \sigma^i )_I{}^J \epsilon_J - \tfrac{1}{16\sqrt{2}} X^{-1} \nabla_\mu F_{\rho\sigma}^i{J_{\nu}}^{\rho\sigma}  \Gamma_7 ( \sigma^i )_I{}^J \epsilon_J \nn \\
&&- \tfrac{1}{32}  g^2 ( \tfrac{1}{9} X^{-6} + \tfrac{2}{3} X^{-2} + X^2 ) \Gamma_\nu \Gamma_\mu \epsilon_I- \tfrac{1}{2304} X^4 H^{\lambda\omega\theta} H^{\rho\sigma\tau} \Gamma_{\lambda\omega\theta} \Gamma_\nu \Gamma_{\rho\sigma\tau} \Gamma_\mu \epsilon_I \nn \\
&&+ \tfrac{1}{512} X^{-2} F_{\omega\theta} F_{\rho\sigma}{J_{\nu}}^{\omega\theta} {J_{\mu}}^{\rho\sigma} \epsilon_I 
+ \tfrac{1}{512} X^{-2} F_{\omega\theta}^i F_{\rho\sigma}^i {J_{\nu}}^{\omega\theta} {J_{\mu}}^{\rho\sigma} \epsilon_I \nn \\
&&+ \tfrac{\ii}{512} X^{-2} \varepsilon_{ijk} F_{\omega\theta}^i F_{\rho\sigma}^j {J_{\nu}}^{\omega\theta}{J_{\mu}}^{\rho\sigma}  ( \sigma^k )_I{}^J \epsilon_J \nn \\
&&+ \tfrac{\ii}{192\sqrt{2}}  g ( X^3 + \tfrac{1}{3} X^{-1} ) H^{\rho\sigma\tau} \Big( \Gamma_\nu \Gamma_{\rho\sigma\tau} \Gamma_\mu - \Gamma_{\rho\sigma\tau} \Gamma_\nu \Gamma_\mu \Big) \epsilon_I \nn \\
&&+ \tfrac{1}{128} g X^{-1} ( X + \tfrac{1}{3} X^{-3} ) F_{\rho\sigma} \Big( \Gamma_\nu{J_{\mu}}^{\rho\sigma}  -{J_{\nu}}^{\rho\sigma} \Gamma_\mu \Big) \Gamma_7 \epsilon_I  \nn \\
&&+ \tfrac{\ii }{128} g X^{-1} ( X + \tfrac{1}{3} X^{-3} ) F_{\rho\sigma}^i \Big( \Gamma_\nu{J_{\mu}}^{\rho\sigma}  + {J_{\nu}}^{\rho\sigma} \Gamma_\mu \Big) ( \sigma^i )_I{}^J \epsilon_J \nn \\
&&+ \tfrac{\ii}{768\sqrt{2}} X F_{\rho\sigma} H^{\lambda\omega\theta} \Big( \Gamma_{\lambda\omega\theta} \Gamma_\nu {J_{\mu}}^{\rho\sigma}  -{J_{\nu}}^{\rho\sigma}  \Gamma_{\lambda\omega\theta} \Gamma_\mu \Big) \Gamma_7 \epsilon_I \nn \\
&&- \tfrac{1}{768\sqrt{2}} X F_{\rho\sigma}^i H^{\lambda\omega\theta} \Big( \Gamma_{\lambda\omega\theta} \Gamma_\nu {J_{\mu}}^{\rho\sigma}  -{J_{\nu}}^{\rho\sigma}  \Gamma_{\lambda\omega\theta} \Gamma_\mu \Big) ( \sigma^i )_I{}^J \epsilon_J \nn \\
&&+\tfrac{\ii}{512} X^{-2} ( F_{\rho\sigma} F_{\omega\theta}^i - F_{\omega\theta} F_{\rho\sigma}^i ) {J_{\nu}}^{\rho\sigma} {J_{\mu}}^{\omega\theta}  \Gamma_7 ( \sigma^i )_I{}^J \epsilon_J - ( \mu \leftrightarrow \nu ) \Big]~,
\eea
and we have defined the Clifford algebra element
\bea
{J_{\mu}}^{\rho\sigma} & \equiv & {\Gamma_\mu}^{\rho\sigma}-6{\delta_{\mu}}^\rho\Gamma^\sigma~.
\eea

Taking the covariant derivative of the dilatino equation \eqref{DilatinoinApp} and contracting with $\Gamma^\mu$ leads to
\bea
&&\Gamma^\mu D_\mu ( \delta\lambda_I ) - \tfrac{\ii}{2\sqrt{2}} g ( X - \tfrac{7}{3} X^{-3} ) \Gamma_7 \delta \lambda_I +\tfrac{1}{24} X^2 H_{\mu\nu\rho} \Gamma^{\mu\nu\rho} \Gamma_7 \delta \lambda_I  \\
&&+ \tfrac{\ii}{8\sqrt{2}} X^{-1} F_{\mu\nu} \Gamma^{\mu\nu} \delta \lambda_I+ \tfrac{1}{8\sqrt{2}} X^{-1} F^i_{\mu\nu} \Gamma^{\mu\nu} \Gamma_7 ( \sigma^i )_I{}^J \delta \lambda_J \nn \\
&=&  \ii \left(E_X\right) \epsilon_I - \tfrac{1}{4\sqrt{2}} X \left( E_A \right)_{\mu} \Gamma^\mu \epsilon_I - \tfrac{\ii}{4\sqrt{2}} X \left( E_{A^i} \right)_{\mu} \Gamma^\mu \Gamma_7 ( \sigma^i )_I{}^J \epsilon_J + \tfrac{\ii}{8} X^{-2} \left( E_{B} \right)_{\mu \nu} \Gamma^{\mu\nu} \Gamma_7 \epsilon_I \nn\\
&&- \tfrac{1}{8\sqrt{2}} X^{-1} \left( B_{F} \right)_{\mu\nu\rho} \Gamma^{\mu\nu\rho} \epsilon_I - \tfrac{\ii}{8\sqrt{2}} X^{-1} \left( B_{F^i} \right)_{\mu\nu\rho} \Gamma^{\mu\nu\rho} \Gamma_7 ( \sigma^i )_I{}^J \epsilon_J \nn\\&&+ \tfrac{\ii}{24} X^2 \left( B_{H} \right)_{\mu\nu\rho\sigma} \Gamma^{\mu\nu\rho\sigma} \Gamma_7 \epsilon_I~.\nn
\eea
We may similarly contract $\mathcal{I}_{\mu\nu I}{}^J \epsilon_J$ with $\Gamma^\nu$. After a very lengthy calculation we find
\bea
&&\Gamma^\nu \mathcal{I}_{\mu\nu I}{}^J \epsilon_J + \tfrac{\ii}{2} \Gamma_\mu \Gamma_\nu D^\nu ( \delta \lambda_I ) + 2 \ii X^{-1} \partial_\mu X \delta \lambda_I + \tfrac{1}{2\sqrt{2}} g ( X - \tfrac{5}{3} X^{-3} ) \Gamma_\mu \Gamma_7 \delta \lambda_I   \nn\\
&&- \tfrac{\ii}{16} X^2 H_{\mu\nu\rho} \Gamma^{\nu\rho} \Gamma_7 \delta \lambda_I + \tfrac{\ii}{16} X^2 H^{\nu\rho\sigma} \Gamma_{\mu\nu\rho\sigma} \Gamma_7 \delta \lambda_I- \tfrac{1}{8 \sqrt{2}} X^{-1} F^{\nu \rho}\Gamma_{\mu \nu \rho} \delta \lambda_I \nn\\
&&+\tfrac{1}{4 \sqrt{2}} X^{-1} F_{\mu \nu}\Gamma^{\nu} \delta \lambda_I  -\tfrac{\ii}{4 \sqrt{2}} X^{-1} F^{i}_{\mu \nu}\Gamma^{\nu} \Gamma_7 ( \sigma^{i} )_{I}{}^{J} \delta \lambda_J + \tfrac{\ii}{8 \sqrt{2}} X^{-1} F^{i  \nu \rho}\Gamma_{\mu \nu \rho} \Gamma_7 ( \sigma^{i} )_{I}{}^{J} \delta \lambda_J\nn \\
&=&  \tfrac{1}{2}\left(E_X\right) \Gamma_\mu \epsilon_ I - \tfrac{1}{2} \left( E_g \right)_{\mu \nu} \Gamma^\nu \epsilon_I  - \tfrac{1}{8} X^{-2} \left( E_B \right)^{\nu\rho} \Gamma_{\mu\nu\rho} \Gamma_7 \epsilon_I \nn \\
&&-\tfrac{\ii}{2 \sqrt{2}} X \left( E_A \right)_{\mu} \epsilon_I+\tfrac{1}{2 \sqrt{2}} X \left( E_{A^{i}} \right)_{\mu} \Gamma_7 ( \sigma^{i} )_{I}{}^{J} \epsilon_J - \tfrac{1}{24} X^2 \left( B_H \right)^{\nu \rho \sigma \tau} \Gamma_{\mu \nu \rho \sigma \tau} \Gamma_7 \epsilon_I\nn \\
&& - \tfrac{3 \ii}{4 \sqrt{2}} X^{-1} \left( B_F \right)_{\mu \nu \rho} \Gamma^{\nu \rho} \epsilon_I + \tfrac{3}{4 \sqrt{2}} X^{-1} \left( B_{F^{i}} \right)_{\mu \nu \rho} \Gamma^{\nu \rho} \Gamma_7 ( \sigma^{i})_{I}{}^{J} \epsilon_J~.
\eea

\section{Supersymmetric supergravity solutions}\label{AppSolutions}

\subsection{The equations}\label{Appeqns}

The solutions found in this paper arise from the following $SU(3) \times U(1)$ symmetric ansatz for the supergravity fields
\begin{eqnarray}\label{ansatzapp}
\dd s^2_6 &=& \alpha^2(r)\dd r^2+\gamma^2(r)(\dd\tau+C)^2+\beta^2(r)\Big[\dd \sigma^2 + \frac{1}{4}\sin^2\sigma(\dd \theta^2+\sin^2\theta \dd \varphi^2)\nn\\
&&+\frac{1}{4}\cos^2\sigma\sin^2\sigma (\dd \psi+\cos\theta \dd\varphi)^2\Big]~,\nn\\
B&=& p(r)\dd r\wedge (\dd\tau+C) + \frac{1}{2}q(r)\dd C~,\nn\\
A^i &=& f^i(r)(\dd\tau +C)~,
\end{eqnarray}
together with $X=X(r)$. The equations of motion for the background $SU(2)_R$ gauge field imply
\begin{equation}
f^i(r) \ = \ \kappa_i f(r)~.
\end{equation}
The equations for the other fields then depend only on the $SU(2)\sim SO(3)$ invariant $\kappa_1^2+\kappa_2^2+\kappa_3^2$, which we can set to one by rescaling $f(r)$. Explicitly, 
one finds that substituting the ansatz (\ref{ansatzapp}) into the equations 
of motion (\ref{FullEOM}) and Einstein equation (\ref{Einstein}) leads to following 
coupled system of ODEs:
\bea
\frac{\lambda\gamma X^4}{\alpha} &=&  \ii f^2 + \ii \frac{q^2}{9} + \frac{p\beta^4}{9\alpha\gamma X^2}~,
\eea
\bea
\left(\frac{\lambda\gamma X^4}{\alpha}\right)' &=&  2 \ii f f' + \ii \left(\frac{2}{3}\right)^2pq +  \left(\frac{2}{3}\right)^2\frac{q\alpha\gamma}{X^2}~,
\eea
\bea
\left(\frac{\beta^4f'}{2\alpha\gamma X^2}\right)'-\frac{4\alpha\gamma f}{X^2} \ = \ - 2 \ii f \lambda~,
\eea
\bea
\frac{\alpha}{\gamma\beta^4}\left(\frac{\gamma\beta^4 X'}{\alpha X}\right)'  & = &  - \frac{1}{8X^2}\left(\frac{f'^2}{\gamma^2}+ \frac{8\alpha^2f^2}{\beta^4}\right) - \left(\frac{2}{3}\right)^2 \frac{1}{8X^2}\left(\frac{p^2}{\gamma^2}+ 2\frac{\alpha^2q^2}{\beta^4}\right)\nn\\
&& + \frac{X^4\lambda^2}{2\beta^4} - \frac{\alpha^2}{6X^6}+\frac{2\alpha^2}{3X^2}-\frac{\alpha^2X^2}{2}~,
\eea
\bea
-\frac{\beta''}{\beta}+\frac{\beta'}{\beta}\frac{(\alpha\gamma)'}{\alpha\gamma} - \frac{(\alpha\gamma)^2}{\beta^4} \ = \ \left(\frac{X'}{X}\right)^2 
+ \frac{X^4\lambda^2}{4\beta^4}~,
\eea
\bea
&&-\frac{\gamma''}{\gamma}+\frac{\beta''}{\beta}+\frac{\alpha'}{\alpha}\left(\frac{\gamma'}{\gamma}-\frac{\beta'}{\beta}\right)
- 3\frac{\beta'}{\beta}\left(\frac{\gamma'}{\gamma}-\frac{\beta'}{\beta}\right)+\frac{6\alpha^2}{\beta^4}(\gamma^2-\beta^2) \nonumber\\
&& \ = \ -\frac{X^4\lambda^2}{2\beta^4} + \frac{1}{2X^2}\left(\frac{f'^2}{\gamma^2}-\frac{4\alpha^2 f^2}{\beta^4}\right)+ \left(\frac{2}{3}\right)^2\frac{1}{2X^2}\left(\frac{p^2}{\gamma^2}-\frac{\alpha^2q^2}{\beta^4}\right)~,
\eea
\bea
-\frac{\gamma''}{\gamma}+\frac{\alpha'}{\alpha}\frac{\gamma'}{\gamma} - 4 \frac{\beta'}{\beta}\frac{\gamma'}{\gamma} + 4\frac{(\alpha\gamma)^2}{\beta^4} & = &  \frac{\alpha^2}{18X^6}-\frac{2\alpha^2}{3X^2}-\frac{\alpha^2X^2}{2} - \frac{X^4\lambda^2}{2\beta^4} \nn\\
&& + \frac{1}{2X^2}\left[\frac{f'^2}{\gamma^2}-\frac{1}{4}\left(\frac{f'^2}{\gamma^2}+ \frac{8\alpha^2 f^2}{\beta^4}\right) \right]\\
&& +\left(\frac{2}{3}\right)^2 \frac{1}{2X^2}\left[\frac{p^2}{\gamma^2}-\frac{1}{4}\left(\frac{p^2}{\gamma^2}+ \frac{2\alpha^2q^2}{\beta^4}\right) \right] \nn~.
\eea
where we have introduced $\lambda=q'-2p$. These are seven equations for seven functions. In addition one can explicitly check that the equations are invariant under changes in the parametrization $r\rightarrow \rho(r)$. 

\subsection{General solutions}

Before writing the general series solutions to the above coupled system of ODEs, 
let us present the solution for Euclidean AdS$_6$ in these coordinates:
\begin{eqnarray}\label{AdS6app}
\alpha(r) & =&  \frac{3\sqrt{3}}{\sqrt{6r^2-1}}~, \qquad 
\beta(r) \ = \ \gamma(r) \ = \  \frac{3 \sqrt{6r^2-1}}{\sqrt{2}}~,\nn\\
p(r) & =& q(r)  \ = \ f(r) \ =\ 0~, \qquad X(r) \ =\ 1~.
\end{eqnarray}
Here only the metric is non-trivial, and the above realizes Euclidean 
AdS$_6$ as a hyperbolic ball with radial coordinate $r\in [\frac{1}{\sqrt{6}},\infty)$, 
with the conformal boundary at infinity $r=\infty$. The point $r=\frac{1}{\sqrt{6}}$ is 
the origin of the ball, where the transverse copies of $S^5$ collapse smoothly to zero. 
Notice in particular that the conformal boundary at $r=\infty$ is equipped 
with a \emph{round} metric on $S^5$, which is conformally flat.
 We would like to find families of solutions that generalize (\ref{AdS6app}) by allowing 
for a squashed five-sphere boundary, keeping the metric 
asymptotically locally Euclidean AdS near $r=\infty$. We define the squashing parameter by:
\begin{equation}
\lim_{r \rightarrow \infty} \frac{\gamma(r)}{r} \ =\  3 \sqrt{3} ~\frac{1}{s}~,
\end{equation}
so that $s=1$ for the round sphere. Even though we did not manage to find solutions in closed form, the solutions can nevertheless be given as expansions around different limits. In general notice that we can use reparametrization invariance to set
\begin{equation}\label{betafixedapp}
\beta(r) \ = \ \frac{3 \sqrt{6r^2-1}}{\sqrt{2}}~,
\end{equation}
which we assume henceforth. In particular we shall only seek solutions 
with the topology of a ball, so that from (\ref{betafixedapp})  necessarily $r=\frac{1}{\sqrt{6}}$ is the origin of the ball. Correspondingly, the fields must 
satisfy certain boundary conditions at this point in order that the full 
solution is smooth at the origin.

\subsubsection{Expansion around the conformal boundary}

When finding gravity duals to a given boundary theory, it is natural to perform an expansion around the conformal boundary at $r=\infty$. This also has the advantage that the squashing parameter can be explicitly seen in the solution. Starting from a general expansion and imposing the equations of motion in section \ref{Appeqns} we find
\begin{eqnarray}
\alpha(r) &=& \frac{3}{\sqrt{2}} \frac{1}{r}+ \frac{486+ q_0^2 s^2}{1944 \sqrt{2} s^2} \frac{1}{r^3}+\ldots~,\nn \\
\gamma(r)&=& \frac{3 \sqrt{3}}{s} r +\frac{-486+\left(243-q_0^2\right) s^2}{324 \sqrt{3} s^3 } \frac{1}{r}+\ldots~, \nn\\
X(r) &=& 1 + \frac{-486 q_0+72 \ii \sqrt{6} q_0^2 s+486 q_0 s^2 +7 q_0^3 s^2+5832 s^2 q_2}{11664 q_0 s^2 } \frac{1}{r^2} + \frac{x_3}{r^3}+\ldots~,\nn\\
p(r)&=& \frac{q_0 \left(54 -\sqrt{6} \ii q_0 s\right)}{162 s^2 } \frac{1}{r^2}+\ldots~, \nn\\
q(r) &=& q_0 r+ \frac{q_2}{r}+\frac{q_3}{r^2}+\ldots~, \nn\\
f(r)&=& f_0 - \frac{f_0 \left(54 -\sqrt{6} \ii q_0 s\right)}{81 s^2 } \frac{1}{r^2}+ \frac{f_3}{r^3}+\ldots~.
\end{eqnarray}
In addition to the squashing parameter $s$, the solution depends on $q_0,f_0,f_3,q_2,q_3,x_3$ and an extra parameter $\alpha_5$, which appears at higher order in the expansion for $\alpha(r)$. All other coefficients in the expansion are fixed in terms of these constants. Of course, some of these parameters will be fixed in the full solution by requiring the correct boundary conditions at the origin $r=\tfrac{1}{\sqrt{6}}$, but at this point they are arbitrary. 

\subsubsection{Expansion around Euclidean AdS}

The family of solutions we seek should approach Euclidean AdS$_6$ (\ref{AdS6app}) as we take the squashing parameter $s\rightarrow 1$. Hence it should be possible to expand the solutions around this limit in terms of a perturbation parameter $\delta$. Thus we make the ansatz
\begin{eqnarray} 
\alpha(r)&=& \frac{3\sqrt{3}}{\sqrt{6r^2-1}} + \delta ~\alpha^{(1)}(r)+ \delta^2 ~\alpha^{(2)}(r)+\ldots~,\nn\\
\gamma(r)&=& \frac{3 \sqrt{6r^2-1}}{\sqrt{2}} + \delta ~\gamma^{(1)}(r)+ \delta^2 ~\gamma^{(2)}(r)+\ldots~,\nn \\
X(r)&=&1+\delta ~X^{(1)}(r)+\delta^2~ X^{(2)}(r)+\ldots~,\nn\\
p(r)&=&\delta~ p^{(1)}(r)+\delta^2~ p^{(2)}(r)+\ldots~,\nn\\
q(r)&=& \delta~ q^{(1)}(r)+\delta^2~ q^{(2)}(r)+\ldots~,\nn\\
f(r)&=& \delta~ f^{(1)}(r)+\delta^2~ f^{(2)}(r)+\ldots~.
\end{eqnarray}
Substituting this expansion into the equations of motion and expanding in powers of $\delta$, at each order we obtain a system of linear differential equations which can be solved in closed form with some effort. For instance, at first order we find
\begin{eqnarray}
\alpha^{(1)}(r)&=& -c_\gamma \frac{\left(1-54 r^2+96 \sqrt{6} r^3-324 r^4+216 r^6\right)}{\sqrt{6} r^2 \left(6 r^2-1\right)^{7/2}}~,\nn\\
\gamma^{(1)}(r)&=&c_\gamma\frac{\left(-5+16 \sqrt{6} r-90 r^2+180 r^4-216 r^6\right)}{\left(6 r^2-1\right)^{5/2}}~,\nn\\
 X^{(1)}(r)&=&c_x \frac{\left(1-2 \sqrt{6} r+6 r^2\right)}{ \left(6 r^2-1\right)^2}~,\nn\\
p^{(1)}(r)&=&c_q \frac{ \left(\sqrt{6}-16 r+12 \sqrt{6} r^2-12 \sqrt{6} r^4\right)}{3 \left(6 r^2-1\right)^3}~,\nn\\
 q^{(1)}(r)&=&-c_q \frac{\left(-4+9 \sqrt{6} r-24 r^2-12 \sqrt{6} r^3+36 \sqrt{6} r^5\right)}{18 \left(6 r^2-1\right)^2}~,\nn\\
 f^{(1)}(r)&=&c_f \frac{\left(-3+8 \sqrt{6} r-36 r^2+36 r^4\right)}{\left(6 r^2-1\right)^2}~.
\end{eqnarray}
The constants of integration have been partially fixed by requiring regularity at the origin $r=\frac{1}{\sqrt{6}}$. In particular we have
\bea
\alpha^{(1)}(r) & \sim & \left(r-\frac{1}{\sqrt{6}}\right)^{1/2}~, \qquad \gamma^{(1)}(r) \ \sim \ \left(r-\frac{1}{\sqrt{6}}\right)^{3/2}~,\nn\\
p^{(1)}(r)& \sim & 1 \ \sim \ X^{(1)}(r)~, \qquad q^{(1)}(r)\ \sim \  \left(r-\frac{1}{\sqrt{6}}\right)\ \sim \ f^{(1)}(r)~.
\eea
Here $\rho\sim (r-\tfrac{1}{\sqrt{6}})^{1/2}$ is geodesic distance from the origin at $\rho=0$. 
 We can furthermore fix an extra constant of integration by fixing a relation between $\delta$ and the squashing parameter $s$ (such that $\delta \rightarrow 0$ as $s \rightarrow 1$). As seen in the next section it will be convenient not to do this uniformly. 

\subsection{Imposing supersymmetry}

We are interested in solutions that preserve some supersymmetry. In order for this to happen, there should exist  non-trivial eight-component Killing spinors $\epsilon_1,\epsilon_2$ 
solving the Killing spinor equation (\ref{KSE}) and dilatino equation (\ref{dilatino}). We choose the frame
\begin{eqnarray}
e^{0} & =& \alpha(r) \dd r~,~\qquad \qquad  \ \, \quad e^{1}\ =\ \gamma(r) (\dd\tau +C)~,~~~e^{2} \ = \ \beta(r) \dd\sigma~,\\
e^{3} & =& \frac{1}{2}\beta(r) \sin\sigma \cos \sigma \tau_3~,~~~e^{4} \ = \ \frac{1}{2}\beta(r) \sin\sigma \tau_2~,~~~e^{5} \ = \ \frac{1}{2}\beta(r) \sin\sigma \tau_1~,\nn
\end{eqnarray}
and the following basis for six-dimensional gamma matrices
\bea
  \Gamma_0 & = &  \left( \begin{array}{cc}  0 & 1_4 \\ 1_4 & 0 \end{array} \right) \, , \quad \Gamma_m \  = \  \left( \begin{array}{cc}  0 & \ii  \gamma_m \\ - \ii \gamma_m & 0 \end{array} \right)\, , \  \ m \, = \, 1, \ldots, 5~, \nn\\ \Gamma_7 &=&  \left( \begin{array}{cc}  - 1_4 & 0 \\ 0 & 1_4 \end{array} \right) \, ,
\eea
where $1_4$ is the $4 \times 4$ unit matrix and $\gamma_m$ are the five-dimensional gamma matrices given explicitly in section 2.1.

 The vanishing of the dilatino variation as well as each component of the integrability condition (\ref{Integrability}) for the Killing spinor equation have the following general structure 
\begin{eqnarray}
P \epsilon_1 +Q \epsilon_2 & =& 0~,\nn\\
R \epsilon_1 +S \epsilon_2 & =& 0~,
\end{eqnarray}
where $P,Q,R,S$ are $8 \times 8$ matrices, whose components are in general complicated functions of the fields. After setting $f_i(r)=\kappa_i f(r)$ we observe the following $SU(2)_R$ structure
\bea
\left(
\begin{array}{cc}
 A + \kappa_3 B &   (\kappa_1 - \ii \kappa_2)B  \\
 (\kappa_1 + \ii \kappa_2)B &   A  -\kappa_3 B  
\end{array}
\right) \left( \begin{array}{c}  \epsilon_1 \\ \epsilon_2 \end{array}\right) \ = \ 0~,
\eea
in terms of $8 \times 8$ matrices $A,B$. We can then diagonalize the block matrix and consider the equivalent problem
\begin{equation}
\left(
\begin{array}{cc}
 A +B &  0  \\
0 &   A  - B  
\end{array}
\right) \left( \begin{array}{c}  \epsilon_1 \\ \epsilon_2 \end{array}\right) \ = \ 0~,
\end{equation}
where we have without loss of generality set $\kappa_1^2+\kappa_2^2+\kappa_3^2=1$. There are four independent conditions. One of these arises from the dilatino variation, whose matrices we denote by $A_0,B_0$, and the other  three conditions arise from integrability 
of the Killing spinor equation, whose matrices we denote by $A_M,B_M$ with $M \in \{12,13,34\}$ (all other components of the integrability condition (\ref{Integrability}) are equivalent to one of these). The dilatino condition as well as $M=12$ and $M=34$ have the following structure:
\bea
A \pm B  & =& 
\left(
\begin{array}{cccccccc}
 * & 0 & 0 & 0 & * & 0 & 0 & 0 \\
 0 & * & 0 & 0 & 0 & * & 0 & 0 \\
 0 & 0 & * & 0 & 0 & 0 & * & 0 \\
 0 & 0 & 0 & * & 0 & 0 & 0 & * \\
 * & 0 & 0 & 0 & * & 0 & 0 & 0 \\
 0 & * & 0 & 0 & 0 & * & 0 & 0 \\
 0 & 0 & * & 0 & 0 & 0 & * & 0 \\
0 & 0 & 0 & * & 0 & 0 & 0 & *\\
\end{array}
\right)~.
\eea
The existence of a non-trivial solution requires, for instance, $\det(A + B)=0$. The above structure implies the determinant factorizes into four factors
\begin{equation}
\det(A + B) \ = \  F_1 F_2 F_3 F_4  \ =\ 0~, 
\end{equation}
where the factors $F_i$ are complicated functions of the supergravity fields 
$\alpha(r)$, $\beta(r)$, $\gamma(r)$, $p(r)$, $q(r)$, $f(r)$, $X(r)$. $F_1$ and $F_3$ differ only by a change of sign in $f(r)$, and the same happens for $F_2$ and $F_4$. We  find two distinct classes of solutions which we describe in the following.

\subsubsection{$3/4$ BPS solutions}
There is a class of solutions that satisfies
\begin{equation}
F_1 \ =\ F_2\ =\ F_3\ =\ 0~, \qquad F_4 \ \neq \ 0~.
\end{equation}
These are a one-parameter family of solutions parametrized by the squashing parameter $s$. The solution expanded around the conformal boundary is given by
\begin{eqnarray}
\alpha(r)&=&\frac{3}{\sqrt{2}} \frac{1}{r}+\frac{8+s^2}{36 \sqrt{2} s^2}\frac{1}{r^3}+\ldots~,~~~\\
\gamma(r)&=&\frac{3 \sqrt{3}}{s} r+\frac{-16+7 s^2}{12 \sqrt{3} s^3 }\frac{1}{r}-\frac{-1280+1120 s^2+241 s^4}{2592 \sqrt{3} s^5}\frac{1}{r^3}+\ldots~,\nn\\
X(r)&=&1+\frac{1-s^2-3  \sqrt{1-s^2}}{54 s^2}\frac{1}{r^2}+\frac{s^2 \sqrt{1-s^2} \kappa}{12 \left(1- s^2+\sqrt{1-s^2}\right)}\frac{1}{r^3}+\ldots~,\nn\\
 p(r) &=& -\frac{ \ii \sqrt{\frac{2}{3}} \left(s^2+3\sqrt{1-s^2}-1\right)}{s^3}\frac{1}{r^2}+\ldots~,\nn\\
 q(r) &=& -\frac{3 \ii \left(\sqrt{6} \sqrt{1-s^2}\right)}{s} r 
 + \frac{\sqrt{\frac{2}{3}} \ii \sqrt{1-s^2} \left(5 s^2+9  \sqrt{1-s^2}-5\right)}{3 s^3}\frac{1}{r}+\ldots~,\nn\\
f(r) &=& \frac{1-s^2+\sqrt{1-s^2}}{s^2} +\frac{2 \left(-2+2 s^2- (2+s^2) \sqrt{1-s^2}\right)}{9 s^4}\frac{1}{r^2}+\frac{\kappa}{r^3}+\ldots~.\nn
\end{eqnarray}
The extra parameter $\kappa$ is fixed by requiring regularity at the origin. The solution expanded around Euclidean AdS$_6$ has $c_\gamma=0$, hence it is convenient to set the relation between the expansion parameter and the squashing parameter to be
\bea
\frac{1}{s} \ =\ 1+\delta^2~.
\eea
With this choice the solution is given by
\begin{eqnarray}
\alpha(r)&=&\frac{3 \sqrt{3}}{\sqrt{6 r^2-1}}+\tfrac{\left(-5 \sqrt{6}+330 \sqrt{6} r^2-3744 r^3+1620 \sqrt{6} r^4+8640 r^5-7560 \sqrt{6} r^6+5184 \sqrt{6} r^8\right)}{9 \sqrt{2} r^2 \left(6 r^2-1\right)^{9/2}} \delta ^2+\ldots~,\nn\\
\gamma(r)&=&\frac{3 \sqrt{6 r^2-1}}{\sqrt{2}}\nn\\
&&-\tfrac{\left(55 \sqrt{2}-384 \sqrt{3} r+1080 \sqrt{2} r^2+768 \sqrt{3} r^3-5400 \sqrt{2} r^4+11232 \sqrt{2} r^6-11664 \sqrt{2} r^8\right) }{6 \left(6 r^2-1\right)^{7/2}}\delta ^2+\ldots~,\nn\\
X(r) &=& 1-\frac{\left(\sqrt{2} \left(1-2 \sqrt{6} r+6 r^2\right)\right)}{3 \left(6 r^2-1\right)^2}\delta +\ldots~,\nn\\
p(r) &=& \frac{18 \ii \sqrt{2} \left(\sqrt{6}-16 r+12 \sqrt{6} r^2-12 \sqrt{6} r^4\right) }{\left(6 r^2-1\right)^3} \delta+\ldots~,\nn\\
q(r) &=& -\frac{3 \ii \sqrt{2} \left(-4+9 \sqrt{6} r-24 r^2-12 \sqrt{6} r^3+36 \sqrt{6} r^5\right)}{\left(6 r^2-1\right)^2} \delta +\ldots~,\nn\\
f(r) &=& \frac{\sqrt{2} \left(-3+8 \sqrt{6} r-36 r^2+36 r^4\right)}{\left(6 r^2-1\right)^2} \delta +\ldots~.
\end{eqnarray}
We have computed the solution up to sixth order in $\delta$. Comparing this expansion with the expansion around the conformal boundary we can compute the coefficient $\kappa$ as a series expansion in $\delta$. We obtain
\begin{equation}
\frac{3\sqrt{3}}{4} \kappa \ = \ \delta +\frac{\sqrt{2}}{3}\delta ^2+\frac{113 }{36}\delta ^3+\frac{25}{9 \sqrt{2}} \delta ^4+\frac{1127}{288}\delta ^5+\frac{35 }{9 \sqrt{2}}\delta ^6+\ldots~.
\end{equation}

\subsubsection{$1/4$ BPS solutions}

There is another class of supersymmetric solutions that satisfies
\begin{equation}
F_1~, \, F_2~, \, F_3 \ \neq \ 0~,\qquad F_4 \ =\  0~.
\end{equation}
These are a two-parameter family of solutions and are parametrized by the squashing parameter $s$ and the background $SU(2)_R$ field at the conformal boundary, which is 
parametrized by $f_0$. The solution 
expanded around the conformal boundary is given by
\begin{eqnarray}
\alpha(r) &=& \frac{3}{\sqrt{2}}\frac{1}{r}-\frac{f_0^2 s^2+9 \left(-2+s^2\right)-6 f_0 \left(-1+s^2\right)}{36 \sqrt{2} }\frac{1}{r^3}+\ldots~,\nn\\
\gamma(r) &=& \frac{3 \sqrt{3}}{s} r+\frac{2 f_0^2 s^2-12 f_0 \left(-1+s^2\right)+9 \left(-3+2 s^2\right)}{12 \sqrt{3} s}\frac{1}{r}+\ldots~,\nn\\
X(r) &=& 1+\frac{18-3 f_0-18 s^2+12 f_0 s^2-2 f_0^2 s^2}{54} \frac{1}{r^2}+\ldots~,\nn\\
p(r) &=& \frac{\ii \sqrt{\frac{2}{3}} (-3+f_0) \left(3+(-3+f_0) s^2\right)}{s}\frac{1}{r^2}+\ldots~,\nn\\
q(r) &=& -\frac{3 \ii \sqrt{6} \left(3+(-3+f_0) s^2\right)}{s} r\nn\\
&&+\frac{\ii \left(3+(-3+f_0) s^2\right) \left(f_0^2 s^2+9 \left(-1+s^2\right)-6 f_0 \left(1+s^2\right)\right)}{6 \sqrt{6} s } \frac{1}{r}+\frac{\xi_1}{r^2}+\ldots~,\nn\\
f(r) &=& f_0+\frac{2 (-3+f_0) f_0}{9}\frac{1}{r^2}+\frac{\xi_2}{r^3}+\ldots~.
\end{eqnarray}
The constants $\xi_1$ and $\xi_2$ are fixed by requiring regularity at the origin. Note that a particular case corresponds to $f_0=0$. In this case the $SU(2)_R$ background field is turned off, but the solution is still supersymmetric with a squashed five-sphere at the
conformal boundary. In this case $F_4=F_2=0$, so we have enhanced supersymmetry; that is, this  one-parameter family of solutions with $f_0=0$ is  $1/2$ BPS.

As an expansion around Euclidean AdS we parametrize the solution in terms of the expansion parameter $\delta$ and an extra parameter $\omega$, related to $s$ and $f_0$ above by
\begin{equation}
\frac{1}{s}\ = \ 1+\delta~, \qquad f_0 \ = \ \delta \,\omega~.
\end{equation}
With this choice the solution is given by
\begin{eqnarray} 
\alpha(r)&=& \frac{3\sqrt{3}}{\sqrt{6r^2-1}} +\frac{\sqrt{3} \left(1-54 r^2+96 \sqrt{6} r^3-324 r^4+216 r^6\right)}{2 r^2 \left(6 r^2-1\right)^{7/2}} \delta+\ldots~,\nn\\
\gamma(r)&=& \frac{3 \sqrt{6r^2-1}}{\sqrt{2}} +\frac{\left(15-48 \sqrt{6} r+270 r^2-540 r^4+648 r^6\right)}{\sqrt{2} \left(6 r^2-1\right)^{5/2}} \delta+\ldots~,\nn\\
X(r)&=&1+\frac{\left(1-2 \sqrt{6} r+6 r^2\right) (4+\omega )}{\left(6 r^2-1\right)^2} \delta +\ldots~,\nn\\
p(r)&=&-\frac{18 \ii \sqrt{2} \left(-\sqrt{3}+8 \sqrt{2} r-12 \sqrt{3} r^2+12 \sqrt{3} r^4\right) (6+\omega )}{\left(6 r^2-1\right)^3} \delta +\ldots~,\nn\\
q(r)&=& -\frac{3 \ii \left(-4+9 \sqrt{6} r-24 r^2-12 \sqrt{6} r^3+36 \sqrt{6} r^5\right) (6+\omega )}{\left(6 r^2-1\right)^2} \delta+\ldots~,\nn\\
f(r)&=& \frac{\left(-3+8 \sqrt{6} r-36 r^2+36 r^4\right) \omega}{\left(6 r^2-1\right)^2} \delta+\ldots~.
\end{eqnarray}
As before it can be  checked explicitly that the solution is regular at $r=\frac{1}{\sqrt{6}}$. We have computed this solution explicitly up to fourth order in $\delta$. Comparing this expansion with the expansion around the conformal boundary we deduce
\begin{eqnarray}
\xi_1&=&2 \ii (6+\omega ) \delta -\frac{1}{5} \ii \left(144+98 \omega +13 \omega ^2\right) \delta ^2 \\ 
&+&\frac{\ii \left(307719+209547 \omega +41094 \omega ^2+1282 \omega ^3\right)}{9450} \delta ^3\nn \\
&-&\frac{\ii \left(26693550+21683700 \omega +6126111 \omega ^2+771474 \omega ^3+51568 \omega ^4\right)}{623700}  \delta ^4+\ldots~,\nn\\
\xi_2&=&\frac{2}{3} \sqrt{\frac{2}{3}} \omega  \delta -\frac{2}{45} \left(-\sqrt{6} \omega +2 \sqrt{6} \omega ^2\right) \delta ^2+\frac{\left(-999 \sqrt{6} \omega -594 \sqrt{6} \omega ^2+244 \sqrt{6} \omega ^3\right)}{42525}\delta ^3 \nn\\
&+&\frac{\left(32724 \sqrt{6} \omega +26082 \sqrt{6} \omega ^2+6105 \sqrt{6} \omega ^3+935 \sqrt{6} \omega ^4\right) }{1403325}\delta ^4+\ldots~.
\end{eqnarray}

\subsection{Killing spinors}

Having found the above supersymmetric solutions we now proceed to solve the dilatino equation (\ref{dilatino}) and Killing spinor equation (\ref{KSE}) for the Killing spinors $\epsilon_I$, $I=1,2$. 

\subsubsection*{$3/4$ BPS solution}

\noindent For the $3/4$ to
\begin{eqnarray}
 \epsilon_1 \ &=& \ a^{(1)}_+  \, \ex^{\ii \frac{\tau}{2}} \left(\begin{array}{c}  k_2 (r) \left[ \cos \sigma + \ii \lambda_+ (s) \ex^{\ii \frac{\psi}{2}} S_{+}^{(1)} \sin \sigma \right] \\ 0 \\ \ii k_3 (r) \left[ \sin \sigma - \ii \lambda_+ (s) \ex^{\ii \frac{\psi}{2}} S_{+}^{(1)} \cos \sigma\right]\\ \ii k_3(r) \lambda_+ (s) \, \ex^{-\ii \frac{\psi}{2}} S_{+}^{(2)} \\ -\ii k_4(r)  \left[ \cos \sigma + \ii \lambda_+ (s) \ex^{\ii \frac{\psi}{2}} S_{+}^{(1)} \sin \sigma \right] \\ 0 \\ k_1(r) \left[ \sin \sigma -  \ii \lambda_+ (s) \ex^{\ii \frac{\psi}{2}} S_{+}^{(1)} \cos \sigma\right] \\ k_1(r) \,  \lambda_+ (s) \ex^{-\ii \frac{\psi}{2}} S_{+}^{(2)} \end{array}\right)~,\\
 \epsilon_2 \ &=& \ a^{(1)}_-  \, \ex^{-\ii \frac{\tau}{2}} \left(\begin{array}{c} 0 \\  \ii k_4 (r) \left[ \cos \sigma - \ii \lambda_- (s) \ex^{-\ii \frac{\psi}{2}} S_{-}^{(1)} \sin \sigma \right] \\ -  k_1(r) \lambda_- (s) \, \ex^{\ii \frac{\psi}{2}} S_{-}^{(2)} \\ k_1 (r) \left[ \sin \sigma + \ii \lambda_- (s) \ex^{-\ii \frac{\psi}{2}} S_{-}^{(1)} \cos \sigma\right] \\ 0 \\  k_2(r)  \left[ \cos \sigma - \ii \lambda_ (s) \ex^{-\ii \frac{\psi}{2}} S_{-}^{(1)} \sin \sigma \right] \\ \ii k_3(r) \,  \lambda_- (s) \ex^{\ii \frac{\psi}{2}} S_{-}^{(2)} \\  - \ii k_3(r) \left[ \sin \sigma + \ii \lambda_- (s) \ex^{-\ii \frac{\psi}{2}} S_{-}^{(1)} \cos \sigma\right] \\ \end{array}\right) \, ~,
\end{eqnarray}
where we have introduced
\bea
 S_{\pm}^{(1)} \ = \ S^{(1)}_{\pm}(\theta, \varphi) &= &a_\pm^{(3)}  \ex^{\pm \ii \frac{\varphi}{2}}\cos \frac{\theta}{2} - a_\pm^{(2)} \ex^{\mp \ii \frac{\varphi}{2}} \sin \frac{\theta}{2}~, \nn\\
S_{\pm}^{(2)}  \ = \ S^{(2)}_{\pm}(\theta, \varphi) &= & a^{(2)}_{\pm} \ex^{\mp \ii \frac{\varphi}{2}} \cos\frac{\theta}{2}+a_{\pm}^{(3)} \ex^{\pm \ii \frac{\varphi}{2}} \sin \frac{\theta}{2}~,\nn \\
\lambda_{\pm} (s) &=& \frac{\pm 1 + \sqrt{1-s^2}}{s}~.
\eea
The Killing spinors contain in total six constants of integration $a^{(i)}_{\pm},~i=1,2,3$. These constants of integration are generically complex, but imposing the symplectic Majorana condition $\mathcal{C}\epsilon_I^* = \varepsilon_I{}^{J}\epsilon_J$ enforces certain reality conditions. The functions $k_i(r)$ are functions of the radial coordinate only and can be expanded either around Euclidean AdS or around the boundary. For instance, expanding around the conformal boundary we obtain
\be
 \begin{aligned}
  k_1(r) &\ =\ \frac{-1+\sqrt{1-s^2}}{s}\sqrt{r} +\frac{1}{2 \sqrt{6}}\, \frac{1}{\sqrt{r}} +\ldots \, ,\\
  k_2(r) &\ =\ \sqrt{r} -\frac{5 \sqrt{1-s^2}-3}{6 \sqrt{6} s} \, \frac{1}{\sqrt{r}}+\ldots\, ,\\
  k_3(r) &\ =\ \frac{-1 + \sqrt{1-s^2}}{s} \sqrt{r}-\frac{1}{2 \sqrt{6}}\, \frac{1}{\sqrt{r}}+\ldots \, ,\\
  k_4(r) &\ =\ \sqrt{r} +\frac{5 \sqrt{1-s^2}-3}{6 \sqrt{6} s}\, \frac{1}{\sqrt{r}}+\ldots \, ,
 \end{aligned}
\ee
Notice that the expansion of the Killing spinor around the boundary is precisely of the form 
\begin{align}
\epsilon_I \ = \ \left(\begin{array}{c} \epsilon_I^+ \\ \epsilon_I^- \end{array}\right) \ = \ \sqrt{r} \left(\begin{array}{c} \chi_I \\ - \ii \chi_I \end{array}\right) + \frac{1}{\sqrt{r}} \left(\begin{array}{c} \varphi_I \\ \ii \varphi_I \end{array}\right) + \cdots\, ,
\end{align}
which arises from the general analysis of section \ref{SecSUSYatBoundary} and should of course hold for our particular solution. This allows us to immediately identify the boundary five-dimensional Killing spinor $\chi_I$ corresponding to our bulk solution. Note that this precisely agrees with (\ref{5dchip}).

\subsubsection*{$1/4$ BPS solution}

\noindent For the $1/4$ BPS solution we find
\be
\epsilon_1 \ = \ c_+  \, \ex^{-\frac{3 \ii \tau}{2}} \left(\begin{array}{c} 0\\  k_2 (r)  \\ 0 \\ 0 \\ 0 \\  -\ii \, k_1 (r) \\ 0 \\ 0 \end{array}\right)\, , \qquad
  \epsilon_2 \ = \ - c_-  \, \ex^{\frac{3 \ii \tau}{2}} \left(\begin{array}{c}  k_1 (r)  \\ 0 \\ 0 \\ 0 \\  - \ii \, k_2 (r) \\ 0 \\ 0 \\ 0 \end{array}\right)~.
\ee
The solution depends now on two constants of integration $c_\pm$. The functions of the radial coordinate admit the following expansion around the conformal boundary
\begin{align}
k_1(r)\ =\ &\sqrt{r} +\frac{(f_0-3) s}{6 \sqrt{6}} \frac{1}{\sqrt{r}}+ \frac{5 (f_0-3)^2 s^2+6 (4 f_0-9)}{432} \left(\frac{1}{r}\right)^{3/2}+\ldots~, \nn \\
k_2(r)\ = \ &\sqrt{r}-\frac{(f_0-3) s}{6 \sqrt{6}}\frac{1}{\sqrt{r}} + \frac{5 (f_0-3)^2 s^2 +6 (4 f_0-9) }{432} \left(\frac{1}{r}\right)^{3/2}+\ldots~.
\end{align}
As before, the corresponding Killing spinors at the boundary can be identified. In this case they are indeed of the form (\ref{5dKS1/4}), as expected. Finally, let us mention that the supersymmetry gets enhanced for the case $f_0=0$ (or equivalently $\omega=0$). In this limit the gauge field vanishes and so the two Killing spinors $\epsilon_I$ for $I=1,2$ decouple and have the same structure. They read
\be
\epsilon_I \ =\ \left(\begin{array}{c} c_I^{(2)}  \,   k_1 (r) \ex^{\frac{3 \ii \tau}{2}} \\  c_I^{(1)} \, k_2 (r)   \ex^{-\frac{3 \ii \tau}{2}}  \\ 0 \\ 0 \\ - \ii \, c_I^{(2)} \, k_2 (r) \ex^{\frac{3 \ii \tau}{2}}  \\  -\ii \, c_I^{(1)} \, k_1 (r)  \ex^{-\frac{3 \ii \tau}{2}}  \\ 0 \\ 0 \end{array}\right)\, ,
\ee
where $c_I^{(j)}$ for $j=1,2$ are the integration constants and where the $r$-dependent functions $k_i(r)$ are the same as in the 1/4 BPS case,  with $f_0= 0$. This solution may thus  be referred to as a $1/2$ BPS solution. 

\section{Asymptotics of multiple sine functions} \label{asymptotes}

Let us start by defining Barnes' multiple zeta function, 
\be
\zeta_\NN \left( s,w \mid \mathbf{a} \right) \ \equiv \  \sum_{m_1, \ldots, m_\NN = 0}^{\infty} \left( w+m_1 a_1 + \cdots m_\NN a_\NN \right)^{-s} \, ,
\ee
where $\mathbf{a}=(a_1,\ldots,a_\NN)$,
$\Real w>0$, $\Real s >\NN$ and $a_1, \ldots, a_\NN>0$. This function is meromorphic in $s$, with simple poles at $s=1, \ldots, \NN$. One can then define the Barnes multiple gamma function $\Gamma_{\NN}(w \mid\mathbf{a}) \equiv \exp \left[ \Psi_\NN\left( w \mid \mathbf{a} \right) \right]$, where
\be
\Psi_\NN\left( w \mid \mathbf{a} \right) \ \equiv \  \frac{\dd}{\dd s} \zeta_\NN \left( s,w \mid \mathbf{a} \right) \mid_{s=0} \ .
\ee
In order to compute the asymptotics of the multiple gamma function, and the closely related multiple sine function, we have to express this function in a more convenient way. In \cite{Ruijsenaars:2000}, it was observed that there is an expansion of $\Psi_\NN (w)$ of the form
\bea
 \Psi_\NN \left( w \mid  \mathbf{a} \right)  & = & \frac{(-1)^{\NN+1}}{\NN!} B_{\NN,\NN}(w) \log w +(-1)^{\NN} \sum_{k=0}^{\NN-1} \frac{B_{\NN,k}(0)w^{\NN-k}}{k! (\NN-k)!}\sum_{\ell=1}^{\NN-k} \frac{1}{\ell}  \nn \\ && +\sum_{k=\NN+1}^{\MM} \frac{(-1)^{k}}{k!}B_{\NN,k}\left( 0 \right) w^{\NN-k} (k-\NN-1)!+\mathcal{R}_{\NN,\MM} (w) \, , \label{psiexpansion}
\eea
where
\be
 \mathcal{R}_{\NN,\MM} (w)  \ \equiv \ \int_{0}^{\infty} \frac{\mathrm{d}t}{t} \ex^{-w t} \left( \prod_{j=1}^{\NN}\left( 1-\ex^{-a_j t} \right)^{-1}-\sum_{k=0}^{\MM}\frac{(-1)^{k}}{k!} B_{\NN,k}(0)t^{k-\NN} \right) \, ,
\ee
and $\MM\geq \NN$ as well as $\Real w>0$. The functions $B_{\NN,\MM}\left( w \right)$ are the so-called multiple Bernoulli polynomials and can be determined by expanding and solving the following relation 
\be
 \frac{t^{\NN} \ex^{x t}}{\prod^{\NN}_{j=1} \left( \ex^{a_j t} -1 \right)} \ = \ \sum_{n=0}^{\infty}\frac{t^{n}}{n!} B_{\NN,n}\left( x \right)~,
\ee
for $B_{\NN,\MM}\left( w \right)$. It was further shown in \cite{Ruijsenaars:2000} that in the asymptotic limit $|w|\rightarrow \infty$ and $|\arg w| < \pi$ the remainder $\mathcal{R}_{\NN,\MM} (w) $ behaves as
\be
 \mathcal{R}_{\NN,\MM} (w) \ = \ \mathcal{O} \left( w^{\NN-\MM-1} \right) \, ,
\ee
and hence  in the asymptotic limit is suppressed by the first three terms in \eqref{psiexpansion}. Similarly, the third term in \eqref{psiexpansion} behaves as
\be
\sum_{k=\NN+1}^{\MM} \frac{(-1)^{k}}{k!}B_{\NN,k}\left( 0 \right) w^{\NN-k} (k-\NN-1)! \ = \ \mathcal{O} \left( w^{-1} \right)~,
\ee
in the asymptotic limit $|w|\rightarrow \infty$. Hence for our purposes we shall only focus on the asymptotics of the first two contributions to $\Psi_\NN$.

We are interested in the asymptotic expansion of the so-called multiple sine function, which is defined in terms of the Gamma function as
\be
 S_\NN (w \mid \mathbf{a}) \ \equiv \ \Gamma_{\NN}(w \mid \mathbf{a})^{-1} \ \Gamma_{\NN} (a_{\mathrm{tot}} - w \mid \mathbf{a})^{(-1)^{\NN}} \, ,
\ee
where $a_\mathrm{tot} = \sum_{i=1}^{\NN} a_i$. To compute the large $N$ limit of the free energy, we are interested in the asymptotics of the logarithm of these functions
\be
\log S_\NN (w \mid\mathbf{a})\ = \ - \Psi_{\NN}(w \mid \mathbf{a})  - \Psi_{\NN} (a_{\mathrm{tot}} - w \mid \mathbf{a})^{(-1)^{\NN}} \, .
\ee
Focusing on the case $\NN=3$, we find the following Bernoulli polynomials
\bea
B_{3,0}(x)&=&\frac{1}{a_1 a_2 a_3}~,\nn\\
B_{3,1}(x)&=&\frac{x}{a_1 a_2 a_3} -\frac{a_{\mathrm{tot}}}{2 a_1 a_2 a_3}~,\nn\\
B_{3,2}(x)&=&\frac{x^2}{a_1 a_2 a_3}-\frac{a_{\mathrm{tot}}}{a_1 a_2 a_3} x+\frac{a_{\mathrm{tot}}^2+ \left(a_1 a_2+a_1 a_3 +a_2 a_3\right)}{6 a_1 a_2 a_3}~,\nn\\
B_{3,3}(x)&=&\frac{x^3}{a_1 a_2 a_3}-\frac{3 a_{\mathrm{tot}}}{2 a_1 a_2 a_3}x^2+\frac{a_{\mathrm{tot}}^2+\left(a_1 a_2+a_1 a_3 +a_2 a_3\right)}{6 a_1 a_2 a_3} x\nn\\&&-\frac{a_{\mathrm{tot}}\left(a_1 a_2+a_1 a_3+a_2 a_3\right)}{4 a_1 a_2 a_3}~.
\eea
We can then compute \eqref{psiexpansion} and take the asymptotic limit of the logarithm of the triple sine function to obtain
\bea
\log S_3 (w \mid \mathbf{a}) & =  & \mathrm{sign}\, \Real w \Bigg[ \frac{\ii \pi}{6 a_1 a_2 a_3} w^{3} - \frac{\ii \pi a_{\mathrm{tot}}}{4 a_1 a_2 a_3} w^{2}
+ \frac{\ii \pi \left( a_{\mathrm{tot}}^{2}+a_1 a_2 + a_1 a_3 +a_2 a_3 \right)}{12 a_1 a_2 a_3}\nn\\&&  -\frac{\ii \pi a_{\mathrm{tot}} \left( a_1 a_2 + a_1 a_3 +a_2 a_3  \right)}{24 a_1 a_2 a_3}+ \mathcal{O}\left( w^{-1} \right)\Bigg] \ .
\eea
This procedure generalizes to any choice of $\NN$, and gives a straightforward method to obtain the asymptotics of these functions.

\bibliography{S5bDualrevised}

\providecommand{\href}[2]{#2}\begingroup\raggedright\begin{thebibliography}{10}

\bibitem{Kallen:2012cs}
J.~K{\"a}ll{\'e}n and M.~Zabzine, ``{Twisted supersymmetric 5D Yang-Mills
  theory and contact geometry},''
  \href{http://dx.doi.org/10.1007/JHEP05(2012)125}{{\em JHEP} {\bfseries 1205}
  (2012) 125},
\href{http://arxiv.org/abs/1202.1956}{{\ttfamily arXiv:1202.1956 [hep-th]}}.

\bibitem{Hosomichi:2012ek}
K.~Hosomichi, R.-K. Seong, and S.~Terashima, ``{Supersymmetric Gauge Theories
  on the Five-Sphere},''
  \href{http://dx.doi.org/10.1016/j.nuclphysb.2012.08.007}{{\em Nucl.Phys.}
  {\bfseries B865} (2012) 376--396},
\href{http://arxiv.org/abs/1203.0371}{{\ttfamily arXiv:1203.0371 [hep-th]}}.

\bibitem{Kallen:2012va}
J.~K{\"a}ll{\'e}n, J.~Qiu, and M.~Zabzine, ``{The perturbative partition
  function of supersymmetric 5D Yang-Mills theory with matter on the
  five-sphere},'' \href{http://dx.doi.org/10.1007/JHEP08(2012)157}{{\em JHEP}
  {\bfseries 1208} (2012) 157},
\href{http://arxiv.org/abs/1206.6008}{{\ttfamily arXiv:1206.6008 [hep-th]}}.

\bibitem{Kim:2012ava}
H.-C. Kim and S.~Kim, ``{M5-branes from gauge theories on the 5-sphere},''
  \href{http://dx.doi.org/10.1007/JHEP05(2013)144}{{\em JHEP} {\bfseries 1305}
  (2013) 144},
\href{http://arxiv.org/abs/1206.6339}{{\ttfamily arXiv:1206.6339 [hep-th]}}.

\bibitem{Jafferis:2012iv}
D.~L. Jafferis and S.~S. Pufu, ``{Exact results for five-dimensional
  superconformal field theories with gravity duals},''
  \href{http://dx.doi.org/10.1007/JHEP05(2014)032}{{\em JHEP} {\bfseries 1405}
  (2014) 032},
\href{http://arxiv.org/abs/1207.4359}{{\ttfamily arXiv:1207.4359 [hep-th]}}.

\bibitem{Imamura:2012xg}
Y.~Imamura, ``{Supersymmetric theories on squashed five-sphere},''
  \href{http://dx.doi.org/10.1093/ptep/pts052}{{\em PTEP} {\bfseries 2013}
  (2013) 013B04},
\href{http://arxiv.org/abs/1209.0561}{{\ttfamily arXiv:1209.0561 [hep-th]}}.

\bibitem{Imamura:2012bm}
Y.~Imamura, ``{Perturbative partition function for squashed $S^5$},''
\href{http://arxiv.org/abs/1210.6308}{{\ttfamily arXiv:1210.6308 [hep-th]}}.

\bibitem{Ferrara:1998gv}
S.~Ferrara, A.~Kehagias, H.~Partouche, and A.~Zaffaroni, ``{AdS(6)
  interpretation of 5-D superconformal field theories},''
  \href{http://dx.doi.org/10.1016/S0370-2693(98)00560-7}{{\em Phys.Lett.}
  {\bfseries B431} (1998) 57--62},
\href{http://arxiv.org/abs/hep-th/9804006}{{\ttfamily arXiv:hep-th/9804006
  [hep-th]}}.

\bibitem{Brandhuber:1999np}
A.~Brandhuber and Y.~Oz, ``{The D-4 - D-8 brane system and five-dimensional
  fixed points},'' \href{http://dx.doi.org/10.1016/S0370-2693(99)00763-7}{{\em
  Phys.Lett.} {\bfseries B460} (1999) 307--312},
\href{http://arxiv.org/abs/hep-th/9905148}{{\ttfamily arXiv:hep-th/9905148
  [hep-th]}}.

\bibitem{Bergman:2012kr}
O.~Bergman and D.~Rodriguez-Gomez, ``{5d quivers and their AdS(6) duals},''
  \href{http://dx.doi.org/10.1007/JHEP07(2012)171}{{\em JHEP} {\bfseries 1207}
  (2012) 171},
\href{http://arxiv.org/abs/1206.3503}{{\ttfamily arXiv:1206.3503 [hep-th]}}.

\bibitem{Romans:1985tw}
L.~Romans, ``{The F(4) Gauged Supergravity in Six-dimensions},''
\href{http://dx.doi.org/10.1016/0550-3213(86)90517-1}{{\em Nucl.Phys.}
  {\bfseries B269} (1986) 691}.

\bibitem{Cvetic:1999un}
M.~Cvetic, H.~Lu, and C.~Pope, ``{Gauged six-dimensional supergravity from
  massive type IIA},''
  \href{http://dx.doi.org/10.1103/PhysRevLett.83.5226}{{\em Phys.Rev.Lett.}
  {\bfseries 83} (1999) 5226--5229},
\href{http://arxiv.org/abs/hep-th/9906221}{{\ttfamily arXiv:hep-th/9906221
  [hep-th]}}.

\bibitem{Farquet:2014kma}
D.~Farquet, J.~Lorenzen, D.~Martelli, and J.~Sparks, ``{Gravity duals of
  supersymmetric gauge theories on three-manifolds},''
\href{http://arxiv.org/abs/1404.0268}{{\ttfamily arXiv:1404.0268 [hep-th]}}.

\bibitem{Lockhart:2012vp}
G.~Lockhart and C.~Vafa, ``{Superconformal Partition Functions and
  Non-perturbative Topological Strings},''
\href{http://arxiv.org/abs/1210.5909}{{\ttfamily arXiv:1210.5909 [hep-th]}}.

\bibitem{Qiu:2013pta}
J.~Qiu and M.~Zabzine, ``{5D Super Yang-Mills on $Y^{p,q}$ Sasaki-Einstein
  manifolds},''
\href{http://arxiv.org/abs/1307.3149}{{\ttfamily arXiv:1307.3149}}.

\bibitem{Herzog:2010hf}
C.~P. Herzog, I.~R. Klebanov, S.~S. Pufu, and T.~Tesileanu, ``{Multi-Matrix
  Models and Tri-Sasaki Einstein Spaces},''
  \href{http://dx.doi.org/10.1103/PhysRevD.83.046001}{{\em Phys.Rev.}
  {\bfseries D83} (2011) 046001},
\href{http://arxiv.org/abs/1011.5487}{{\ttfamily arXiv:1011.5487 [hep-th]}}.

\bibitem{Martelli:2012sz}
D.~Martelli, A.~Passias, and J.~Sparks, ``{The supersymmetric NUTs and bolts of
  holography},'' \href{http://dx.doi.org/10.1016/j.nuclphysb.2013.04.026}{{\em
  Nucl.Phys.} {\bfseries B876} (2013) 810--870},
\href{http://arxiv.org/abs/1212.4618}{{\ttfamily arXiv:1212.4618 [hep-th]}}.

\bibitem{2007arXiv0704.3373A}
M.~T. {Anderson}, ``{Extension of symmetries on Einstein manifolds with
  boundary},'' {\em ArXiv e-prints} (Apr., 2007) ,
  \href{http://arxiv.org/abs/0704.3373}{{\ttfamily arXiv:0704.3373 [math.DG]}}.

\bibitem{Emparan:1999pm}
R.~Emparan, C.~V. Johnson, and R.~C. Myers, ``{Surface terms as counterterms in
  the AdS / CFT correspondence},''
  \href{http://dx.doi.org/10.1103/PhysRevD.60.104001}{{\em Phys.Rev.}
  {\bfseries D60} (1999) 104001},
\href{http://arxiv.org/abs/hep-th/9903238}{{\ttfamily arXiv:hep-th/9903238
  [hep-th]}}.

\bibitem{deHaro2000xn}
S.~de~Haro, S.~N. Solodukhin, and K.~Skenderis, ``Holographic reconstruction of
  space-time and renormalization in the ads / cft correspondence,'' {\em
  Commun.Math.Phys.} {\bfseries 217} (2001) 595--622.

\bibitem{Taylor:2000xw}
M.~Taylor, ``{More on counterterms in the gravitational action and
  anomalies},''
\href{http://arxiv.org/abs/hep-th/0002125}{{\ttfamily arXiv:hep-th/0002125
  [hep-th]}}.

\bibitem{Gibbons:1976ue}
G.~Gibbons and S.~Hawking, ``{Action Integrals and Partition Functions in
  Quantum Gravity},''
\href{http://dx.doi.org/10.1103/PhysRevD.15.2752}{{\em Phys.Rev.} {\bfseries
  D15} (1977) 2752--2756}.

\bibitem{Alday:2014rxa}
L.~F. Alday, M.~Fluder, P.~Richmond, and J.~Sparks, ``{The gravity dual of
  supersymmetric gauge theories on a squashed five-sphere},''
\href{http://arxiv.org/abs/1404.1925}{{\ttfamily arXiv:1404.1925 [hep-th]}}.

\bibitem{Fefferman}
C.~Fefferman and R.~Graham, ``Conformal invariants,'' {\em In \'Elie Cartan et
  les Math\'ematiques d'aujourd'hui, Ast\'erisque (1985), 95.} .

\bibitem{2007arXiv0710.0919F}
C.~{Fefferman} and C.~R. {Graham}, ``{The ambient metric},'' {\em ArXiv
  e-prints} (Oct., 2007) , \href{http://arxiv.org/abs/0710.0919}{{\ttfamily
  arXiv:0710.0919 [math.DG]}}.

\bibitem{Balasubramanian:1999re}
V.~Balasubramanian and P.~Kraus, ``{A Stress tensor for Anti-de Sitter
  gravity},'' \href{http://dx.doi.org/10.1007/s002200050764}{{\em
  Commun.Math.Phys.} {\bfseries 208} (1999) 413--428},
\href{http://arxiv.org/abs/hep-th/9902121}{{\ttfamily arXiv:hep-th/9902121
  [hep-th]}}.

\bibitem{Nishimura:2000wj}
M.~Nishimura, ``{Conformal supergravity from the AdS / CFT correspondence},''
  \href{http://dx.doi.org/10.1016/S0550-3213(00)00472-7}{{\em Nucl.Phys.}
  {\bfseries B588} (2000) 471--482},
\href{http://arxiv.org/abs/hep-th/0004179}{{\ttfamily arXiv:hep-th/0004179
  [hep-th]}}.

\bibitem{Klare:2012gn}
C.~Klare, A.~Tomasiello, and A.~Zaffaroni, ``{Supersymmetry on Curved Spaces
  and Holography},'' \href{http://dx.doi.org/10.1007/JHEP08(2012)061}{{\em
  JHEP} {\bfseries 1208} (2012) 061},
\href{http://arxiv.org/abs/1205.1062}{{\ttfamily arXiv:1205.1062 [hep-th]}}.

\bibitem{Hristov:2013spa}
K.~Hristov, A.~Tomasiello, and A.~Zaffaroni, ``{Supersymmetry on
  Three-dimensional Lorentzian Curved Spaces and Black Hole Holography},''
  \href{http://dx.doi.org/10.1007/JHEP05(2013)057}{{\em JHEP} {\bfseries 1305}
  (2013) 057},
\href{http://arxiv.org/abs/1302.5228}{{\ttfamily arXiv:1302.5228 [hep-th]}}.

\bibitem{Cassani:2013dba}
D.~Cassani and D.~Martelli, ``{Supersymmetry on curved spaces and
  superconformal anomalies},''
  \href{http://dx.doi.org/10.1007/JHEP10(2013)025}{{\em JHEP} {\bfseries 1310}
  (2013) 025},
\href{http://arxiv.org/abs/1307.6567}{{\ttfamily arXiv:1307.6567 [hep-th]}}.

\bibitem{Klare:2013dka}
C.~Klare and A.~Zaffaroni, ``{Extended Supersymmetry on Curved Spaces},''
  \href{http://dx.doi.org/10.1007/JHEP10(2013)218}{{\em JHEP} {\bfseries 1310}
  (2013) 218},
\href{http://arxiv.org/abs/1308.1102}{{\ttfamily arXiv:1308.1102 [hep-th]}}.

\bibitem{2014arXiv1403.2311L}
A.~{Lischewski}, ``{Charged Conformal Killing Spinors},'' {\em ArXiv e-prints}
  (Mar., 2014) , \href{http://arxiv.org/abs/1403.2311}{{\ttfamily
  arXiv:1403.2311 [math.DG]}}.

\bibitem{Zucker:1999ej}
M.~Zucker, ``{Minimal off-shell supergravity in five-dimensions},''
  \href{http://dx.doi.org/10.1016/S0550-3213(99)00750-6}{{\em Nucl.Phys.}
  {\bfseries B570} (2000) 267--283},
\href{http://arxiv.org/abs/hep-th/9907082}{{\ttfamily arXiv:hep-th/9907082
  [hep-th]}}.

\bibitem{Kugo:2000hn}
T.~Kugo and K.~Ohashi, ``{Supergravity tensor calculus in 5-D from 6-D},''
  \href{http://dx.doi.org/10.1143/PTP.104.835}{{\em Prog.Theor.Phys.}
  {\bfseries 104} (2000) 835--865},
\href{http://arxiv.org/abs/hep-ph/0006231}{{\ttfamily arXiv:hep-ph/0006231
  [hep-ph]}}.

\bibitem{Pan:2013uoa}
Y.~Pan, ``{Rigid Supersymmetry on 5-dimensional Riemannian Manifolds and
  Contact Geometry},'' \href{http://dx.doi.org/10.1007/JHEP05(2014)041}{{\em
  JHEP} {\bfseries 1405} (2014) 041},
\href{http://arxiv.org/abs/1308.1567}{{\ttfamily arXiv:1308.1567 [hep-th]}}.

\bibitem{Imamura:2014ima}
Y.~Imamura and H.~Matsuno, ``{Supersymmetric backgrounds from 5d $\mathcal N=$
  1 supergravity},'' \href{http://dx.doi.org/10.1007/JHEP07(2014)055}{{\em
  JHEP} {\bfseries 1407} (2014) 055},
\href{http://arxiv.org/abs/1404.0210}{{\ttfamily arXiv:1404.0210 [hep-th]}}.

\bibitem{Assel:2012nf}
B.~Assel, J.~Estes, and M.~Yamazaki, ``{Wilson Loops in 5d N=1 SCFTs and
  AdS/CFT},'' \href{http://dx.doi.org/10.1007/s00023-013-0249-5}{{\em Annales
  Henri Poincare} {\bfseries 15} (2014) 589--632},
\href{http://arxiv.org/abs/1212.1202}{{\ttfamily arXiv:1212.1202 [hep-th]}}.

\bibitem{Alday:2013lba}
L.~F. Alday, D.~Martelli, P.~Richmond, and J.~Sparks, ``{Localization on
  Three-Manifolds},'' \href{http://dx.doi.org/10.1007/JHEP10(2013)095}{{\em
  JHEP} {\bfseries 1310} (2013) 095},
\href{http://arxiv.org/abs/1307.6848}{{\ttfamily arXiv:1307.6848 [hep-th]}}.

\bibitem{Farquet:2014bda}
D.~Farquet and J.~Sparks, ``{Wilson loops on three-manifolds and their M2-brane
  duals},''
\href{http://arxiv.org/abs/1406.2493}{{\ttfamily arXiv:1406.2493 [hep-th]}}.

\bibitem{Closset:2012ru}
C.~Closset, T.~T. Dumitrescu, G.~Festuccia, and Z.~Komargodski,
  ``{Supersymmetric Field Theories on Three-Manifolds},''
  \href{http://dx.doi.org/10.1007/JHEP05(2013)017}{{\em JHEP} {\bfseries 1305}
  (2013) 017},
\href{http://arxiv.org/abs/1212.3388}{{\ttfamily arXiv:1212.3388 [hep-th]}}.

\bibitem{Closset:2013vra}
C.~Closset, T.~T. Dumitrescu, G.~Festuccia, and Z.~Komargodski, ``{The Geometry
  of Supersymmetric Partition Functions},''
  \href{http://dx.doi.org/10.1007/JHEP01(2014)124}{{\em JHEP} {\bfseries 1401}
  (2014) 124},
\href{http://arxiv.org/abs/1309.5876}{{\ttfamily arXiv:1309.5876 [hep-th]}}.

\bibitem{Ruijsenaars:2000}
S.~N.~M. Ruijsenaars, ``On barnes' multiple zeta and gamma functions,'' {\em
  Advances in Mathematics} {\bfseries 156} (2000) 107--132.

\end{thebibliography}\endgroup

\end{document}